\documentclass[aps,prx,onecolumn,superscriptaddress,11pt]{revtex4-2} 
\usepackage{graphicx}
\usepackage{units}
\usepackage{amsmath}
\usepackage{amssymb}
\usepackage{braket}
\usepackage[colorlinks=true, linkcolor=black, citecolor=black, urlcolor=black]{hyperref}
\usepackage[T1]{fontenc}

\newcommand{\spu}{\uparrow} 
\newcommand{\spd}{\downarrow}
\newcommand{\Jexp}{J_\parallel}
\newcommand{\eps}{\epsilon}

\begin{document}

\title{Two-qubit logic with anisotropic exchange in a fin field-effect transistor}

\author{Simon Geyer}
\email{e-mail: \mbox{simon.geyer@unibas.ch; dominik.zumbuhl@unibas.ch; andreas.kuhlmann@unibas.ch}}
\affiliation{Department of Physics, University of Basel, Klingelbergstrasse 82, CH-4056 Basel, Switzerland}

\author{Bence Hetényi}
\affiliation{Department of Physics, University of Basel, Klingelbergstrasse 82, CH-4056 Basel, Switzerland}
\affiliation{IBM Research Europe-Zurich, S\"aumerstrasse 4, CH-8803 R\"uschlikon, Switzerland}

\author{Stefano Bosco}

\author{Leon C. Camenzind}
\altaffiliation{Current address: RIKEN, Center for Emergent Matter Science (CEMS), Wako-shi, Saitama 351-0198, Japan}

\author{Rafael S. Eggli}
\affiliation{Department of Physics, University of Basel, Klingelbergstrasse 82, CH-4056 Basel, Switzerland}

\author{Andreas Fuhrer} 
\affiliation{IBM Research Europe-Zurich, S\"aumerstrasse 4, CH-8803 R\"uschlikon, Switzerland}

\author{Daniel Loss}

\author{Richard J. Warburton}

\author{Dominik M. Zumb\"uhl}
\email{e-mail: \mbox{simon.geyer@unibas.ch; dominik.zumbuhl@unibas.ch; andreas.kuhlmann@unibas.ch}}

\author{Andreas V. Kuhlmann}
\email{e-mail: \mbox{simon.geyer@unibas.ch; dominik.zumbuhl@unibas.ch; andreas.kuhlmann@unibas.ch}}
\affiliation{Department of Physics, University of Basel, Klingelbergstrasse 82, CH-4056 Basel, Switzerland}

\begin{abstract}
Semiconductor spin qubits offer a unique opportunity for scalable quantum computation by leveraging classical transistor technology. Hole spin qubits benefit from fast all-electrical qubit control and sweet spots to counteract charge and nuclear spin noise. The demonstration of a two-qubit quantum gate in a silicon fin field-effect transistor, that is, the workhorse device of today’s semiconductor industry, has remained an open challenge. Here, we demonstrate a controlled rotation two-qubit gate on hole spins in an industry-compatible device. A short gate time of \unit[24]{ns} is achieved. The quantum logic exploits an exchange interaction that can be tuned from above \unit[500]{MHz} to close-to-off. Significantly, the exchange is strikingly anisotropic. By developing a general theory, we show that the anisotropy arises as a consequence of a strong spin-orbit interaction. Upon tunnelling from one quantum dot to the other, the spin is rotated by almost \unit[90]{degrees}. The exchange Hamiltonian no longer has Heisenberg form and is engineered in such a way that there is no trade-off between speed and fidelity of the two-qubit gate. This ideal behaviour applies over a wide range of magnetic field orientations rendering the concept robust with respect to variations from qubit to qubit. Our work brings hole spin qubits in silicon transistors a step closer to the realization of a large-scale quantum computer. 
\end{abstract}

\date{\today}

\maketitle

Semiconductor quantum dot (QD) spin qubits are prime candidates for future implementations of large-scale quantum circuits \cite{Loss1998,Vandersypen2017,Veldhorst2017}. Currently, the most advanced spin-based quantum processor allows for universal control of six electron spin qubits in silicon (Si) \cite{Philips2022}, closely followed by a four-qubit demonstration with holes in germanium \cite{Hendrickx2021}. In comparison to electron spins, hole spins have the advantage that they can be controlled all-electrically, without the added complexity of on-chip micromagnets \cite{Tokura2006,PioroLadriere2008}, thanks to their intrinsic spin-orbit interaction (SOI). Moreover, holes benefit from a reduced hyperfine interaction \cite{Prechtel2016} and the absence of valleys \cite{Zwanenburg2013}. 

Holes in quasi-one-dimensional (1D) nanostructures are highly attractive for implementing fast and coherent qubits. The mixing of heavy- and light-hole states on account of the 1D-confinement results in an unusually strong and electrically tunable direct Rashba spin-orbit interaction (DRSOI) with sweet spots for charge and hyperfine noise \cite{Kloeffel2018,Bosco2021,Bosco2021a}, enabling ultra-fast hole spin qubits \cite{Froning2021,Wang2022} hidden from the noise \cite{Piot2022}. Conveniently, such a 1D-system can be realized using today's industry standard transistor design known as the fin field-effect transistor (FinFET) \cite{Auth2012}. Adapting FinFETs for QD integration \cite{Maurand2016,Piot2022,Kuhlmann2018,Geyer2021,Zwerver2022,Camenzind2022} may potentially facilitate quantum computer scale-up by leveraging decades of technology development in the semiconductor industry \cite{GonzalezZalba2021}. Furthermore, it has recently been shown that hole spin qubits in a bulk-Si FinFET can be operated at temperatures above \unit[4]{K} \cite{Camenzind2022}, paving the way for FinFET-based quantum integrated circuits that host both the qubit array and its classical control electronics on the same chip \cite{Petit2020,Yang2020,Xue2021}.

Universal quantum computation requires both single-qubit control and two-qubit interactions. Native two-qubit gates for spins such as the $\sqrt{\mathrm{SWAP}}$ \cite{Loss1998,Petta2005}, the controlled phase (CPHASE) \cite{Veldhorst2015,Watson2018,Mills2022,Xue2022} or the controlled rotation (CROT) \cite{Zajac2018,Watson2018,Huang2019,Petit2020,Noiri2022,Philips2022} rely on the exchange interaction, which arises from the wavefunction overlap between two adjacent QDs. For electrons in Si, two-qubit gate fidelities have recently surpassed the fault-tolerance threshold of 99\% \cite{Noiri2022,Xue2022,Mills2022}, but for holes in Si or FinFETs the demonstration of two-qubit logic is still missing due to the challenges in obtaining a controllable exchange interaction \cite{Fang2022}. 

We make this important step towards a FinFET-based quantum processor by demonstrating a CROT for holes in a Si FinFET. The strong SOI in combination with a large and highly tunable exchange splitting enables the execution of a controlled spin-flip in just $\unit[{\simeq}\,{24}]{ns}$. While the exchange interaction is crucial for implementing high-fidelity two-qubit gates, it is, in particular for hole spins, still largely unexplored. We measure the dependence of the exchange splitting on the magnetic field direction and find large values in some directions, close-to-zero values in other directions. In addition, we develop a general theoretical framework, applicable to a wide range of devices, and identify the SOI as the main reason for the exchange anisotropy. From our measurements we can extract the full exchange matrix and hence accurately determine the Hamiltonian of the two coupled spins, allowing us to predict the optimum operating points for the gates. For holes unlike electrons, the strong exchange anisotropy facilitates CROTs with both high fidelity and high speed for an experimental setting that is robust against device variations. 

\begin{figure*}[t]
\centering
\includegraphics[width=\columnwidth]{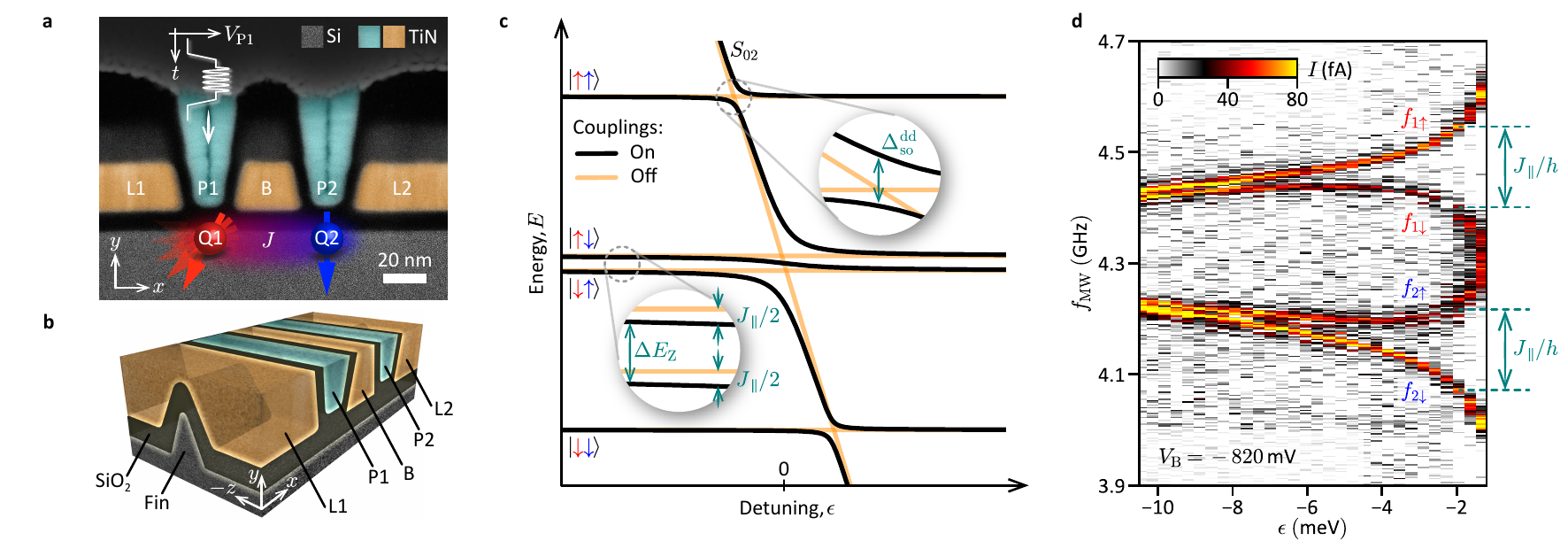}
\caption{\textbf{Two-qubit system in a Si FinFET.} \textbf{a}, False-colour transmission electron microscope image of a co-fabricated device showing the cross-section along the fin. The qubits (Q1, Q2) are located underneath the plunger gates (P1, P2) and are manipulated by applying microwaves to the P1-gate. The barrier gate (B) controls the inter-dot tunnelling; the lead gates (L1, L2) accumulate the hole reservoirs. Measurements are performed on a device with $\unit[{\simeq}\,{20}]{nm}$-wide B- and P-gates. \textbf{b}, A 3D render of the device illustrating the triangular-shaped fin covered by the wrap-around gates. \textbf{c}, Two-spin energy-level diagram close to the (1,1)-(0,2) charge transition with (black) and without (orange) interactions. The singlet state $S_{02}$ hybridizes with the antiparallel (parallel) two-spin states on account of spin-conserving tunnelling (SOI). A finite exchange splitting $\Jexp$ lowers the energy of the antiparallel two-spin states with respect to the parallel ones. \textbf{d}, Exchange spin funnel measurements for both qubits, revealing an increase (decrease) in $f_{1\uparrow}$, $f_{2\uparrow}$ ($f_{1\downarrow}$, $f_{2\downarrow}$) at the upper (lower) branch. Data was taken at $V_\mathrm{B}=\unit[-820]{mV}$ and $|\mathbf{B}|=\unit[0.146]{T}$ with orientation $\alpha=30^\circ$, $\beta=0^\circ$.}
\label{fig1}
\end{figure*}

Fig.\ \ref{fig1}a shows the device cross-section along the triangular-shaped fin, revealing ultrashort lengths, highly uniform profiles and perfect alignment of the gate electrodes \cite{Kuhlmann2018,Geyer2021}; Fig.\ \ref{fig1}b presents a 3D illustration of the device. The double quantum dot (DQD) hosting qubits Q1 and Q2 is formed beneath plunger gates P1 and P2, and the barrier gate B provides control over the inter-dot tunnel coupling $t_c$ \cite{Camenzind2022}. The distance between the QDs was chosen to match the spin-orbit length \cite{Camenzind2022,Geyer2021}. Taking advantage of the strong SOI, all-electrical spin control is implemented by electric-dipole spin resonance (EDSR) \cite{Golovach2006, Nowack2007}. For this purpose, fast voltage pulses and microwave (MW) bursts are applied to P1 and a spin-flip is detected in the form of an increased spin blockade leakage current. The device is tuned close to the (1,1)-(0,2) charge transition, where ($n$,$m$) denotes a state with $n$ ($m$) excess holes on the left (right) QD. In Fig.\ \ref{fig1}c the eigenenergies of the two-spin states ($\ket{\uparrow\uparrow}$, $\ket{\uparrow\downarrow}$, $\ket{\downarrow\uparrow}$, $\ket{\downarrow\downarrow}$) in the (1,1) and the singlet ground state $S_{02}$ in the (0,2) charge region are plotted as a function of the detuning $\epsilon$, which describes the energy difference between the (1,1) and (0,2) charge states. While spin-conserving tunnelling causes an anticrossing between the $S_{02}$ and the antiparallel two-spin states, spin-non-conserving tunnelling on account of the SOI results in an anticrossing between the $S_{02}$ and the parallel two-spin states. As a consequence of the anticrossing with the singlet state, the energy of the antiparallel states decreases by $\Jexp(\epsilon)/2$, where $\Jexp(\epsilon)$ is the exchange coupling between the two spins. The energy level structure of the two-hole system can be probed by performing MW spectroscopy (Fig.\ \ref{fig1}d): at large negative $\epsilon$, the resonance frequencies of both qubits differ due to the individual $g$-tensor $g_i$ for each QD, and are independent of each other. At more positive detunings, closer to the (0,2) region, the exchange interaction splits both resonances by $\Jexp/h$, resulting in four conditional transitions. The corresponding EDSR frequencies are denoted by $f_{i\sigma}$, where $i$ is the index of the target qubit and $\sigma$ the control qubit state $\ket{\uparrow}$ or $\ket{\downarrow}$.

\begin{figure}[t]
\centering
\includegraphics[width=0.5\columnwidth]{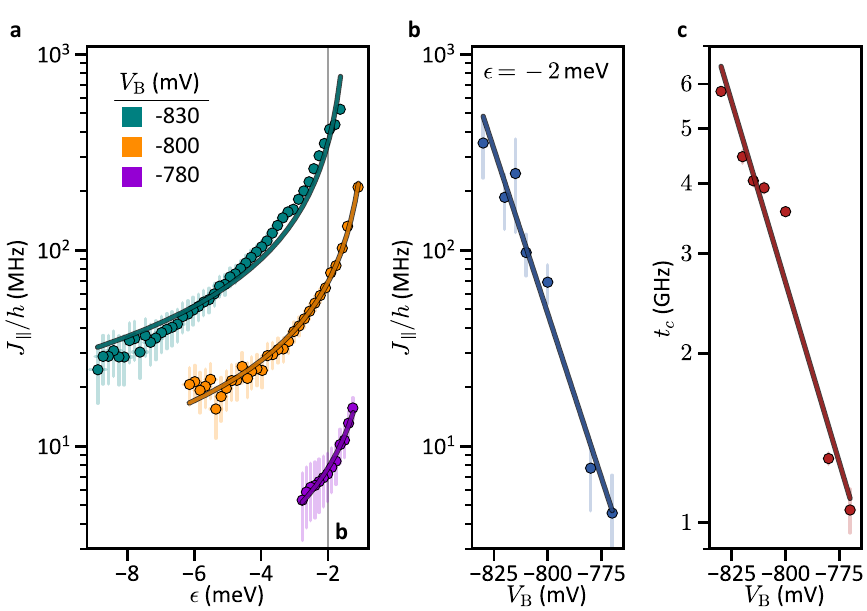}
\caption{\textbf{Tunable exchange coupling.} \textbf{a}, Detuning dependence of the exchange frequency for ${V_\mathrm{B}}\,{=}\,{-830}$, $-800$ and $\unit[-780]{mV}$. The solid curves represent fits to Eq.\ (\ref{eq:stepanenko_approx}) and errors the width of the EDSR resonance. \textbf{b}, $\Jexp/h$ determined for $\eps= \unit[-2]{meV}$ and \textbf{c} fitted tunnel coupling as a function of $V_\mathrm{B}$. The solid lines show exponential function fits to the data. The error bars represent in \textbf{b} the estimated errors due to a detuning uncertainty, and in \textbf{c} the standard errors for the best-fit values.}
\label{fig2}
\end{figure}

We map out the $\epsilon$-dependence of $\Jexp$ that, as shown in Fig.\ \ref{fig2}a, is well described by
\begin{equation}
\label{eq:stepanenko_approx}
    \Jexp = J_0 \cos(2\tilde \theta) =  \frac{2t_c^2}{U_0-\eps} \cos(2\tilde \theta),
\end{equation} 
valid in the limit of $t_c \ll U_0-\eps$ \cite{Stepanenko2012, Russ2018, Hendrickx2020a}. Here $U_0$ is an energy offset of the $\eps$-axis, $J_0$ the bare exchange, and $\cos(2\tilde{\theta})$ a SOI-induced correction factor, which is independent of detuning and discussed later. The exchange splitting shows an exponential dependence on the barrier gate voltage $V_\mathrm{B}$ (Fig.\ \ref{fig2}b) and reaches values of up to $\unit[{\simeq}\,{525}]{MHz}$. At the same time, exchange can be turned off within the resolution limit of our spectroscopy experiment that is given by the EDSR linewidth of $\unit[{\simeq}\,{2}]{MHz}$ \cite{Huang2019,Watson2018,Hendrickx2020a}. This means, using the two control knobs $\epsilon$ and $V_\mathrm{B}$, we achieve excellent control over the exchange coupling. Since $t_c \propto \Jexp^{1/2}$ the tunnel coupling is also exponentially dependent on $V_\mathrm{B}$ and tunable by almost one order of magnitude (Fig.\ \ref{fig2}c). 

\begin{figure*}[t]
\centering
\includegraphics[width=\columnwidth]{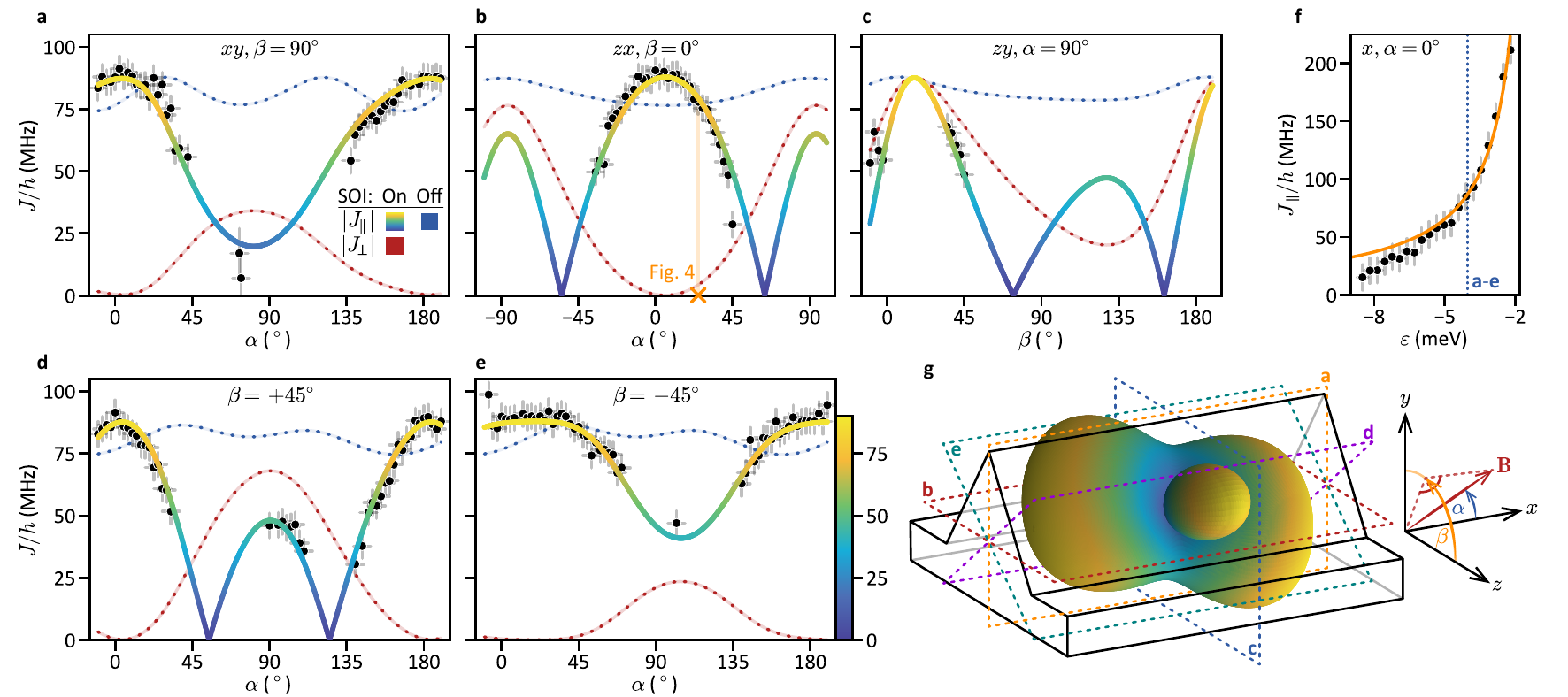}
\caption{\textbf{Anisotropic exchange.} \textbf{a-e}, Exchange frequency as a function of magnetic field direction, which is expressed with the angles $\alpha$ and $\beta$ (see coordinate system in \textbf{g}), for five different planes at $\eps=\unit[-4.03]{meV}$. For certain $\mathbf{B}$-orientations the qubits could not be read out via Pauli spin blockade and hence $\Jexp/h$ (black points) could not be determined. \textbf{f,} Detuning dependence of $\Jexp/h$ for $\mathbf{B}$ applied in $x$-direction. The multicolored curves in \textbf{a}-\textbf{e} and the orange one in \textbf{f} represent a common fit of Eq.\ \eqref{eq:Jexp} to all the data presented in this figure. While the red dashed curves in \textbf{a}-\textbf{e} visualize $|J_\perp|/h$, the blue dashed ones illustrate the exchange modulation due to the different and anisotropic $g$-tensors in the absence of SOI. \textbf{g}, Schematic representation of the fin structure (black and grey lines) overlaid by a three-dimensional surface plot of $|\Jexp|/h$. The coloured dashed rectangles indicate the planes of \textbf{a-e}. The data presented in this figure are taken at $V_\mathrm{B}=\unit[-820]{mV}$ and the error bars account for the EDSR linewidth and uncertainties in $\mathbf{B}$-field due to magnetic flux trapping.}
\label{fig3}
\end{figure*}

In Figs.\ \ref{fig3}a-e the dependence of $\Jexp$ on the magnetic field orientation is shown, revealing a striking anisotropy with vanishing splittings. The highly anisotropic exchange frequency is mainly due to the strong SOI and can be qualitatively understood from the gap size $\Delta_\mathrm{so}^\mathrm{dd}$ of the anticrossing between the $S_{02}$ and the parallel two-spin states. $\Delta_\mathrm{so}^\mathrm{dd}$ is proportional to $|\boldsymbol{n}_\mathrm{so} \times \mathrm{\textbf{B}}|$, where $\boldsymbol{n}_\mathrm{so}$ is a unit vector pointing in the direction of the spin-orbit field and \textbf{B} the external magnetic field \cite{NadjPerge2012}. Therefore, $\Delta_\mathrm{so}^\mathrm{dd}$ changes with magnetic field orientation and so do the two-hole energy levels (see Fig.\ \ref{fig1}c). However, we remark that from the dependence of $\Delta_\mathrm{so}^\mathrm{dd}$ on $\textbf{B}/|\textbf{B}|$ the exchange matrix $\mathcal J$ cannot be extracted. 

We derive an equation for $\mathcal J$ starting from a Fermi-Hubbard model and including both the SOI as well as the anisotropic and differing hole $g$-factors (Methods and Supplementary Section 5). Tuned deep into the (1,1) charge regime where spin manipulation takes place, the system is approximated by the Hamiltonian
\begin{equation}
H_{(1,1)} = \frac 1 2 \mu_\mathrm{B} {\bf B}\cdot g_1\boldsymbol \sigma_1 + \frac 1 2 \mu_\mathrm{B} {\bf B}\cdot g_2\boldsymbol \sigma_2 + \frac 1 4 \boldsymbol \sigma_1 \cdot \mathcal J \boldsymbol \sigma_2\,.
    \label{eq:Heff11lab}
\end{equation}
Here $\mu_\mathrm{B}$ is Bohr's magneton and $\boldsymbol \sigma_{i}$ the vector of Pauli matrices for each QD. The exchange matrix is given by $\mathcal J = J_0 R_\text{so}(-2d/\lambda_\text{so})$, where $R_\text{so}(\varphi)$ is the counterclockwise rotation matrix around $\mathbf{n}_\mathrm{so}$ by an angle $\varphi$, $\lambda_\mathrm{so}$ the spin-orbit length, and $d$ the inter-dot distance. The experimentally observed exchange splitting is given by (Methods and Supplementary Section 5)
\begin{equation}
\Jexp = {\bf n}_1\cdot \mathcal J {\bf n}_2 = J_0 {\bf n}_1\cdot R_\mathrm{so}(-2d/\lambda_\mathrm{so}) {\bf n}_2\,,
\label{eq:Jexp}
\end{equation}
 where $\mathbf{n}_i = g_i\mathbf{B}/|g_i\mathbf{B}|$ denotes the Zeeman field direction. On comparing Eqs.\ \eqref{eq:stepanenko_approx} and \eqref{eq:Jexp} we find for the previously introduced correction factor $\cos (2\tilde \theta) = {\bf n}_1\cdot R_\text{so}(-2d/\lambda_\text{so}) {\bf n}_2$. Finally, by describing the magnetic field direction using the two angles $\alpha$ and $\beta$ (Fig.\ \ref{fig3}), we obtain a fit equation $\Jexp(\alpha, \beta)$ with five fitting parameters, namely $t_c$, $U_0$, $\mathbf{n}_\mathrm{so}$ and $\lambda_\mathrm{so}$. 

Next, we apply this model to the data (black points) shown in Figs.\ \ref{fig3}a-f and perform a common fit to the full data set, consisting of measurements of $\Jexp(\alpha, \beta)$ in five different planes (visualized in Fig.\ \ref{fig3}g) at constant detuning, and $\Jexp(\epsilon)$ for $\mathbf{B}$ pointing in $x$-direction. There is excellent agreement between theory and experiment for the best-fit parameters: $\lambda_\mathrm{so}=\unit[31]{nm}$, $\mathbf{n}_\mathrm{so} = (-0.06, 0.41, 0.91)$, $t_c=\unit[5.61]{GHz}$ and $U_0 = \unit[1.07]{meV}$. The spin-orbit length coincides with the values reported before \cite{Geyer2021, Camenzind2022}, and corresponds to a spin rotation angle of $\theta_\mathrm{so}=d/\lambda_\mathrm{so} \simeq 0.41\pi$ for a hole tunneling from one QD to the other over $\unit[d\,{\simeq}\,{40}]{nm}$. The direction of the spin-orbit field, represented by ($\alpha_\mathrm{so}=93^\circ$, $\beta_\text{so} =23^\circ$), is as expected perpendicular to the long axis of the fin and thus orthogonal to the hole momentum \cite{Kloeffel2018,Bosco2021a}. The small out-of-the-substrate-plane tilt can arise on account of strain or electric fields not being perfectly aligned along the $y$-direction. Using the five best-fit parameter values we can, for the first time, reconstruct the full exchange matrix
\begin{eqnarray}
\mathcal J = J_0 
\left(\begin{array}{rrr}
 -0.87 &  0.41 & -0.28 \\
 -0.49 & -0.60 &  0.64 \\
  0.10 &  0.69 &  0.72 \\
\end{array}\right).
\label{eq:Jmatexp}
\end{eqnarray}
Because we also find the $g$-tensors when measuring $\Jexp(\alpha, \beta)$ by means of MW spectroscopy, the two-spin Hamiltonian \eqref{eq:Heff11lab} is fully characterized, thus allowing us to optimize two-qubit gate operations as discussed later. Furthermore, we can analyze the different contributions to the exchange anisotropy with Eq.\ \eqref{eq:Jexp}: by setting $\theta_\mathrm{so}$ to zero, we are left with the effect of the anisotropic $g$-tensors. We find that the $g$-factor contribution to the $\Jexp$-anisotropy was minor (dashed blue curves in Figs.\ \ref{fig3}a-e). Finally, we remark that the observed rotational exchange anisotropy relies on a strong SOI and the presence of an external magnetic field \cite{Kavokin2001,Kavokin2004}, as opposed to a weaker Ising-like anisotropy that can be found in inversion symmetric hole DQDs \cite{Hetenyi2020} or at zero magnetic field \cite{Hetenyi2022,Katsaros2020}.

\begin{figure*}[t]
\centering
\includegraphics[width=0.5\columnwidth]{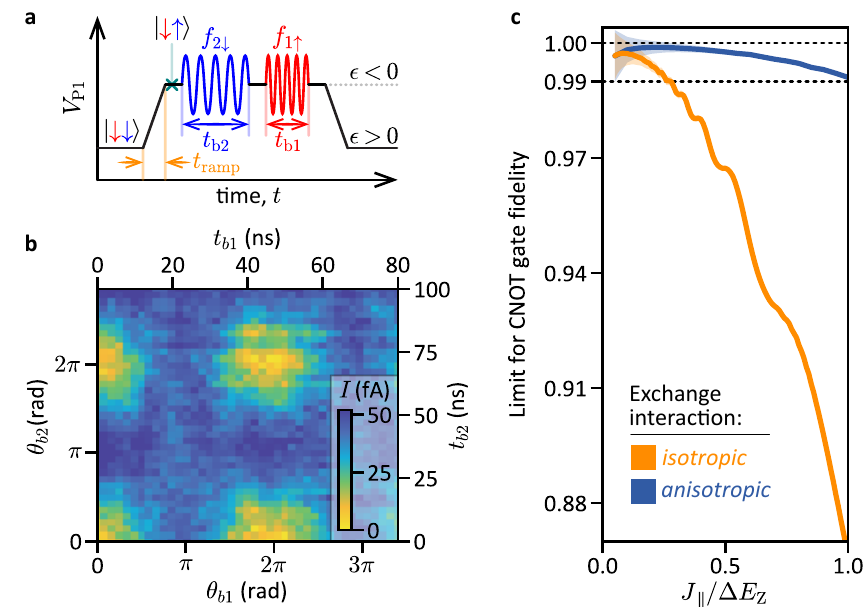}
\caption{\textbf{Fast two-qubit logic with Si hole spin qubits.} \textbf{a}, Pulse sequence for the CROT gate operation. A trapezoidal waveform with a ramp time of $\unit[20]{ns}$ is used to initialize the spins in the $\ket{\downarrow\uparrow}$-state and to readout their state after applying two microwave bursts resonant with Q2 ($f_{2\downarrow} = \unit[4.25]{GHz}$) and Q1 ($f_{1\uparrow} = \unit[4.66]{GHz}$). \textbf{b}, Parity measurement of the qubits demonstrating a conditional rotation of Q1 controlled by the state of Q2. This data is taken at ${\Jexp/h}\,\unit[{\simeq}\,{80}]{MHz}$, $V_\mathrm{B}=\unit[-810]{mV}$, $|\mathbf{B}|=\unit[0.146]{T}$, $\alpha=25^\circ$, $\beta=0^\circ$ and $\eps=\unit[-2.9]{meV}$. \textbf{c}, Numerically calculated CNOT gate fidelity versus exchange splitting $\Jexp$ (in units of $\Delta E_\mathrm{Z}$) for anisotropic (blue, with parameters of \textbf{b}) and isotropic (orange) exchange. The shaded regions indicate the precision of the numerics.}
\label{fig4}
\end{figure*}

We make use of the large exchange splitting to demonstrate a fast two-qubit gate for holes in Si, namely a controlled rotation \cite{Watson2018,Huang2019,Hendrickx2020,Hendrickx2021,Noiri2022}. This gate operation is naturally implemented by driving just one of the four EDSR transitions (see Fig.\ \ref{fig1}d), resulting in a rotation of the target qubit conditional on the state of the control qubit. First, we initialize $\ket{\mathrm{Q1},\mathrm{Q2}}$ in the $\ket{\downarrow\uparrow}$-state by adiabatically pulsing from ${\epsilon}\,{>}\,{0}$, where the spin-blockaded $\ket{\downarrow\downarrow}$-state is occupied, to $\unit[{\epsilon}\,{=}\,{-2.9}]{meV}$, where $\unit[{\Jexp/h}\,{\simeq}\,{80}]{MHz}$ and MW-induced state leakage is suppressed \cite{Watson2018} (Supplementary Section~4). Subsequently, the state of the control qubit Q2 is prepared by a MW burst of length $t_{b2}$ and frequency $f_{2\downarrow}$, and finally a controlled rotation of the target qubit Q1 is triggered by the following pulse with $t_{b1}$ and $f_{1\uparrow}$ (Fig.\ \ref{fig4}a). The measurement outcome is presented in Fig.\ \ref{fig4}b, revealing the characteristic fading in and out of the target qubit's Rabi oscillations as a function of $t_{b2}$, that is, the spin state of the control qubit \cite{Hendrickx2021,Hendrickx2020}. A controlled spin flip for Q1 is executed in $\unit[{\simeq}\,{24}]{ns}$, which is short compared to other realizations with electrons in Si \cite{Noiri2022} or holes in Ge \cite{Hendrickx2020,Hendrickx2021}. We remark that our transport-based readout scheme severely limits the duration of the qubits' manipulation stage \cite{Camenzind2022}, such that randomized benchmarking to determine a two-qubit gate fidelity could not be performed \cite{Knill2008}.

Two key requirements need to be fulfilled for high-fidelity CROT gates. First, in order to prevent a mixing of the antiparallel spin states ($\ket{\uparrow\downarrow}$, $\ket{\downarrow\uparrow}$), the Zeeman energy difference between the qubits $\Delta E_\mathrm{Z}$ must be much larger than the ``perpendicular'' exchange coupling $J_{\perp}$ (see Methods). Second, either ${\Jexp}\,{\gg}\,{hf_\mathrm{Rabi}}$ or ${\Jexp/\sqrt{15}}\,{=}\,{hf_\mathrm{Rabi}}$ to avoid unwanted rotations of the off-resonant states \cite{Russ2018,Noiri2022}. Hence, for electrons with isotropic exchange (${\Jexp}\,{=}\,{J_\perp}\,{=}\,{J}$) the speed of high-fidelity CROT gates is limited by ${hf_\mathrm{Rabi}}\,{\ll}\,{J}\,{\ll}\,{\Delta E_\mathrm{Z}}$. However, for hole spins with highly anisotropic exchange interaction this limit can be overcome. In fact, ${\Jexp}\,{=}\,{J_0}$ while ${J_\perp}\,{=}\,{0}$ is possible, for instance, if the $g$-tensors are isotropic, for $\theta_\mathrm{so}=\pi/2$ and $\mathbf{B}$ perpendicular to $\mathbf{n}_\mathrm{so}$; we remark that the latter condition also ensures fast single qubit rotations. Consequently, our theory predicts that for holes in comparison to electrons a controlled-NOT (CNOT) gate (differing from a CROT by a phase factor) with fidelity above the fault-tolerance threshold of 99\% can be realized with much shorter gate times (Fig.\ \ref{fig4}c). For the CROT experiment presented in Fig.\ \ref{fig4}b the magnetic field orientation (marked by the vertical orange line in Fig.\ \ref{fig3}b) was chosen such that $|\Jexp| = 0.90\,J_0$ and ${|J_\perp|}\,{=}\,{0.05}\,{J_0}$. In Figs.\ \ref{fig3}a-e the red dashed curves show the dependence of $J_\perp$ on $\mathbf{B}/|\mathbf{B}|$, highlighting that the ideal configuration (${\Jexp}\,{\simeq}\,{J_0}$, ${J_\perp}\,{\simeq}\,{0}$) is stretched over a wide range of directions. The CROT sweet spot is consequently robust against device variations, making it highly suitable for large qubit arrays.

In summary, we investigated the exchange coupling between two hole spins in a Si FinFET and found it to be both highly anisotropic and tunable, allowing for an interaction strength \unit[${>}{0.5}$]{GHz}. We identify the strong SOI as the main microscopic origin of this anisotropy and propose a simple procedure for determining the exchange matrix. This measurement and analysis scheme applies to a wide variety of devices, for instance also to electron spin qubits with synthetic SOI in the presence of a magnetic field gradient (Supplementary Section 6) \cite{Noiri2022,Watson2018,Philips2022}. By fully characterizing the Hamiltonian of the two coupled spins, the best possible configuration for implementing two-qubit gates can be identified. A strongly anisotropic exchange results in extended sweet spots in magnetic field orientation, where both fast and fault-tolerant CROTs can be performed. The robustness of these sweet spots against device variations makes CROT gate operations with anisotropic exchange highly attractive for large-scale qubit arrays. Finally, by choosing a close-to-ideal configuration we realize a CROT gate in just $\unit[{\simeq}\,{24}]{ns}$. The advance reported here in combination with fast readout \cite{Huang2021} and high-fidelity single-qubit operations at temperatures above \unit[1]{K} \cite{Camenzind2022} demonstrate that industrial FinFET technology has great potential for realizing a universal quantum processor with all-around high-performance fidelities, integrated on the same chip with the classical control electronics. 


\newpage
\noindent\textbf{Methods}\\

\noindent\textbf{Device fabrication.} The fin structures are orientated along the $[110]$ crystal direction on a near-intrinsic, natural Si substrate ($\unit[{\rho}\,{>}\,{10}]{k\Omega\,cm}$ and (100) surface), and are covered by a $\unit[{\simeq}\,{7}]{nm}$-thick, thermally-grown silicon dioxide (SiO$_2$) layer. Two layers of titanium nitride (TiN) gate electrodes, which are electrically isolated by a $\unit[{\simeq}\,{4.5}]{nm}$-thick SiO$_2$ layer deposited by atomic layer deposition, are used for DQD formation. The second gate layer is integrated by means of a self-aligned process, resulting in a perfect layer-to-layer alignment. The p-type source and drain regions are made of platinum silicide. Finally, the devices are embedded in a $\unit[{\simeq}\,{100}]{nm}$-thick SiO$_2$ layer and are measured through contact vias filled with tungsten. Further details on the device fabrication are provided in Refs. \cite{Geyer2021, Kuhlmann2018}.\\

\noindent\textbf{Experimental setup.} All measurements are performed using a Bluefors dry dilution refrigerator with a base temperature of $\unit[{\sim}\,{40}]{mK}$ and a three-axis magnet that provides arbitrary control of the magnetic field vector $\mathbf{B}$. The DC voltages are supplied by a low-noise voltage source (BasPI SP927) and the fast pulses applied to the P1-gate (Fig.\ \ref{fig1}a) by an arbitrary waveform generator (Tektronix AWG5208), which also controls the I and Q inputs of a vector signal generator (Rohde$\,$\&$\,$Schwarz SGS100A) for generating sideband-modulated EDSR microwave pulses. The source-to-drain current is measured with a current-to-voltage amplifier (BasPI SP983c) and a lock-in amplifier (Signal Recovery 7265), chopping the microwave signal at a frequency of \unit[89.2]{Hz} for better noise rejection. Further details are provided by Supplementary Section 1.\\

\noindent\textbf{Derivation of the fit function for the exchange matrix.} Using a Fermi-Hubbard model with a single orbital state $\ket i$ per site $i=\{1,2\}$, our DQD system is described by the Hamiltonian
\begin{equation}
H_\mathrm{FH} = \sum \limits_{i,j\in\{1,2\}} \sum \limits_{ss'\in\{\spu,\spd\}} \tilde H_{ij}^{ss'} a^\dagger_{is} a_{js'} + U \sum \limits_{i\in\{1,2\}} n_{i\spu} n_{i\spd}\,.
\label{eq:HFH}
\end{equation}
Here $a^\dagger_{is}$ ($a_{is}$) creates (removes) a hole on site $i$ and spin $s=\{\ket\spu, \ket\spd\}$, $n_{is} = a^\dagger_{is} a_{is}$ is the occupation number operator, and $U$ is the charging energy. The single-particle Hamiltonian $\tilde H$ is given by
\begin{equation}
\tilde H = \frac{\tilde\eps}{2}\tau_z + t_c\cos(\theta_\text{so})\tau_x + t_c\sin(\theta_\text{so})\tau_y\mathbf{n}_\text{so}\cdot\boldsymbol\sigma + \frac12\mu_B\mathbf{B}\cdot\left[\frac{1+\tau_z}{2}g_1\boldsymbol\sigma + \frac{1-\tau_z}{2}g_2\boldsymbol\sigma\right],
\label{eq:HDQDsp}
\end{equation}
and contains spin-conserving inter-dot tunnelling $t_c \cos (\theta_\text{so}) \tau_x $ and a SOI-induced spin-flip hopping term $t_c \sin (\theta_\text{so}) \tau_y \mathbf{n}_\text{so} \cdot \boldsymbol \sigma$. Here $(\tau_x,\tau_y,\tau_z)$ are the Pauli matrices for the orbital degree of freedom, e.g.\ $\tau_z={\ket{1}}{\bra{1}}-{\ket{2}}{\bra{2}}$, $\boldsymbol\sigma$ is the vector of Pauli matrices acting on the spin degree of freedom, ${\bf n}_\text{so}$ the direction of the spin-orbit field and $\theta_{\text{so}}=d/\lambda_\text{so}$ is the spin rotation angle due to tunnelling~\cite{Froning2021b}. In the lab frame, as defined in Fig.\ \ref{fig1}, the $g$-tensors $g_1$ and $g_2$ are symmetric (Supplementary Section~3). Finally, $\tilde\eps$ is the energy difference for a hole occupying the left or the right QD, and is expressed in terms of the detuning energy $\eps$ between the (1,1) and (0,2) charge states by $\tilde\eps=\eps+U-U_0$.

We perform a transformation from the lab frame to the so-called ``spin-orbit frame'' and find
\begin{equation}
    \tilde H^\text{so} = U^\dagger_\text{so} \tilde H U_\text{so} = \frac{\tilde\eps}{2}\tau_z + t_c\tau_x + \frac{1}{2}\mu_B\mathbf{B}\cdot \left[ \frac{1+\tau_z}{2}g^\text{so}_1\boldsymbol\sigma + \frac{1-\tau_z}{2} g^\text{so}_2\boldsymbol\sigma \right].
\end{equation}
In the spin-orbit frame non-spin-conserving tunnelling is gauged away by the unitary transformation $U_\text{so} = \exp(-i\theta_\text{so}\tau_z\mathbf{n}_\text{so}\cdot \boldsymbol\sigma/2)$, and the $g$-tensors are given by $g^\text{so}_1 = g_1 R_\text{so} (\theta_\text{so})$ and $g^\text{so}_2 = g_2 R_\text{so} (-\theta_\text{so})$. Here $R_\text{so} (\varphi)$ denotes a counterclockwise rotation around ${\bf n}_\text{so}$ by an angle $\varphi$. Since our DQD system is operated close to the $\ket{S_{02}}$-$\ket S$ anticrossing, the Hamiltonian $H_\mathrm{FH}$ can be represented in the basis $\{\ket{S_{02}}, \ket S, \ket{T_-}, \ket{T_+}, \ket{T_0}\}$
\begin{equation}
H_{5\times 5} = 
\left(
\begin{array}{ccccc}
 U_0-\eps & \sqrt{2}t_c & 0 & 0 & 0 \\
 \sqrt{2}t_c & 0 & -\frac{\text{$\delta $b}_x+i\text{$\delta $b}_y}{\sqrt{2}} & \frac{\text{$\delta $b}_x-i\text{$\delta $b}_y}{\sqrt{2}} & \text{$\delta $b}_z \\
 0 & -\frac{\text{$\delta$b}_x-i\text{$\delta$b}_y}{\sqrt{2}} & \bar{b}_z & 0 & \frac{\bar{b}_x-i\bar{b}_y}{\sqrt{2}} \\
 0 & \frac{\text{$\delta$b}_x+i\text{$\delta$b}_y}{\sqrt{2}} & 0 & -\bar{b}_z & \frac{\bar{b}_x+i\bar{b}_y}{\sqrt{2}} \\
 0 & \text{$\delta$b}_z & \frac{\bar{b}_x+i\bar{b}_y}{\sqrt{2}} & \frac{\bar{b}_x-i\bar{b}_y}{\sqrt{2}} & 0 \\
\end{array}
\right),
\label{eq:H5x5}
\end{equation}
where the average and gradient Zeeman fields $\mathbf{\bar b} = \mu_B \mathbf B(g^\text{so}_1+g^\text{so}_2)/2$ and $\delta \mathbf{b} = \mu_B \mathbf B(g^\text{so}_1-g^\text{so}_2)/2$ were introduced. In the spin-orbit frame, the singlet subspace $\{\ket{S_{02}}, \ket S\}$ is coupled by the total tunnel coupling $t_c$ and the hybridized singlets $S_\pm$ have energies $E_{S_+}= U_0-\eps+J_0$ and $E_{S_-}=-J_0$ with $J_0 = \sqrt 2 \tan (\gamma/2) = -(U_0 - \eps)[1-\sqrt{1+8t_c^2/(U_0-\eps)^2}]/2$ and mixing angle $\gamma = \arctan[\sqrt 8 t_c/(U_0-\eps) ]$. Furthermore, we remark that $J_0\simeq2t_c^2/(U_0-\eps)$ in the limit of $t_c/(U_0-~\eps)\ll~1$. Because $S_+$ couples only weakly to the triplet states, our Hilbert space can be restricted to the four levels $\{ \ket{S_-}, \ket{T_-}, \ket{T_+}, \ket{T_0}\}$ and we obtain
\begin{equation}
    H_{4\times 4} = \left(
\begin{array}{cccc}
 -J_0 
 & -\frac{\text{$\delta $b}_x+i \text{$\delta $b}_y}{\sqrt{2}} \cos(\frac \gamma 2)
 & \frac{\text{$\delta $b}_x-i \text{$\delta $b}_y}{\sqrt{2}} \cos(\frac \gamma 2)
 & \text{$\delta $b}_z \cos(\frac \gamma 2)
 \\
 -\frac{\text{$\delta $b}_x-i \text{$\delta $b}_y}{\sqrt{2}} \cos(\frac \gamma 2)
 & \bar{b}_z & 0 
 & \frac{\bar{b}_x-i \bar{b}_y}{\sqrt{2}} 
 \\
 \frac{\text{$\delta $b}_x+i \text{$\delta $b}_y}{\sqrt{2}} \cos(\frac \gamma 2)
 & 0 
 & -\bar{b}_z 
 & \frac{\bar{b}_x+i \bar{b}_y}{\sqrt{2}} 
 \\
 \text{$\delta $b}_z \cos(\frac \gamma 2)
 & \frac{\bar{b}_x+i \bar{b}_y}{\sqrt{2}} 
 & \frac{\bar{b}_x-i \bar{b}_y}{\sqrt{2}} 
 & 0 \\
\end{array}
\right).
\label{eq:Heff4x4}
\end{equation}
Hole spin manipulation is performed deep in the (1,1) charge stability region, allowing us to introduce the localized spin operators $\boldsymbol \sigma^\text{so}_1$ and $\boldsymbol \sigma^\text{so}_2$. The Hamiltonian \eqref{eq:Heff4x4} can then be written as
\begin{equation}
    H^\text{so}_{(1,1)} = \frac 1 2 \mu_B {\bf B}\cdot g^\text{so}_1\boldsymbol \sigma^\text{so}_1 + \frac 1 2 \mu_B {\bf B}\cdot g^\text{so}_2 \boldsymbol \sigma^\text{so}_2 + \frac 1 4 J_0 \boldsymbol \sigma^\text{so}_1 \cdot \boldsymbol\sigma^\text{so}_2\, ,
    \label{eq:Heff4x4so}
\end{equation}
revealing that the exchange interaction is isotropic in the spin-orbit frame. To find an expression for the experimentally measured values, we first rewrite Eq.\ \eqref{eq:Heff4x4so} in the lab frame
\begin{equation}
    H_{(1,1)}^\text{lab} = \frac 1 2 \mu_B {\bf B}\cdot g_1\boldsymbol \sigma_1 + \frac 1 2 \mu_B {\bf B}\cdot g_2\boldsymbol \sigma_2 + \frac 1 4 \boldsymbol \sigma_1 \cdot \mathcal J \boldsymbol \sigma_2\, .
    \label{eq:Heff4x4lab}
\end{equation}
Here $\mathcal J =J_0 R_\text{so} (-2\theta_\text{so})$ represents the exchange matrix in the lab frame, $\boldsymbol \sigma_1 = R^\text{so}(-\theta_\text{so}) \boldsymbol \sigma^\text{so}_1$ and $\boldsymbol \sigma_2 = R^\text{so}(\theta_\text{so}) \boldsymbol \sigma^\text{so}_2 $. In addition, independent rotations $R_1$ and $R_2$ are applied to Q1 and Q2, such that the single particle terms of the Hamiltonian \eqref{eq:Heff4x4lab} become diagonal:
\begin{equation}
    H^\text{Q}_{(1,1)} = \frac 1 2 E_{Z,1} \sigma^\text{Q}_{z,1} + \frac 1 2  E_{Z,2} \sigma^\text{Q}_{z,2}  + \frac 1 4 \boldsymbol \sigma^\text{Q}_{1} \cdot  \mathcal J^\text{Q} \boldsymbol \sigma^\text{Q}_{2}\, ,
    \label{eq:Heff4x4Q}
\end{equation}
where $E_{Z,i} {\bf e}^\text{Q}_{z}=\mu_B R_i g_i {\bf B}$ is the $i$-th site's Zeeman splitting, ${\bf e}^\text{Q}_{z}$ the spin quantization axis and $\mathcal J^\text{Q} = J_0 R_1 R_\text{so} (-2\theta_\text{so}) R^{T}_2$ the exchange matrix in the so-called ``qubit frame'', wherein the exchange splitting $\Jexp$ is experimentally observed. To obtain an expression for $\Jexp$ we rewrite the Hamiltonian of Eq.~\eqref{eq:Heff4x4Q} in matrix form using the two-qubit basis $\{\ket{\spu\spu}, \ket{\spu\spd},\ket{\spd\spu},\ket{\spd\spd} \}$
\begin{equation}
     H^\text{Q}_{(1,1)} = \left(
\begin{array}{cccc}
 E_Z + \frac 1 4 J^\text{Q}_{zz} & 0 & 0 & 0\\
 0 & \frac 1 2 \Delta E_Z - \frac 1 4 J^\text{Q}_{zz} & \frac 1 2 J_\perp & 0\\
 0 & \frac 1 2 (J_\perp)^* & -\frac 1 2 \Delta E_Z - \frac 1 4 J^\text{Q}_{zz} & 0\\
 0 & 0 & 0 & -E_Z + \frac 1 4 J^\text{Q}_{zz}\\
\end{array}
\right) .
\label{eq:Heff4x4pert}
\end{equation}
Here we neglect every coupling that would contribute to the eigenvalues in $\mathcal{O} (J_0^2/E_Z)$ and introduce $J_\perp = [J^\text{Q}_{xx}+J^\text{Q}_{yy} + i (J^\text{Q}_{xy}-J^\text{Q}_{yx})]/2$, $E_Z=(E_{Z,1}+E_{Z,2})/2$ and $\Delta E_Z=E_{Z,1}-E_{Z,2}$. The eigenenergies of Eq.~\eqref{eq:Heff4x4pert} are
\begin{subequations}
\begin{equation}
    E_{\spu\spu} = E_Z + \frac 1 4 J^\text{Q}_{zz}\, , \hspace{1cm}    E_{\spd\spd} = -E_Z + \frac 1 4 J^\text{Q}_{zz}\, ,
\end{equation}
\begin{equation}
    E_{\widetilde{\spu\spd}} = \frac 1 2 \Delta \tilde E_Z - \frac 1 4 J^\text{Q}_{zz}\, , \hspace{1cm}  
    E_{\widetilde{\spd\spu}} = -\frac 1 2 \Delta \tilde E_Z - \frac 1 4 J^\text{Q}_{zz}\, ,
\end{equation}
\end{subequations}
with $\Delta \tilde{E}_Z = \sqrt{\Delta E_Z^2 + |J_\perp|^2}$. We thus find for the exchange splitting, which is defined as the energy difference between the two transitions flipping the same spin, $\Jexp = E_{\spu\spu}-E_{\widetilde{\spu\spd}} - (E_{\widetilde{\spd\spu}}-E_{\spd\spd}) = J^\text{Q}_{zz}$. The matrix element $J^\text{Q}_{zz}$ in turn is given by
\begin{equation}
 J^\text{Q}_{zz}= \Jexp =  {\bf e}_{z}^\text{Q} \cdot \mathcal J^\text{Q} {\bf e}^\text{Q}_{z} = \mathbf{n}_1 \cdot \mathcal J \mathbf{n}_2 = J_0\, \mathbf{n}_1\cdot R_\text{so} (-2  \theta_\text{so})\mathbf{n}_2\, .
 \label{eq:Jfit}
\end{equation}
Eq.\ \eqref{eq:Jfit} is the fit function employed to describe the observed exchange anisotropy, where the effect of both spin-orbit interaction and the anisotropy of the $g$-tensors is accounted for. We note that an explicit dependence on the magnetic field direction arises from $\mathbf{n}_i = g_i {\bf B}/|g_i {\bf B}|$. Further details of the derivation are found in Supplementary Section~5.\\

\noindent\textbf{Numerical calculation of the CNOT gate fidelity.} The CROT gate operation is modeled by numerically evaluating the Hamiltonian's time evolution 
\begin{equation}
\text{CROT}_\mathrm{num} = \mathcal{T} \exp{\left[-\frac{i}{\hbar} \int_{0}^{t_\pi} dt \, H^\mathrm{Q}_{(1,1)}(t) \right]}.
\label{eq:CROTnum}
\end{equation}
Here $\mathcal{T}$ denotes time-ordering, $t_\pi$ is the spin-flip time, and the time-dependent Hamiltonian $H^\mathrm{Q}_{(1,1)}(t)$ results from Eq.\ \eqref{eq:Heff4x4pert} after adding the drive ${hf_\mathrm{Rabi}}\,{\sin(2\pi f_{1\uparrow}\,t)}\,{\sigma_{x,1}}$, where the Rabi frequency fulfils the condition $hf_\mathrm{Rabi}=J_\parallel/\sqrt{15}$ in order to suppress off-resonant driving \cite{Noiri2022,Russ2018}. Finally, the CNOT gate fidelity is determined by $\mathcal{F}=\frac{1}{4}\text{Tr}\left[\text{CNOT}_\mathrm{num}\,\text{CNOT}\right]$, where CNOT is the ideal gate matrix and $\text{CNOT}_\mathrm{num}$ is obtained by applying single-qubit phase corrections to Eq.\ \eqref{eq:CROTnum}. For more details see Supplementary Section 7.\\

\noindent\textbf{Acknowledgments}\\
We acknowledge support by the cleanroom operation team, particularly U.\ Drechsler, A.\ Olziersky and D.\ D.\ Pineda, at the IBM Binnig and Rohrer Nanotechnology Center, and technical support at the University of Basel by S.\ Martin and M.\ Steinacher. In addition, we thank T.\ Berger for providing us with a 3D render of the FinFET device. This work was partially supported by the NCCR SPIN, the Swiss NSF (grant no. 179024), and the EU H2020 European Microkelvin Platform EMP (grant no. 824109). L.C.C. acknowledges support by a Swiss NSF mobility fellowship (P2BSP2\_200127).\\

\noindent\textbf{Author contributions}\\
S.G.\ and A.V.K.\ conceived and performed the experiments with inputs from L.C.C., R.E., R.J.W.,\ A.F.\ and D.M.Z. A.V.K.\ and S.G.\ designed and fabricated the device with support by A.F. B.H., S.B.\ and D.L.\ developed the theory model. S.G., A.V.K., B.H.\ and S.B.\ analysed the data and wrote the manuscript with inputs from all the authors.  A.V.K.\ managed the project with support from R.J.W.\ and D.M.Z.\\

\noindent\textbf{Competing interests}\\
The authors declare no competing interests.\\

\bibliography{2-qubit-paper}

\begin{thebibliography}{49}%
\makeatletter
\providecommand \@ifxundefined [1]{%
 \@ifx{#1\undefined}
}%
\providecommand \@ifnum [1]{%
 \ifnum #1\expandafter \@firstoftwo
 \else \expandafter \@secondoftwo
 \fi
}%
\providecommand \@ifx [1]{%
 \ifx #1\expandafter \@firstoftwo
 \else \expandafter \@secondoftwo
 \fi
}%
\providecommand \natexlab [1]{#1}%
\providecommand \enquote  [1]{``#1''}%
\providecommand \bibnamefont  [1]{#1}%
\providecommand \bibfnamefont [1]{#1}%
\providecommand \citenamefont [1]{#1}%
\providecommand \href@noop [0]{\@secondoftwo}%
\providecommand \href [0]{\begingroup \@sanitize@url \@href}%
\providecommand \@href[1]{\@@startlink{#1}\@@href}%
\providecommand \@@href[1]{\endgroup#1\@@endlink}%
\providecommand \@sanitize@url [0]{\catcode `\\12\catcode `\$12\catcode
  `\&12\catcode `\#12\catcode `\^12\catcode `\_12\catcode `\%12\relax}%
\providecommand \@@startlink[1]{}%
\providecommand \@@endlink[0]{}%
\providecommand \url  [0]{\begingroup\@sanitize@url \@url }%
\providecommand \@url [1]{\endgroup\@href {#1}{\urlprefix }}%
\providecommand \urlprefix  [0]{URL }%
\providecommand \Eprint [0]{\href }%
\providecommand \doibase [0]{https://doi.org/}%
\providecommand \selectlanguage [0]{\@gobble}%
\providecommand \bibinfo  [0]{\@secondoftwo}%
\providecommand \bibfield  [0]{\@secondoftwo}%
\providecommand \translation [1]{[#1]}%
\providecommand \BibitemOpen [0]{}%
\providecommand \bibitemStop [0]{}%
\providecommand \bibitemNoStop [0]{.\EOS\space}%
\providecommand \EOS [0]{\spacefactor3000\relax}%
\providecommand \BibitemShut  [1]{\csname bibitem#1\endcsname}%
\let\auto@bib@innerbib\@empty
\bibitem [{\citenamefont {Loss}\ and\ \citenamefont
  {DiVincenzo}(1998)}]{Loss1998}%
  \BibitemOpen
  \bibfield  {author} {\bibinfo {author} {\bibfnamefont {D.}~\bibnamefont
  {Loss}}\ and\ \bibinfo {author} {\bibfnamefont {D.~P.}\ \bibnamefont
  {DiVincenzo}},\ }\bibfield  {title} {\bibinfo {title} {{Quantum computation
  with quantum dots}},\ }\href {https://doi.org/10.1103/PhysRevA.57.120}
  {\bibfield  {journal} {\bibinfo  {journal} {Physical Review A}\ }\textbf
  {\bibinfo {volume} {57}},\ \bibinfo {pages} {120} (\bibinfo {year}
  {1998})}\BibitemShut {NoStop}%
\bibitem [{\citenamefont {Vandersypen}\ \emph {et~al.}(2017)\citenamefont
  {Vandersypen}, \citenamefont {Bluhm}, \citenamefont {Clarke}, \citenamefont
  {Dzurak}, \citenamefont {Ishihara}, \citenamefont {Morello}, \citenamefont
  {Reilly}, \citenamefont {Schreiber},\ and\ \citenamefont
  {Veldhorst}}]{Vandersypen2017}%
  \BibitemOpen
  \bibfield  {author} {\bibinfo {author} {\bibfnamefont {L.~M.~K.}\
  \bibnamefont {Vandersypen}}, \bibinfo {author} {\bibfnamefont
  {H.}~\bibnamefont {Bluhm}}, \bibinfo {author} {\bibfnamefont {J.~S.}\
  \bibnamefont {Clarke}}, \bibinfo {author} {\bibfnamefont {A.~S.}\
  \bibnamefont {Dzurak}}, \bibinfo {author} {\bibfnamefont {R.}~\bibnamefont
  {Ishihara}}, \bibinfo {author} {\bibfnamefont {A.}~\bibnamefont {Morello}},
  \bibinfo {author} {\bibfnamefont {D.~J.}\ \bibnamefont {Reilly}}, \bibinfo
  {author} {\bibfnamefont {L.~R.}\ \bibnamefont {Schreiber}},\ and\ \bibinfo
  {author} {\bibfnamefont {M.}~\bibnamefont {Veldhorst}},\ }\bibfield  {title}
  {\bibinfo {title} {{Interfacing spin qubits in quantum dots and donors-hot,
  dense, and coherent}},\ }\href {https://doi.org/10.1038/s41534-017-0038-y}
  {\bibfield  {journal} {\bibinfo  {journal} {npj Quantum Information}\
  }\textbf {\bibinfo {volume} {3}},\ \bibinfo {pages} {34} (\bibinfo {year}
  {2017})}\BibitemShut {NoStop}%
\bibitem [{\citenamefont {Veldhorst}\ \emph {et~al.}(2017)\citenamefont
  {Veldhorst}, \citenamefont {Eenink}, \citenamefont {Yang},\ and\
  \citenamefont {Dzurak}}]{Veldhorst2017}%
  \BibitemOpen
  \bibfield  {author} {\bibinfo {author} {\bibfnamefont {M.}~\bibnamefont
  {Veldhorst}}, \bibinfo {author} {\bibfnamefont {H.~G.~J.}\ \bibnamefont
  {Eenink}}, \bibinfo {author} {\bibfnamefont {C.~H.}\ \bibnamefont {Yang}},\
  and\ \bibinfo {author} {\bibfnamefont {A.~S.}\ \bibnamefont {Dzurak}},\
  }\bibfield  {title} {\bibinfo {title} {Silicon {CMOS} architecture for a
  spin-based quantum computer},\ }\href
  {https://doi.org/10.1038/s41467-017-01905-6} {\bibfield  {journal} {\bibinfo
  {journal} {Nature Communications}\ }\textbf {\bibinfo {volume} {8}},\
  \bibinfo {pages} {1766} (\bibinfo {year} {2017})}\BibitemShut {NoStop}%
\bibitem [{\citenamefont {Philips}\ \emph {et~al.}(2022)\citenamefont
  {Philips}, \citenamefont {M{\k{a}}dzik}, \citenamefont {Amitonov},
  \citenamefont {de~Snoo}, \citenamefont {Russ}, \citenamefont {Kalhor},
  \citenamefont {Volk}, \citenamefont {Lawrie}, \citenamefont {Brousse},
  \citenamefont {Tryputen}, \citenamefont {Wuetz}, \citenamefont {Sammak},
  \citenamefont {Veldhorst}, \citenamefont {Scappucci},\ and\ \citenamefont
  {Vandersypen}}]{Philips2022}%
  \BibitemOpen
  \bibfield  {author} {\bibinfo {author} {\bibfnamefont {S.~G.~J.}\
  \bibnamefont {Philips}}, \bibinfo {author} {\bibfnamefont {M.~T.}\
  \bibnamefont {M{\k{a}}dzik}}, \bibinfo {author} {\bibfnamefont {S.~V.}\
  \bibnamefont {Amitonov}}, \bibinfo {author} {\bibfnamefont {S.~L.}\
  \bibnamefont {de~Snoo}}, \bibinfo {author} {\bibfnamefont {M.}~\bibnamefont
  {Russ}}, \bibinfo {author} {\bibfnamefont {N.}~\bibnamefont {Kalhor}},
  \bibinfo {author} {\bibfnamefont {C.}~\bibnamefont {Volk}}, \bibinfo {author}
  {\bibfnamefont {W.~I.~L.}\ \bibnamefont {Lawrie}}, \bibinfo {author}
  {\bibfnamefont {D.}~\bibnamefont {Brousse}}, \bibinfo {author} {\bibfnamefont
  {L.}~\bibnamefont {Tryputen}}, \bibinfo {author} {\bibfnamefont {B.~P.}\
  \bibnamefont {Wuetz}}, \bibinfo {author} {\bibfnamefont {A.}~\bibnamefont
  {Sammak}}, \bibinfo {author} {\bibfnamefont {M.}~\bibnamefont {Veldhorst}},
  \bibinfo {author} {\bibfnamefont {G.}~\bibnamefont {Scappucci}},\ and\
  \bibinfo {author} {\bibfnamefont {L.~M.~K.}\ \bibnamefont {Vandersypen}},\
  }\bibfield  {title} {\bibinfo {title} {Universal control of a six-qubit
  quantum processor in silicon},\ }\href
  {https://doi.org/10.1038/s41586-022-05117-x} {\bibfield  {journal} {\bibinfo
  {journal} {Nature}\ }\textbf {\bibinfo {volume} {609}},\ \bibinfo {pages}
  {919} (\bibinfo {year} {2022})}\BibitemShut {NoStop}%
\bibitem [{\citenamefont {Hendrickx}\ \emph {et~al.}(2021)\citenamefont
  {Hendrickx}, \citenamefont {Lawrie}, \citenamefont {Russ}, \citenamefont {van
  Riggelen}, \citenamefont {de~Snoo}, \citenamefont {Schouten}, \citenamefont
  {Sammak}, \citenamefont {Scappucci},\ and\ \citenamefont
  {Veldhorst}}]{Hendrickx2021}%
  \BibitemOpen
  \bibfield  {author} {\bibinfo {author} {\bibfnamefont {N.~W.}\ \bibnamefont
  {Hendrickx}}, \bibinfo {author} {\bibfnamefont {W.~I.~L.}\ \bibnamefont
  {Lawrie}}, \bibinfo {author} {\bibfnamefont {M.}~\bibnamefont {Russ}},
  \bibinfo {author} {\bibfnamefont {F.}~\bibnamefont {van Riggelen}}, \bibinfo
  {author} {\bibfnamefont {S.~L.}\ \bibnamefont {de~Snoo}}, \bibinfo {author}
  {\bibfnamefont {R.~N.}\ \bibnamefont {Schouten}}, \bibinfo {author}
  {\bibfnamefont {A.}~\bibnamefont {Sammak}}, \bibinfo {author} {\bibfnamefont
  {G.}~\bibnamefont {Scappucci}},\ and\ \bibinfo {author} {\bibfnamefont
  {M.}~\bibnamefont {Veldhorst}},\ }\bibfield  {title} {\bibinfo {title} {{A
  four-qubit germanium quantum processor}},\ }\href
  {https://doi.org/10.1038/s41586-021-03332-6} {\bibfield  {journal} {\bibinfo
  {journal} {Nature}\ }\textbf {\bibinfo {volume} {591}},\ \bibinfo {pages}
  {580} (\bibinfo {year} {2021})}\BibitemShut {NoStop}%
\bibitem [{\citenamefont {Tokura}\ \emph {et~al.}(2006)\citenamefont {Tokura},
  \citenamefont {van~der Wiel}, \citenamefont {Obata},\ and\ \citenamefont
  {Tarucha}}]{Tokura2006}%
  \BibitemOpen
  \bibfield  {author} {\bibinfo {author} {\bibfnamefont {Y.}~\bibnamefont
  {Tokura}}, \bibinfo {author} {\bibfnamefont {W.~G.}\ \bibnamefont {van~der
  Wiel}}, \bibinfo {author} {\bibfnamefont {T.}~\bibnamefont {Obata}},\ and\
  \bibinfo {author} {\bibfnamefont {S.}~\bibnamefont {Tarucha}},\ }\bibfield
  {title} {\bibinfo {title} {{Coherent Single Electron Spin Control in a
  Slanting Zeeman Field}},\ }\href
  {https://doi.org/10.1103/physrevlett.96.047202} {\bibfield  {journal}
  {\bibinfo  {journal} {Physical Review Letters}\ }\textbf {\bibinfo {volume}
  {96}},\ \bibinfo {pages} {047202} (\bibinfo {year} {2006})}\BibitemShut
  {NoStop}%
\bibitem [{\citenamefont {Pioro-Ladri{\`{e}}re}\ \emph
  {et~al.}(2008)\citenamefont {Pioro-Ladri{\`{e}}re}, \citenamefont {Obata},
  \citenamefont {Tokura}, \citenamefont {Shin}, \citenamefont {Kubo},
  \citenamefont {Yoshida}, \citenamefont {Taniyama},\ and\ \citenamefont
  {Tarucha}}]{PioroLadriere2008}%
  \BibitemOpen
  \bibfield  {author} {\bibinfo {author} {\bibfnamefont {M.}~\bibnamefont
  {Pioro-Ladri{\`{e}}re}}, \bibinfo {author} {\bibfnamefont {T.}~\bibnamefont
  {Obata}}, \bibinfo {author} {\bibfnamefont {Y.}~\bibnamefont {Tokura}},
  \bibinfo {author} {\bibfnamefont {Y.-S.}\ \bibnamefont {Shin}}, \bibinfo
  {author} {\bibfnamefont {T.}~\bibnamefont {Kubo}}, \bibinfo {author}
  {\bibfnamefont {K.}~\bibnamefont {Yoshida}}, \bibinfo {author} {\bibfnamefont
  {T.}~\bibnamefont {Taniyama}},\ and\ \bibinfo {author} {\bibfnamefont
  {S.}~\bibnamefont {Tarucha}},\ }\bibfield  {title} {\bibinfo {title}
  {{Electrically driven single-electron spin resonance in a slanting Zeeman
  field}},\ }\href {https://doi.org/10.1038/nphys1053} {\bibfield  {journal}
  {\bibinfo  {journal} {Nature Physics}\ }\textbf {\bibinfo {volume} {4}},\
  \bibinfo {pages} {776} (\bibinfo {year} {2008})}\BibitemShut {NoStop}%
\bibitem [{\citenamefont {Prechtel}\ \emph {et~al.}(2016)\citenamefont
  {Prechtel}, \citenamefont {Kuhlmann}, \citenamefont {Houel}, \citenamefont
  {Ludwig}, \citenamefont {Valentin}, \citenamefont {Wieck},\ and\
  \citenamefont {Warburton}}]{Prechtel2016}%
  \BibitemOpen
  \bibfield  {author} {\bibinfo {author} {\bibfnamefont {J.~H.}\ \bibnamefont
  {Prechtel}}, \bibinfo {author} {\bibfnamefont {A.~V.}\ \bibnamefont
  {Kuhlmann}}, \bibinfo {author} {\bibfnamefont {J.}~\bibnamefont {Houel}},
  \bibinfo {author} {\bibfnamefont {A.}~\bibnamefont {Ludwig}}, \bibinfo
  {author} {\bibfnamefont {S.~R.}\ \bibnamefont {Valentin}}, \bibinfo {author}
  {\bibfnamefont {A.~D.}\ \bibnamefont {Wieck}},\ and\ \bibinfo {author}
  {\bibfnamefont {R.~J.}\ \bibnamefont {Warburton}},\ }\bibfield  {title}
  {\bibinfo {title} {{Decoupling a hole spin qubit from the nuclear~spins}},\
  }\href {https://doi.org/10.1038/nmat4704} {\bibfield  {journal} {\bibinfo
  {journal} {Nature Materials}\ }\textbf {\bibinfo {volume} {15}},\ \bibinfo
  {pages} {981} (\bibinfo {year} {2016})}\BibitemShut {NoStop}%
\bibitem [{\citenamefont {Zwanenburg}\ \emph {et~al.}(2013)\citenamefont
  {Zwanenburg}, \citenamefont {Dzurak}, \citenamefont {Morello}, \citenamefont
  {Simmons}, \citenamefont {Hollenberg}, \citenamefont {Klimeck}, \citenamefont
  {Rogge}, \citenamefont {Coppersmith},\ and\ \citenamefont
  {Eriksson}}]{Zwanenburg2013}%
  \BibitemOpen
  \bibfield  {author} {\bibinfo {author} {\bibfnamefont {F.~A.}\ \bibnamefont
  {Zwanenburg}}, \bibinfo {author} {\bibfnamefont {A.~S.}\ \bibnamefont
  {Dzurak}}, \bibinfo {author} {\bibfnamefont {A.}~\bibnamefont {Morello}},
  \bibinfo {author} {\bibfnamefont {M.~Y.}\ \bibnamefont {Simmons}}, \bibinfo
  {author} {\bibfnamefont {L.~C.~L.}\ \bibnamefont {Hollenberg}}, \bibinfo
  {author} {\bibfnamefont {G.}~\bibnamefont {Klimeck}}, \bibinfo {author}
  {\bibfnamefont {S.}~\bibnamefont {Rogge}}, \bibinfo {author} {\bibfnamefont
  {S.~N.}\ \bibnamefont {Coppersmith}},\ and\ \bibinfo {author} {\bibfnamefont
  {M.~A.}\ \bibnamefont {Eriksson}},\ }\bibfield  {title} {\bibinfo {title}
  {{Silicon quantum electronics}},\ }\href
  {https://doi.org/10.1103/revmodphys.85.961} {\bibfield  {journal} {\bibinfo
  {journal} {Reviews of Modern Physics}\ }\textbf {\bibinfo {volume} {85}},\
  \bibinfo {pages} {961} (\bibinfo {year} {2013})}\BibitemShut {NoStop}%
\bibitem [{\citenamefont {Kloeffel}\ \emph {et~al.}(2018)\citenamefont
  {Kloeffel}, \citenamefont {Ran{\v{c}}i{\'{c}}},\ and\ \citenamefont
  {Loss}}]{Kloeffel2018}%
  \BibitemOpen
  \bibfield  {author} {\bibinfo {author} {\bibfnamefont {C.}~\bibnamefont
  {Kloeffel}}, \bibinfo {author} {\bibfnamefont {M.~J.}\ \bibnamefont
  {Ran{\v{c}}i{\'{c}}}},\ and\ \bibinfo {author} {\bibfnamefont
  {D.}~\bibnamefont {Loss}},\ }\bibfield  {title} {\bibinfo {title} {{Direct
  Rashba spin-orbit interaction in Si and Ge nanowires with different growth
  directions}},\ }\href {https://doi.org/10.1103/physrevb.97.235422} {\bibfield
   {journal} {\bibinfo  {journal} {Physical Review B}\ }\textbf {\bibinfo
  {volume} {97}},\ \bibinfo {pages} {235422} (\bibinfo {year}
  {2018})}\BibitemShut {NoStop}%
\bibitem [{\citenamefont {Bosco}\ and\ \citenamefont {Loss}(2021)}]{Bosco2021}%
  \BibitemOpen
  \bibfield  {author} {\bibinfo {author} {\bibfnamefont {S.}~\bibnamefont
  {Bosco}}\ and\ \bibinfo {author} {\bibfnamefont {D.}~\bibnamefont {Loss}},\
  }\bibfield  {title} {\bibinfo {title} {{Fully Tunable Hyperfine Interactions
  of Hole Spin Qubits in Si and Ge Quantum Dots}},\ }\href
  {https://doi.org/10.1103/physrevlett.127.190501} {\bibfield  {journal}
  {\bibinfo  {journal} {Physical Review Letters}\ }\textbf {\bibinfo {volume}
  {127}},\ \bibinfo {pages} {190501} (\bibinfo {year} {2021})}\BibitemShut
  {NoStop}%
\bibitem [{\citenamefont {Bosco}\ \emph {et~al.}(2021)\citenamefont {Bosco},
  \citenamefont {Het{\'{e}}nyi},\ and\ \citenamefont {Loss}}]{Bosco2021a}%
  \BibitemOpen
  \bibfield  {author} {\bibinfo {author} {\bibfnamefont {S.}~\bibnamefont
  {Bosco}}, \bibinfo {author} {\bibfnamefont {B.}~\bibnamefont
  {Het{\'{e}}nyi}},\ and\ \bibinfo {author} {\bibfnamefont {D.}~\bibnamefont
  {Loss}},\ }\bibfield  {title} {\bibinfo {title} {{Hole Spin Qubits in Si
  FinFETs With Fully Tunable Spin-Orbit Coupling and Sweet Spots for Charge
  Noise}},\ }\href {https://doi.org/10.1103/prxquantum.2.010348} {\bibfield
  {journal} {\bibinfo  {journal} {{PRX} {Quantum}}\ }\textbf {\bibinfo {volume}
  {2}},\ \bibinfo {pages} {010348} (\bibinfo {year} {2021})}\BibitemShut
  {NoStop}%
\bibitem [{\citenamefont {Froning}\ \emph
  {et~al.}(2021{\natexlab{a}})\citenamefont {Froning}, \citenamefont
  {Camenzind}, \citenamefont {van~der Molen}, \citenamefont {Li}, \citenamefont
  {Bakkers}, \citenamefont {Zumbühl},\ and\ \citenamefont
  {Braakman}}]{Froning2021}%
  \BibitemOpen
  \bibfield  {author} {\bibinfo {author} {\bibfnamefont {F.~N.~M.}\
  \bibnamefont {Froning}}, \bibinfo {author} {\bibfnamefont {L.~C.}\
  \bibnamefont {Camenzind}}, \bibinfo {author} {\bibfnamefont {O.~A.~H.}\
  \bibnamefont {van~der Molen}}, \bibinfo {author} {\bibfnamefont
  {A.}~\bibnamefont {Li}}, \bibinfo {author} {\bibfnamefont {E.~P. A.~M.}\
  \bibnamefont {Bakkers}}, \bibinfo {author} {\bibfnamefont {D.~M.}\
  \bibnamefont {Zumbühl}},\ and\ \bibinfo {author} {\bibfnamefont {F.~R.}\
  \bibnamefont {Braakman}},\ }\bibfield  {title} {\bibinfo {title} {{Ultrafast
  hole spin qubit with gate-tunable spin{\textendash}orbit switch
  functionality}},\ }\href {https://doi.org/10.1038/s41565-020-00828-6}
  {\bibfield  {journal} {\bibinfo  {journal} {Nature Nanotechnology}\ }\textbf
  {\bibinfo {volume} {16}},\ \bibinfo {pages} {308} (\bibinfo {year}
  {2021}{\natexlab{a}})}\BibitemShut {NoStop}%
\bibitem [{\citenamefont {Wang}\ \emph {et~al.}(2022)\citenamefont {Wang},
  \citenamefont {Xu}, \citenamefont {Gao}, \citenamefont {Liu}, \citenamefont
  {Ma}, \citenamefont {Zhang}, \citenamefont {Wang}, \citenamefont {Cao},
  \citenamefont {Wang}, \citenamefont {Zhang}, \citenamefont {Culcer},
  \citenamefont {Hu}, \citenamefont {Jiang}, \citenamefont {Li}, \citenamefont
  {Guo},\ and\ \citenamefont {Guo}}]{Wang2022}%
  \BibitemOpen
  \bibfield  {author} {\bibinfo {author} {\bibfnamefont {K.}~\bibnamefont
  {Wang}}, \bibinfo {author} {\bibfnamefont {G.}~\bibnamefont {Xu}}, \bibinfo
  {author} {\bibfnamefont {F.}~\bibnamefont {Gao}}, \bibinfo {author}
  {\bibfnamefont {H.}~\bibnamefont {Liu}}, \bibinfo {author} {\bibfnamefont
  {R.-L.}\ \bibnamefont {Ma}}, \bibinfo {author} {\bibfnamefont
  {X.}~\bibnamefont {Zhang}}, \bibinfo {author} {\bibfnamefont
  {Z.}~\bibnamefont {Wang}}, \bibinfo {author} {\bibfnamefont {G.}~\bibnamefont
  {Cao}}, \bibinfo {author} {\bibfnamefont {T.}~\bibnamefont {Wang}}, \bibinfo
  {author} {\bibfnamefont {J.-J.}\ \bibnamefont {Zhang}}, \bibinfo {author}
  {\bibfnamefont {D.}~\bibnamefont {Culcer}}, \bibinfo {author} {\bibfnamefont
  {X.}~\bibnamefont {Hu}}, \bibinfo {author} {\bibfnamefont {H.-W.}\
  \bibnamefont {Jiang}}, \bibinfo {author} {\bibfnamefont {H.-O.}\ \bibnamefont
  {Li}}, \bibinfo {author} {\bibfnamefont {G.-C.}\ \bibnamefont {Guo}},\ and\
  \bibinfo {author} {\bibfnamefont {G.-P.}\ \bibnamefont {Guo}},\ }\bibfield
  {title} {\bibinfo {title} {{Ultrafast coherent control of a hole spin qubit
  in a germanium quantum dot}},\ }\href
  {https://doi.org/10.1038/s41467-021-27880-7} {\bibfield  {journal} {\bibinfo
  {journal} {Nature Communications}\ }\textbf {\bibinfo {volume} {13}},\
  \bibinfo {pages} {206} (\bibinfo {year} {2022})}\BibitemShut {NoStop}%
\bibitem [{\citenamefont {Piot}\ \emph {et~al.}(2022)\citenamefont {Piot},
  \citenamefont {Brun}, \citenamefont {Schmitt}, \citenamefont {Zihlmann},
  \citenamefont {Michal}, \citenamefont {Apra}, \citenamefont {Abadillo-Uriel},
  \citenamefont {Jehl}, \citenamefont {Bertrand}, \citenamefont {Niebojewski},
  \citenamefont {Hutin}, \citenamefont {Vinet}, \citenamefont {Urdampilleta},
  \citenamefont {Meunier}, \citenamefont {Niquet}, \citenamefont {Maurand},\
  and\ \citenamefont {Franceschi}}]{Piot2022}%
  \BibitemOpen
  \bibfield  {author} {\bibinfo {author} {\bibfnamefont {N.}~\bibnamefont
  {Piot}}, \bibinfo {author} {\bibfnamefont {B.}~\bibnamefont {Brun}}, \bibinfo
  {author} {\bibfnamefont {V.}~\bibnamefont {Schmitt}}, \bibinfo {author}
  {\bibfnamefont {S.}~\bibnamefont {Zihlmann}}, \bibinfo {author}
  {\bibfnamefont {V.~P.}\ \bibnamefont {Michal}}, \bibinfo {author}
  {\bibfnamefont {A.}~\bibnamefont {Apra}}, \bibinfo {author} {\bibfnamefont
  {J.~C.}\ \bibnamefont {Abadillo-Uriel}}, \bibinfo {author} {\bibfnamefont
  {X.}~\bibnamefont {Jehl}}, \bibinfo {author} {\bibfnamefont {B.}~\bibnamefont
  {Bertrand}}, \bibinfo {author} {\bibfnamefont {H.}~\bibnamefont
  {Niebojewski}}, \bibinfo {author} {\bibfnamefont {L.}~\bibnamefont {Hutin}},
  \bibinfo {author} {\bibfnamefont {M.}~\bibnamefont {Vinet}}, \bibinfo
  {author} {\bibfnamefont {M.}~\bibnamefont {Urdampilleta}}, \bibinfo {author}
  {\bibfnamefont {T.}~\bibnamefont {Meunier}}, \bibinfo {author} {\bibfnamefont
  {Y.-M.}\ \bibnamefont {Niquet}}, \bibinfo {author} {\bibfnamefont
  {R.}~\bibnamefont {Maurand}},\ and\ \bibinfo {author} {\bibfnamefont {S.~D.}\
  \bibnamefont {Franceschi}},\ }\bibfield  {title} {\bibinfo {title} {A single
  hole spin with enhanced coherence in natural silicon},\ }\href
  {https://doi.org/10.1038/s41565-022-01196-z} {\bibfield  {journal} {\bibinfo
  {journal} {Nature Nanotechnology}\ }\textbf {\bibinfo {volume} {17}},\
  \bibinfo {pages} {1072} (\bibinfo {year} {2022})}\BibitemShut {NoStop}%
\bibitem [{\citenamefont {Auth}\ \emph {et~al.}(2012)\citenamefont {Auth},
  \citenamefont {Allen}, \citenamefont {Blattner}, \citenamefont {Bergstrom},
  \citenamefont {Brazier}, \citenamefont {Bost}, \citenamefont {Buehler},
  \citenamefont {Chikarmane}, \citenamefont {Ghani}, \citenamefont {Glassman},
  \citenamefont {Grover}, \citenamefont {Han}, \citenamefont {Hanken},
  \citenamefont {Hattendorf}, \citenamefont {Hentges}, \citenamefont
  {Heussner}, \citenamefont {Hicks}, \citenamefont {Ingerly}, \citenamefont
  {Jain}, \citenamefont {Jaloviar}, \citenamefont {James}, \citenamefont
  {Jones}, \citenamefont {Jopling}, \citenamefont {Joshi}, \citenamefont
  {Kenyon}, \citenamefont {Liu}, \citenamefont {McFadden}, \citenamefont
  {McIntyre}, \citenamefont {Neirynck}, \citenamefont {Parker}, \citenamefont
  {Pipes}, \citenamefont {Post}, \citenamefont {Pradhan}, \citenamefont
  {Prince}, \citenamefont {Ramey}, \citenamefont {Reynolds}, \citenamefont
  {Roesler}, \citenamefont {Sandford}, \citenamefont {Seiple}, \citenamefont
  {Smith}, \citenamefont {Thomas}, \citenamefont {Towner}, \citenamefont
  {Troeger}, \citenamefont {Weber}, \citenamefont {Yashar}, \citenamefont
  {Zawadzki},\ and\ \citenamefont {Mistry}}]{Auth2012}%
  \BibitemOpen
  \bibfield  {author} {\bibinfo {author} {\bibfnamefont {C.}~\bibnamefont
  {Auth}}, \bibinfo {author} {\bibfnamefont {C.}~\bibnamefont {Allen}},
  \bibinfo {author} {\bibfnamefont {A.}~\bibnamefont {Blattner}}, \bibinfo
  {author} {\bibfnamefont {D.}~\bibnamefont {Bergstrom}}, \bibinfo {author}
  {\bibfnamefont {M.}~\bibnamefont {Brazier}}, \bibinfo {author} {\bibfnamefont
  {M.}~\bibnamefont {Bost}}, \bibinfo {author} {\bibfnamefont {M.}~\bibnamefont
  {Buehler}}, \bibinfo {author} {\bibfnamefont {V.}~\bibnamefont {Chikarmane}},
  \bibinfo {author} {\bibfnamefont {T.}~\bibnamefont {Ghani}}, \bibinfo
  {author} {\bibfnamefont {T.}~\bibnamefont {Glassman}}, \bibinfo {author}
  {\bibfnamefont {R.}~\bibnamefont {Grover}}, \bibinfo {author} {\bibfnamefont
  {W.}~\bibnamefont {Han}}, \bibinfo {author} {\bibfnamefont {D.}~\bibnamefont
  {Hanken}}, \bibinfo {author} {\bibfnamefont {M.}~\bibnamefont {Hattendorf}},
  \bibinfo {author} {\bibfnamefont {P.}~\bibnamefont {Hentges}}, \bibinfo
  {author} {\bibfnamefont {R.}~\bibnamefont {Heussner}}, \bibinfo {author}
  {\bibfnamefont {J.}~\bibnamefont {Hicks}}, \bibinfo {author} {\bibfnamefont
  {D.}~\bibnamefont {Ingerly}}, \bibinfo {author} {\bibfnamefont
  {P.}~\bibnamefont {Jain}}, \bibinfo {author} {\bibfnamefont {S.}~\bibnamefont
  {Jaloviar}}, \bibinfo {author} {\bibfnamefont {R.}~\bibnamefont {James}},
  \bibinfo {author} {\bibfnamefont {D.}~\bibnamefont {Jones}}, \bibinfo
  {author} {\bibfnamefont {J.}~\bibnamefont {Jopling}}, \bibinfo {author}
  {\bibfnamefont {S.}~\bibnamefont {Joshi}}, \bibinfo {author} {\bibfnamefont
  {C.}~\bibnamefont {Kenyon}}, \bibinfo {author} {\bibfnamefont
  {H.}~\bibnamefont {Liu}}, \bibinfo {author} {\bibfnamefont {R.}~\bibnamefont
  {McFadden}}, \bibinfo {author} {\bibfnamefont {B.}~\bibnamefont {McIntyre}},
  \bibinfo {author} {\bibfnamefont {J.}~\bibnamefont {Neirynck}}, \bibinfo
  {author} {\bibfnamefont {C.}~\bibnamefont {Parker}}, \bibinfo {author}
  {\bibfnamefont {L.}~\bibnamefont {Pipes}}, \bibinfo {author} {\bibfnamefont
  {I.}~\bibnamefont {Post}}, \bibinfo {author} {\bibfnamefont {S.}~\bibnamefont
  {Pradhan}}, \bibinfo {author} {\bibfnamefont {M.}~\bibnamefont {Prince}},
  \bibinfo {author} {\bibfnamefont {S.}~\bibnamefont {Ramey}}, \bibinfo
  {author} {\bibfnamefont {T.}~\bibnamefont {Reynolds}}, \bibinfo {author}
  {\bibfnamefont {J.}~\bibnamefont {Roesler}}, \bibinfo {author} {\bibfnamefont
  {J.}~\bibnamefont {Sandford}}, \bibinfo {author} {\bibfnamefont
  {J.}~\bibnamefont {Seiple}}, \bibinfo {author} {\bibfnamefont
  {P.}~\bibnamefont {Smith}}, \bibinfo {author} {\bibfnamefont
  {C.}~\bibnamefont {Thomas}}, \bibinfo {author} {\bibfnamefont
  {D.}~\bibnamefont {Towner}}, \bibinfo {author} {\bibfnamefont
  {T.}~\bibnamefont {Troeger}}, \bibinfo {author} {\bibfnamefont
  {C.}~\bibnamefont {Weber}}, \bibinfo {author} {\bibfnamefont
  {P.}~\bibnamefont {Yashar}}, \bibinfo {author} {\bibfnamefont
  {K.}~\bibnamefont {Zawadzki}},\ and\ \bibinfo {author} {\bibfnamefont
  {K.}~\bibnamefont {Mistry}},\ }\bibfield  {title} {\bibinfo {title} {{A 22nm
  high performance and low-power {CMOS} technology featuring fully-depleted
  tri-gate transistors, self-aligned contacts and high density {MIM}
  capacitors}},\ }in\ \href {https://doi.org/10.1109/vlsit.2012.6242496} {\emph
  {\bibinfo {booktitle} {{2012 Symposium on {VLSI} Technology ({VLSIT})}}}}\
  (\bibinfo  {publisher} {{IEEE}},\ \bibinfo {year} {2012})\BibitemShut
  {NoStop}%
\bibitem [{\citenamefont {Maurand}\ \emph {et~al.}(2016)\citenamefont
  {Maurand}, \citenamefont {Jehl}, \citenamefont {Kotekar-Patil}, \citenamefont
  {Corna}, \citenamefont {Bohuslavskyi}, \citenamefont {Lavi{\'{e}}ville},
  \citenamefont {Hutin}, \citenamefont {Barraud}, \citenamefont {Vinet},
  \citenamefont {Sanquer},\ and\ \citenamefont {Franceschi}}]{Maurand2016}%
  \BibitemOpen
  \bibfield  {author} {\bibinfo {author} {\bibfnamefont {R.}~\bibnamefont
  {Maurand}}, \bibinfo {author} {\bibfnamefont {X.}~\bibnamefont {Jehl}},
  \bibinfo {author} {\bibfnamefont {D.}~\bibnamefont {Kotekar-Patil}}, \bibinfo
  {author} {\bibfnamefont {A.}~\bibnamefont {Corna}}, \bibinfo {author}
  {\bibfnamefont {H.}~\bibnamefont {Bohuslavskyi}}, \bibinfo {author}
  {\bibfnamefont {R.}~\bibnamefont {Lavi{\'{e}}ville}}, \bibinfo {author}
  {\bibfnamefont {L.}~\bibnamefont {Hutin}}, \bibinfo {author} {\bibfnamefont
  {S.}~\bibnamefont {Barraud}}, \bibinfo {author} {\bibfnamefont
  {M.}~\bibnamefont {Vinet}}, \bibinfo {author} {\bibfnamefont
  {M.}~\bibnamefont {Sanquer}},\ and\ \bibinfo {author} {\bibfnamefont {S.~D.}\
  \bibnamefont {Franceschi}},\ }\bibfield  {title} {\bibinfo {title} {{A {CMOS}
  silicon spin qubit}},\ }\href {https://doi.org/10.1038/ncomms13575}
  {\bibfield  {journal} {\bibinfo  {journal} {Nature Communications}\ }\textbf
  {\bibinfo {volume} {7}},\ \bibinfo {pages} {13575} (\bibinfo {year}
  {2016})}\BibitemShut {NoStop}%
\bibitem [{\citenamefont {Kuhlmann}\ \emph {et~al.}(2018)\citenamefont
  {Kuhlmann}, \citenamefont {Deshpande}, \citenamefont {Camenzind},
  \citenamefont {Zumb{\" u}hl},\ and\ \citenamefont {Fuhrer}}]{Kuhlmann2018}%
  \BibitemOpen
  \bibfield  {author} {\bibinfo {author} {\bibfnamefont {A.~V.}\ \bibnamefont
  {Kuhlmann}}, \bibinfo {author} {\bibfnamefont {V.}~\bibnamefont {Deshpande}},
  \bibinfo {author} {\bibfnamefont {L.~C.}\ \bibnamefont {Camenzind}}, \bibinfo
  {author} {\bibfnamefont {D.~M.}\ \bibnamefont {Zumb{\" u}hl}},\ and\ \bibinfo
  {author} {\bibfnamefont {A.}~\bibnamefont {Fuhrer}},\ }\bibfield  {title}
  {\bibinfo {title} {{Ambipolar quantum dots in undoped silicon fin
  field-effect transistors}},\ }\href {https://doi.org/10.1063/1.5048097}
  {\bibfield  {journal} {\bibinfo  {journal} {Applied Physics Letters}\
  }\textbf {\bibinfo {volume} {113}},\ \bibinfo {pages} {122107} (\bibinfo
  {year} {2018})}\BibitemShut {NoStop}%
\bibitem [{\citenamefont {Geyer}\ \emph {et~al.}(2021)\citenamefont {Geyer},
  \citenamefont {Camenzind}, \citenamefont {Czornomaz}, \citenamefont
  {Deshpande}, \citenamefont {Fuhrer}, \citenamefont {Warburton}, \citenamefont
  {Zumbühl},\ and\ \citenamefont {Kuhlmann}}]{Geyer2021}%
  \BibitemOpen
  \bibfield  {author} {\bibinfo {author} {\bibfnamefont {S.}~\bibnamefont
  {Geyer}}, \bibinfo {author} {\bibfnamefont {L.~C.}\ \bibnamefont
  {Camenzind}}, \bibinfo {author} {\bibfnamefont {L.}~\bibnamefont
  {Czornomaz}}, \bibinfo {author} {\bibfnamefont {V.}~\bibnamefont
  {Deshpande}}, \bibinfo {author} {\bibfnamefont {A.}~\bibnamefont {Fuhrer}},
  \bibinfo {author} {\bibfnamefont {R.~J.}\ \bibnamefont {Warburton}}, \bibinfo
  {author} {\bibfnamefont {D.~M.}\ \bibnamefont {Zumbühl}},\ and\ \bibinfo
  {author} {\bibfnamefont {A.~V.}\ \bibnamefont {Kuhlmann}},\ }\bibfield
  {title} {\bibinfo {title} {{Self-aligned gates for scalable silicon quantum
  computing}},\ }\href {https://doi.org/10.1063/5.0036520} {\bibfield
  {journal} {\bibinfo  {journal} {Applied Physics Letters}\ }\textbf {\bibinfo
  {volume} {118}},\ \bibinfo {pages} {104004} (\bibinfo {year}
  {2021})}\BibitemShut {NoStop}%
\bibitem [{\citenamefont {Zwerver}\ \emph {et~al.}(2022)\citenamefont
  {Zwerver}, \citenamefont {Krähenmann}, \citenamefont {Watson}, \citenamefont
  {Lampert}, \citenamefont {George}, \citenamefont {Pillarisetty},
  \citenamefont {Bojarski}, \citenamefont {Amin}, \citenamefont {Amitonov},
  \citenamefont {Boter}, \citenamefont {Caudillo}, \citenamefont
  {Correas-Serrano}, \citenamefont {Dehollain}, \citenamefont {Droulers},
  \citenamefont {Henry}, \citenamefont {Kotlyar}, \citenamefont {Lodari},
  \citenamefont {Lüthi}, \citenamefont {Michalak}, \citenamefont {Mueller},
  \citenamefont {Neyens}, \citenamefont {Roberts}, \citenamefont {Samkharadze},
  \citenamefont {Zheng}, \citenamefont {Zietz}, \citenamefont {Scappucci},
  \citenamefont {Veldhorst}, \citenamefont {Vandersypen},\ and\ \citenamefont
  {Clarke}}]{Zwerver2022}%
  \BibitemOpen
  \bibfield  {author} {\bibinfo {author} {\bibfnamefont {A.~M.~J.}\
  \bibnamefont {Zwerver}}, \bibinfo {author} {\bibfnamefont {T.}~\bibnamefont
  {Krähenmann}}, \bibinfo {author} {\bibfnamefont {T.~F.}\ \bibnamefont
  {Watson}}, \bibinfo {author} {\bibfnamefont {L.}~\bibnamefont {Lampert}},
  \bibinfo {author} {\bibfnamefont {H.~C.}\ \bibnamefont {George}}, \bibinfo
  {author} {\bibfnamefont {R.}~\bibnamefont {Pillarisetty}}, \bibinfo {author}
  {\bibfnamefont {S.~A.}\ \bibnamefont {Bojarski}}, \bibinfo {author}
  {\bibfnamefont {P.}~\bibnamefont {Amin}}, \bibinfo {author} {\bibfnamefont
  {S.~V.}\ \bibnamefont {Amitonov}}, \bibinfo {author} {\bibfnamefont {J.~M.}\
  \bibnamefont {Boter}}, \bibinfo {author} {\bibfnamefont {R.}~\bibnamefont
  {Caudillo}}, \bibinfo {author} {\bibfnamefont {D.}~\bibnamefont
  {Correas-Serrano}}, \bibinfo {author} {\bibfnamefont {J.~P.}\ \bibnamefont
  {Dehollain}}, \bibinfo {author} {\bibfnamefont {G.}~\bibnamefont {Droulers}},
  \bibinfo {author} {\bibfnamefont {E.~M.}\ \bibnamefont {Henry}}, \bibinfo
  {author} {\bibfnamefont {R.}~\bibnamefont {Kotlyar}}, \bibinfo {author}
  {\bibfnamefont {M.}~\bibnamefont {Lodari}}, \bibinfo {author} {\bibfnamefont
  {F.}~\bibnamefont {Lüthi}}, \bibinfo {author} {\bibfnamefont {D.~J.}\
  \bibnamefont {Michalak}}, \bibinfo {author} {\bibfnamefont {B.~K.}\
  \bibnamefont {Mueller}}, \bibinfo {author} {\bibfnamefont {S.}~\bibnamefont
  {Neyens}}, \bibinfo {author} {\bibfnamefont {J.}~\bibnamefont {Roberts}},
  \bibinfo {author} {\bibfnamefont {N.}~\bibnamefont {Samkharadze}}, \bibinfo
  {author} {\bibfnamefont {G.}~\bibnamefont {Zheng}}, \bibinfo {author}
  {\bibfnamefont {O.~K.}\ \bibnamefont {Zietz}}, \bibinfo {author}
  {\bibfnamefont {G.}~\bibnamefont {Scappucci}}, \bibinfo {author}
  {\bibfnamefont {M.}~\bibnamefont {Veldhorst}}, \bibinfo {author}
  {\bibfnamefont {L.~M.~K.}\ \bibnamefont {Vandersypen}},\ and\ \bibinfo
  {author} {\bibfnamefont {J.~S.}\ \bibnamefont {Clarke}},\ }\bibfield  {title}
  {\bibinfo {title} {{Qubits made by advanced semiconductor manufacturing}},\
  }\href {https://doi.org/10.1038/s41928-022-00727-9} {\bibfield  {journal}
  {\bibinfo  {journal} {Nature Electronics}\ }\textbf {\bibinfo {volume} {5}},\
  \bibinfo {pages} {184} (\bibinfo {year} {2022})}\BibitemShut {NoStop}%
\bibitem [{\citenamefont {Camenzind}\ \emph {et~al.}(2022)\citenamefont
  {Camenzind}, \citenamefont {Geyer}, \citenamefont {Fuhrer}, \citenamefont
  {Warburton}, \citenamefont {Zumbühl},\ and\ \citenamefont
  {Kuhlmann}}]{Camenzind2022}%
  \BibitemOpen
  \bibfield  {author} {\bibinfo {author} {\bibfnamefont {L.~C.}\ \bibnamefont
  {Camenzind}}, \bibinfo {author} {\bibfnamefont {S.}~\bibnamefont {Geyer}},
  \bibinfo {author} {\bibfnamefont {A.}~\bibnamefont {Fuhrer}}, \bibinfo
  {author} {\bibfnamefont {R.~J.}\ \bibnamefont {Warburton}}, \bibinfo {author}
  {\bibfnamefont {D.~M.}\ \bibnamefont {Zumbühl}},\ and\ \bibinfo {author}
  {\bibfnamefont {A.~V.}\ \bibnamefont {Kuhlmann}},\ }\bibfield  {title}
  {\bibinfo {title} {{A hole spin qubit in a fin field-effect transistor above
  4{\hspace{0.167em}}kelvin}},\ }\href
  {https://doi.org/10.1038/s41928-022-00722-0} {\bibfield  {journal} {\bibinfo
  {journal} {Nature Electronics}\ }\textbf {\bibinfo {volume} {5}},\ \bibinfo
  {pages} {178} (\bibinfo {year} {2022})}\BibitemShut {NoStop}%
\bibitem [{\citenamefont {Gonzalez-Zalba}\ \emph {et~al.}(2021)\citenamefont
  {Gonzalez-Zalba}, \citenamefont {de~Franceschi}, \citenamefont {Charbon},
  \citenamefont {Meunier}, \citenamefont {Vinet},\ and\ \citenamefont
  {Dzurak}}]{GonzalezZalba2021}%
  \BibitemOpen
  \bibfield  {author} {\bibinfo {author} {\bibfnamefont {M.~F.}\ \bibnamefont
  {Gonzalez-Zalba}}, \bibinfo {author} {\bibfnamefont {S.}~\bibnamefont
  {de~Franceschi}}, \bibinfo {author} {\bibfnamefont {E.}~\bibnamefont
  {Charbon}}, \bibinfo {author} {\bibfnamefont {T.}~\bibnamefont {Meunier}},
  \bibinfo {author} {\bibfnamefont {M.}~\bibnamefont {Vinet}},\ and\ \bibinfo
  {author} {\bibfnamefont {A.~S.}\ \bibnamefont {Dzurak}},\ }\bibfield  {title}
  {\bibinfo {title} {{Scaling silicon-based quantum computing using {CMOS}
  technology}},\ }\href {https://doi.org/10.1038/s41928-021-00681-y} {\bibfield
   {journal} {\bibinfo  {journal} {Nature Electronics}\ }\textbf {\bibinfo
  {volume} {4}},\ \bibinfo {pages} {872} (\bibinfo {year} {2021})}\BibitemShut
  {NoStop}%
\bibitem [{\citenamefont {Petit}\ \emph {et~al.}(2020)\citenamefont {Petit},
  \citenamefont {Eenink}, \citenamefont {Russ}, \citenamefont {Lawrie},
  \citenamefont {Hendrickx}, \citenamefont {Philips}, \citenamefont {Clarke},
  \citenamefont {Vandersypen},\ and\ \citenamefont {Veldhorst}}]{Petit2020}%
  \BibitemOpen
  \bibfield  {author} {\bibinfo {author} {\bibfnamefont {L.}~\bibnamefont
  {Petit}}, \bibinfo {author} {\bibfnamefont {H.~G.~J.}\ \bibnamefont
  {Eenink}}, \bibinfo {author} {\bibfnamefont {M.}~\bibnamefont {Russ}},
  \bibinfo {author} {\bibfnamefont {W.~I.~L.}\ \bibnamefont {Lawrie}}, \bibinfo
  {author} {\bibfnamefont {N.~W.}\ \bibnamefont {Hendrickx}}, \bibinfo {author}
  {\bibfnamefont {S.~G.~J.}\ \bibnamefont {Philips}}, \bibinfo {author}
  {\bibfnamefont {J.~S.}\ \bibnamefont {Clarke}}, \bibinfo {author}
  {\bibfnamefont {L.~M.~K.}\ \bibnamefont {Vandersypen}},\ and\ \bibinfo
  {author} {\bibfnamefont {M.}~\bibnamefont {Veldhorst}},\ }\bibfield  {title}
  {\bibinfo {title} {{Universal quantum logic in hot silicon qubits}},\ }\href
  {https://doi.org/10.1038/s41586-020-2170-7} {\bibfield  {journal} {\bibinfo
  {journal} {Nature}\ }\textbf {\bibinfo {volume} {580}},\ \bibinfo {pages}
  {355} (\bibinfo {year} {2020})}\BibitemShut {NoStop}%
\bibitem [{\citenamefont {Yang}\ \emph {et~al.}(2020)\citenamefont {Yang},
  \citenamefont {Leon}, \citenamefont {Hwang}, \citenamefont {Saraiva},
  \citenamefont {Tanttu}, \citenamefont {Huang}, \citenamefont {Lemyre},
  \citenamefont {Chan}, \citenamefont {Tan}, \citenamefont {Hudson},
  \citenamefont {Itoh}, \citenamefont {Morello}, \citenamefont
  {Pioro-Ladri{\`{e}}re}, \citenamefont {Laucht},\ and\ \citenamefont
  {Dzurak}}]{Yang2020}%
  \BibitemOpen
  \bibfield  {author} {\bibinfo {author} {\bibfnamefont {C.~H.}\ \bibnamefont
  {Yang}}, \bibinfo {author} {\bibfnamefont {R.~C.~C.}\ \bibnamefont {Leon}},
  \bibinfo {author} {\bibfnamefont {J.~C.~C.}\ \bibnamefont {Hwang}}, \bibinfo
  {author} {\bibfnamefont {A.}~\bibnamefont {Saraiva}}, \bibinfo {author}
  {\bibfnamefont {T.}~\bibnamefont {Tanttu}}, \bibinfo {author} {\bibfnamefont
  {W.}~\bibnamefont {Huang}}, \bibinfo {author} {\bibfnamefont {J.~C.}\
  \bibnamefont {Lemyre}}, \bibinfo {author} {\bibfnamefont {K.~W.}\
  \bibnamefont {Chan}}, \bibinfo {author} {\bibfnamefont {K.~Y.}\ \bibnamefont
  {Tan}}, \bibinfo {author} {\bibfnamefont {F.~E.}\ \bibnamefont {Hudson}},
  \bibinfo {author} {\bibfnamefont {K.~M.}\ \bibnamefont {Itoh}}, \bibinfo
  {author} {\bibfnamefont {A.}~\bibnamefont {Morello}}, \bibinfo {author}
  {\bibfnamefont {M.}~\bibnamefont {Pioro-Ladri{\`{e}}re}}, \bibinfo {author}
  {\bibfnamefont {A.}~\bibnamefont {Laucht}},\ and\ \bibinfo {author}
  {\bibfnamefont {A.~S.}\ \bibnamefont {Dzurak}},\ }\bibfield  {title}
  {\bibinfo {title} {{Operation of a silicon quantum processor unit cell above
  one kelvin}},\ }\href {https://doi.org/10.1038/s41586-020-2171-6} {\bibfield
  {journal} {\bibinfo  {journal} {Nature}\ }\textbf {\bibinfo {volume} {580}},\
  \bibinfo {pages} {350} (\bibinfo {year} {2020})}\BibitemShut {NoStop}%
\bibitem [{\citenamefont {Xue}\ \emph {et~al.}(2021)\citenamefont {Xue},
  \citenamefont {Patra}, \citenamefont {van Dijk}, \citenamefont {Samkharadze},
  \citenamefont {Subramanian}, \citenamefont {Corna}, \citenamefont {Wuetz},
  \citenamefont {Jeon}, \citenamefont {Sheikh}, \citenamefont
  {Juarez-Hernandez}, \citenamefont {Esparza}, \citenamefont {Rampurawala},
  \citenamefont {Carlton}, \citenamefont {Ravikumar}, \citenamefont {Nieva},
  \citenamefont {Kim}, \citenamefont {Lee}, \citenamefont {Sammak},
  \citenamefont {Scappucci}, \citenamefont {Veldhorst}, \citenamefont
  {Sebastiano}, \citenamefont {Babaie}, \citenamefont {Pellerano},
  \citenamefont {Charbon},\ and\ \citenamefont {Vandersypen}}]{Xue2021}%
  \BibitemOpen
  \bibfield  {author} {\bibinfo {author} {\bibfnamefont {X.}~\bibnamefont
  {Xue}}, \bibinfo {author} {\bibfnamefont {B.}~\bibnamefont {Patra}}, \bibinfo
  {author} {\bibfnamefont {J.~P.~G.}\ \bibnamefont {van Dijk}}, \bibinfo
  {author} {\bibfnamefont {N.}~\bibnamefont {Samkharadze}}, \bibinfo {author}
  {\bibfnamefont {S.}~\bibnamefont {Subramanian}}, \bibinfo {author}
  {\bibfnamefont {A.}~\bibnamefont {Corna}}, \bibinfo {author} {\bibfnamefont
  {B.~P.}\ \bibnamefont {Wuetz}}, \bibinfo {author} {\bibfnamefont
  {C.}~\bibnamefont {Jeon}}, \bibinfo {author} {\bibfnamefont {F.}~\bibnamefont
  {Sheikh}}, \bibinfo {author} {\bibfnamefont {E.}~\bibnamefont
  {Juarez-Hernandez}}, \bibinfo {author} {\bibfnamefont {B.~P.}\ \bibnamefont
  {Esparza}}, \bibinfo {author} {\bibfnamefont {H.}~\bibnamefont
  {Rampurawala}}, \bibinfo {author} {\bibfnamefont {B.}~\bibnamefont
  {Carlton}}, \bibinfo {author} {\bibfnamefont {S.}~\bibnamefont {Ravikumar}},
  \bibinfo {author} {\bibfnamefont {C.}~\bibnamefont {Nieva}}, \bibinfo
  {author} {\bibfnamefont {S.}~\bibnamefont {Kim}}, \bibinfo {author}
  {\bibfnamefont {H.-J.}\ \bibnamefont {Lee}}, \bibinfo {author} {\bibfnamefont
  {A.}~\bibnamefont {Sammak}}, \bibinfo {author} {\bibfnamefont
  {G.}~\bibnamefont {Scappucci}}, \bibinfo {author} {\bibfnamefont
  {M.}~\bibnamefont {Veldhorst}}, \bibinfo {author} {\bibfnamefont
  {F.}~\bibnamefont {Sebastiano}}, \bibinfo {author} {\bibfnamefont
  {M.}~\bibnamefont {Babaie}}, \bibinfo {author} {\bibfnamefont
  {S.}~\bibnamefont {Pellerano}}, \bibinfo {author} {\bibfnamefont
  {E.}~\bibnamefont {Charbon}},\ and\ \bibinfo {author} {\bibfnamefont
  {L.~M.~K.}\ \bibnamefont {Vandersypen}},\ }\bibfield  {title} {\bibinfo
  {title} {{{CMOS}-based cryogenic control of silicon quantum circuits}},\
  }\href {https://doi.org/10.1038/s41586-021-03469-4} {\bibfield  {journal}
  {\bibinfo  {journal} {Nature}\ }\textbf {\bibinfo {volume} {593}},\ \bibinfo
  {pages} {205} (\bibinfo {year} {2021})}\BibitemShut {NoStop}%
\bibitem [{\citenamefont {Petta}\ \emph {et~al.}(2005)\citenamefont {Petta},
  \citenamefont {Johnson}, \citenamefont {Taylor}, \citenamefont {Laird},
  \citenamefont {Yacoby}, \citenamefont {Lukin}, \citenamefont {Marcus},
  \citenamefont {Hanson},\ and\ \citenamefont {Gossard}}]{Petta2005}%
  \BibitemOpen
  \bibfield  {author} {\bibinfo {author} {\bibfnamefont {J.~R.}\ \bibnamefont
  {Petta}}, \bibinfo {author} {\bibfnamefont {A.~C.}\ \bibnamefont {Johnson}},
  \bibinfo {author} {\bibfnamefont {J.~M.}\ \bibnamefont {Taylor}}, \bibinfo
  {author} {\bibfnamefont {E.~A.}\ \bibnamefont {Laird}}, \bibinfo {author}
  {\bibfnamefont {A.}~\bibnamefont {Yacoby}}, \bibinfo {author} {\bibfnamefont
  {M.~D.}\ \bibnamefont {Lukin}}, \bibinfo {author} {\bibfnamefont {C.~M.}\
  \bibnamefont {Marcus}}, \bibinfo {author} {\bibfnamefont {M.~P.}\
  \bibnamefont {Hanson}},\ and\ \bibinfo {author} {\bibfnamefont {A.~C.}\
  \bibnamefont {Gossard}},\ }\bibfield  {title} {\bibinfo {title} {{Coherent
  Manipulation of Coupled Electron Spins in Semiconductor Quantum Dots}},\
  }\href {https://doi.org/10.1126/science.1116955} {\bibfield  {journal}
  {\bibinfo  {journal} {Science}\ }\textbf {\bibinfo {volume} {309}},\ \bibinfo
  {pages} {2180} (\bibinfo {year} {2005})}\BibitemShut {NoStop}%
\bibitem [{\citenamefont {Veldhorst}\ \emph {et~al.}(2015)\citenamefont
  {Veldhorst}, \citenamefont {Yang}, \citenamefont {Hwang}, \citenamefont
  {Huang}, \citenamefont {Dehollain}, \citenamefont {Muhonen}, \citenamefont
  {Simmons}, \citenamefont {Laucht}, \citenamefont {Hudson}, \citenamefont
  {Itoh}, \citenamefont {Morello},\ and\ \citenamefont
  {Dzurak}}]{Veldhorst2015}%
  \BibitemOpen
  \bibfield  {author} {\bibinfo {author} {\bibfnamefont {M.}~\bibnamefont
  {Veldhorst}}, \bibinfo {author} {\bibfnamefont {C.~H.}\ \bibnamefont {Yang}},
  \bibinfo {author} {\bibfnamefont {J.~C.~C.}\ \bibnamefont {Hwang}}, \bibinfo
  {author} {\bibfnamefont {W.}~\bibnamefont {Huang}}, \bibinfo {author}
  {\bibfnamefont {J.~P.}\ \bibnamefont {Dehollain}}, \bibinfo {author}
  {\bibfnamefont {J.~T.}\ \bibnamefont {Muhonen}}, \bibinfo {author}
  {\bibfnamefont {S.}~\bibnamefont {Simmons}}, \bibinfo {author} {\bibfnamefont
  {A.}~\bibnamefont {Laucht}}, \bibinfo {author} {\bibfnamefont {F.~E.}\
  \bibnamefont {Hudson}}, \bibinfo {author} {\bibfnamefont {K.~M.}\
  \bibnamefont {Itoh}}, \bibinfo {author} {\bibfnamefont {A.}~\bibnamefont
  {Morello}},\ and\ \bibinfo {author} {\bibfnamefont {A.~S.}\ \bibnamefont
  {Dzurak}},\ }\bibfield  {title} {\bibinfo {title} {{A two-qubit logic gate in
  silicon}},\ }\href {https://doi.org/10.1038/nature15263} {\bibfield
  {journal} {\bibinfo  {journal} {Nature}\ }\textbf {\bibinfo {volume} {526}},\
  \bibinfo {pages} {410} (\bibinfo {year} {2015})}\BibitemShut {NoStop}%
\bibitem [{\citenamefont {Watson}\ \emph {et~al.}(2018)\citenamefont {Watson},
  \citenamefont {Philips}, \citenamefont {Kawakami}, \citenamefont {Ward},
  \citenamefont {Scarlino}, \citenamefont {Veldhorst}, \citenamefont {Savage},
  \citenamefont {Lagally}, \citenamefont {Friesen}, \citenamefont
  {Coppersmith}, \citenamefont {Eriksson},\ and\ \citenamefont
  {Vandersypen}}]{Watson2018}%
  \BibitemOpen
  \bibfield  {author} {\bibinfo {author} {\bibfnamefont {T.~F.}\ \bibnamefont
  {Watson}}, \bibinfo {author} {\bibfnamefont {S.~G.~J.}\ \bibnamefont
  {Philips}}, \bibinfo {author} {\bibfnamefont {E.}~\bibnamefont {Kawakami}},
  \bibinfo {author} {\bibfnamefont {D.~R.}\ \bibnamefont {Ward}}, \bibinfo
  {author} {\bibfnamefont {P.}~\bibnamefont {Scarlino}}, \bibinfo {author}
  {\bibfnamefont {M.}~\bibnamefont {Veldhorst}}, \bibinfo {author}
  {\bibfnamefont {D.~E.}\ \bibnamefont {Savage}}, \bibinfo {author}
  {\bibfnamefont {M.~G.}\ \bibnamefont {Lagally}}, \bibinfo {author}
  {\bibfnamefont {M.}~\bibnamefont {Friesen}}, \bibinfo {author} {\bibfnamefont
  {S.~N.}\ \bibnamefont {Coppersmith}}, \bibinfo {author} {\bibfnamefont
  {M.~A.}\ \bibnamefont {Eriksson}},\ and\ \bibinfo {author} {\bibfnamefont
  {L.~M.~K.}\ \bibnamefont {Vandersypen}},\ }\bibfield  {title} {\bibinfo
  {title} {{A programmable two-qubit quantum processor in silicon}},\ }\href
  {https://doi.org/10.1038/nature25766} {\bibfield  {journal} {\bibinfo
  {journal} {Nature}\ }\textbf {\bibinfo {volume} {555}},\ \bibinfo {pages}
  {633} (\bibinfo {year} {2018})}\BibitemShut {NoStop}%
\bibitem [{\citenamefont {Mills}\ \emph {et~al.}(2022)\citenamefont {Mills},
  \citenamefont {Guinn}, \citenamefont {Gullans}, \citenamefont {Sigillito},
  \citenamefont {Feldman}, \citenamefont {Nielsen},\ and\ \citenamefont
  {Petta}}]{Mills2022}%
  \BibitemOpen
  \bibfield  {author} {\bibinfo {author} {\bibfnamefont {A.~R.}\ \bibnamefont
  {Mills}}, \bibinfo {author} {\bibfnamefont {C.~R.}\ \bibnamefont {Guinn}},
  \bibinfo {author} {\bibfnamefont {M.~J.}\ \bibnamefont {Gullans}}, \bibinfo
  {author} {\bibfnamefont {A.~J.}\ \bibnamefont {Sigillito}}, \bibinfo {author}
  {\bibfnamefont {M.~M.}\ \bibnamefont {Feldman}}, \bibinfo {author}
  {\bibfnamefont {E.}~\bibnamefont {Nielsen}},\ and\ \bibinfo {author}
  {\bibfnamefont {J.~R.}\ \bibnamefont {Petta}},\ }\bibfield  {title} {\bibinfo
  {title} {{Two-qubit silicon quantum processor with operation fidelity
  exceeding 99{\%}}},\ }\href {https://doi.org/10.1126/sciadv.abn5130}
  {\bibfield  {journal} {\bibinfo  {journal} {Science Advances}\ }\textbf
  {\bibinfo {volume} {8}},\ \bibinfo {pages} {14} (\bibinfo {year}
  {2022})}\BibitemShut {NoStop}%
\bibitem [{\citenamefont {Xue}\ \emph {et~al.}(2022)\citenamefont {Xue},
  \citenamefont {Russ}, \citenamefont {Samkharadze}, \citenamefont {Undseth},
  \citenamefont {Sammak}, \citenamefont {Scappucci},\ and\ \citenamefont
  {Vandersypen}}]{Xue2022}%
  \BibitemOpen
  \bibfield  {author} {\bibinfo {author} {\bibfnamefont {X.}~\bibnamefont
  {Xue}}, \bibinfo {author} {\bibfnamefont {M.}~\bibnamefont {Russ}}, \bibinfo
  {author} {\bibfnamefont {N.}~\bibnamefont {Samkharadze}}, \bibinfo {author}
  {\bibfnamefont {B.}~\bibnamefont {Undseth}}, \bibinfo {author} {\bibfnamefont
  {A.}~\bibnamefont {Sammak}}, \bibinfo {author} {\bibfnamefont
  {G.}~\bibnamefont {Scappucci}},\ and\ \bibinfo {author} {\bibfnamefont
  {L.~M.~K.}\ \bibnamefont {Vandersypen}},\ }\bibfield  {title} {\bibinfo
  {title} {{Quantum logic with spin qubits crossing the surface code
  threshold}},\ }\href {https://doi.org/10.1038/s41586-021-04273-w} {\bibfield
  {journal} {\bibinfo  {journal} {Nature}\ }\textbf {\bibinfo {volume} {601}},\
  \bibinfo {pages} {343} (\bibinfo {year} {2022})}\BibitemShut {NoStop}%
\bibitem [{\citenamefont {Zajac}\ \emph {et~al.}(2018)\citenamefont {Zajac},
  \citenamefont {Sigillito}, \citenamefont {Russ}, \citenamefont {Borjans},
  \citenamefont {Taylor}, \citenamefont {Burkard},\ and\ \citenamefont
  {Petta}}]{Zajac2018}%
  \BibitemOpen
  \bibfield  {author} {\bibinfo {author} {\bibfnamefont {D.~M.}\ \bibnamefont
  {Zajac}}, \bibinfo {author} {\bibfnamefont {A.~J.}\ \bibnamefont
  {Sigillito}}, \bibinfo {author} {\bibfnamefont {M.}~\bibnamefont {Russ}},
  \bibinfo {author} {\bibfnamefont {F.}~\bibnamefont {Borjans}}, \bibinfo
  {author} {\bibfnamefont {J.~M.}\ \bibnamefont {Taylor}}, \bibinfo {author}
  {\bibfnamefont {G.}~\bibnamefont {Burkard}},\ and\ \bibinfo {author}
  {\bibfnamefont {J.~R.}\ \bibnamefont {Petta}},\ }\bibfield  {title} {\bibinfo
  {title} {{Resonantly driven {CNOT} gate for electron spins}},\ }\href
  {https://doi.org/10.1126/science.aao5965} {\bibfield  {journal} {\bibinfo
  {journal} {Science}\ }\textbf {\bibinfo {volume} {359}},\ \bibinfo {pages}
  {439} (\bibinfo {year} {2018})}\BibitemShut {NoStop}%
\bibitem [{\citenamefont {Huang}\ \emph {et~al.}(2019)\citenamefont {Huang},
  \citenamefont {Yang}, \citenamefont {Chan}, \citenamefont {Tanttu},
  \citenamefont {Hensen}, \citenamefont {Leon}, \citenamefont {Fogarty},
  \citenamefont {Hwang}, \citenamefont {Hudson}, \citenamefont {Itoh},
  \citenamefont {Morello}, \citenamefont {Laucht},\ and\ \citenamefont
  {Dzurak}}]{Huang2019}%
  \BibitemOpen
  \bibfield  {author} {\bibinfo {author} {\bibfnamefont {W.}~\bibnamefont
  {Huang}}, \bibinfo {author} {\bibfnamefont {C.~H.}\ \bibnamefont {Yang}},
  \bibinfo {author} {\bibfnamefont {K.~W.}\ \bibnamefont {Chan}}, \bibinfo
  {author} {\bibfnamefont {T.}~\bibnamefont {Tanttu}}, \bibinfo {author}
  {\bibfnamefont {B.}~\bibnamefont {Hensen}}, \bibinfo {author} {\bibfnamefont
  {R.~C.~C.}\ \bibnamefont {Leon}}, \bibinfo {author} {\bibfnamefont {M.~A.}\
  \bibnamefont {Fogarty}}, \bibinfo {author} {\bibfnamefont {J.~C.~C.}\
  \bibnamefont {Hwang}}, \bibinfo {author} {\bibfnamefont {F.~E.}\ \bibnamefont
  {Hudson}}, \bibinfo {author} {\bibfnamefont {K.~M.}\ \bibnamefont {Itoh}},
  \bibinfo {author} {\bibfnamefont {A.}~\bibnamefont {Morello}}, \bibinfo
  {author} {\bibfnamefont {A.}~\bibnamefont {Laucht}},\ and\ \bibinfo {author}
  {\bibfnamefont {A.~S.}\ \bibnamefont {Dzurak}},\ }\bibfield  {title}
  {\bibinfo {title} {{Fidelity benchmarks for two-qubit gates in silicon}},\
  }\href {https://doi.org/10.1038/s41586-019-1197-0} {\bibfield  {journal}
  {\bibinfo  {journal} {Nature}\ }\textbf {\bibinfo {volume} {569}},\ \bibinfo
  {pages} {532} (\bibinfo {year} {2019})}\BibitemShut {NoStop}%
\bibitem [{\citenamefont {Noiri}\ \emph {et~al.}(2022)\citenamefont {Noiri},
  \citenamefont {Takeda}, \citenamefont {Nakajima}, \citenamefont {Kobayashi},
  \citenamefont {Sammak}, \citenamefont {Scappucci},\ and\ \citenamefont
  {Tarucha}}]{Noiri2022}%
  \BibitemOpen
  \bibfield  {author} {\bibinfo {author} {\bibfnamefont {A.}~\bibnamefont
  {Noiri}}, \bibinfo {author} {\bibfnamefont {K.}~\bibnamefont {Takeda}},
  \bibinfo {author} {\bibfnamefont {T.}~\bibnamefont {Nakajima}}, \bibinfo
  {author} {\bibfnamefont {T.}~\bibnamefont {Kobayashi}}, \bibinfo {author}
  {\bibfnamefont {A.}~\bibnamefont {Sammak}}, \bibinfo {author} {\bibfnamefont
  {G.}~\bibnamefont {Scappucci}},\ and\ \bibinfo {author} {\bibfnamefont
  {S.}~\bibnamefont {Tarucha}},\ }\bibfield  {title} {\bibinfo {title} {{Fast
  universal quantum gate above the fault-tolerance threshold in silicon}},\
  }\href {https://doi.org/10.1038/s41586-021-04182-y} {\bibfield  {journal}
  {\bibinfo  {journal} {Nature}\ }\textbf {\bibinfo {volume} {601}},\ \bibinfo
  {pages} {338} (\bibinfo {year} {2022})}\BibitemShut {NoStop}%
\bibitem [{\citenamefont {Fang}\ \emph {et~al.}(2022)\citenamefont {Fang},
  \citenamefont {Philippopoulos}, \citenamefont {Culcer}, \citenamefont
  {Coish},\ and\ \citenamefont {Chesi}}]{Fang2022}%
  \BibitemOpen
  \bibfield  {author} {\bibinfo {author} {\bibfnamefont {Y.}~\bibnamefont
  {Fang}}, \bibinfo {author} {\bibfnamefont {P.}~\bibnamefont
  {Philippopoulos}}, \bibinfo {author} {\bibfnamefont {D.}~\bibnamefont
  {Culcer}}, \bibinfo {author} {\bibfnamefont {W.~A.}\ \bibnamefont {Coish}},\
  and\ \bibinfo {author} {\bibfnamefont {S.}~\bibnamefont {Chesi}},\
  }\href@noop {} {\bibinfo {title} {Recent advances in hole-spin qubits}}
  (\bibinfo {year} {2022}),\ \Eprint {https://arxiv.org/abs/2210.13725}
  {arXiv:2210.13725} \BibitemShut {NoStop}%
\bibitem [{\citenamefont {Golovach}\ \emph {et~al.}(2006)\citenamefont
  {Golovach}, \citenamefont {Borhani},\ and\ \citenamefont
  {Loss}}]{Golovach2006}%
  \BibitemOpen
  \bibfield  {author} {\bibinfo {author} {\bibfnamefont {V.~N.}\ \bibnamefont
  {Golovach}}, \bibinfo {author} {\bibfnamefont {M.}~\bibnamefont {Borhani}},\
  and\ \bibinfo {author} {\bibfnamefont {D.}~\bibnamefont {Loss}},\ }\bibfield
  {title} {\bibinfo {title} {{Electric-dipole-induced spin resonance in quantum
  dots}},\ }\href {https://doi.org/10.1103/physrevb.74.165319} {\bibfield
  {journal} {\bibinfo  {journal} {Physical Review B}\ }\textbf {\bibinfo
  {volume} {74}},\ \bibinfo {pages} {165319} (\bibinfo {year}
  {2006})}\BibitemShut {NoStop}%
\bibitem [{\citenamefont {Nowack}\ \emph {et~al.}(2007)\citenamefont {Nowack},
  \citenamefont {Koppens}, \citenamefont {Nazarov},\ and\ \citenamefont
  {Vandersypen}}]{Nowack2007}%
  \BibitemOpen
  \bibfield  {author} {\bibinfo {author} {\bibfnamefont {K.~C.}\ \bibnamefont
  {Nowack}}, \bibinfo {author} {\bibfnamefont {F.~H.~L.}\ \bibnamefont
  {Koppens}}, \bibinfo {author} {\bibfnamefont {Y.~V.}\ \bibnamefont
  {Nazarov}},\ and\ \bibinfo {author} {\bibfnamefont {L.~M.~K.}\ \bibnamefont
  {Vandersypen}},\ }\bibfield  {title} {\bibinfo {title} {{Coherent Control of
  a Single Electron Spin with Electric Fields}},\ }\href
  {https://doi.org/10.1126/science.1148092} {\bibfield  {journal} {\bibinfo
  {journal} {Science}\ }\textbf {\bibinfo {volume} {318}},\ \bibinfo {pages}
  {1430} (\bibinfo {year} {2007})}\BibitemShut {NoStop}%
\bibitem [{\citenamefont {Stepanenko}\ \emph {et~al.}(2012)\citenamefont
  {Stepanenko}, \citenamefont {Rudner}, \citenamefont {Halperin},\ and\
  \citenamefont {Loss}}]{Stepanenko2012}%
  \BibitemOpen
  \bibfield  {author} {\bibinfo {author} {\bibfnamefont {D.}~\bibnamefont
  {Stepanenko}}, \bibinfo {author} {\bibfnamefont {M.}~\bibnamefont {Rudner}},
  \bibinfo {author} {\bibfnamefont {B.~I.}\ \bibnamefont {Halperin}},\ and\
  \bibinfo {author} {\bibfnamefont {D.}~\bibnamefont {Loss}},\ }\bibfield
  {title} {\bibinfo {title} {{Singlet-triplet splitting in double quantum dots
  due to spin-orbit and hyperfine interactions}},\ }\href
  {https://doi.org/10.1103/physrevb.85.075416} {\bibfield  {journal} {\bibinfo
  {journal} {Physical Review B}\ }\textbf {\bibinfo {volume} {85}},\ \bibinfo
  {pages} {075416} (\bibinfo {year} {2012})}\BibitemShut {NoStop}%
\bibitem [{\citenamefont {Russ}\ \emph {et~al.}(2018)\citenamefont {Russ},
  \citenamefont {Zajac}, \citenamefont {Sigillito}, \citenamefont {Borjans},
  \citenamefont {Taylor}, \citenamefont {Petta},\ and\ \citenamefont
  {Burkard}}]{Russ2018}%
  \BibitemOpen
  \bibfield  {author} {\bibinfo {author} {\bibfnamefont {M.}~\bibnamefont
  {Russ}}, \bibinfo {author} {\bibfnamefont {D.~M.}\ \bibnamefont {Zajac}},
  \bibinfo {author} {\bibfnamefont {A.~J.}\ \bibnamefont {Sigillito}}, \bibinfo
  {author} {\bibfnamefont {F.}~\bibnamefont {Borjans}}, \bibinfo {author}
  {\bibfnamefont {J.~M.}\ \bibnamefont {Taylor}}, \bibinfo {author}
  {\bibfnamefont {J.~R.}\ \bibnamefont {Petta}},\ and\ \bibinfo {author}
  {\bibfnamefont {G.}~\bibnamefont {Burkard}},\ }\bibfield  {title} {\bibinfo
  {title} {{High-fidelity quantum gates in {Si}/{SiGe} double quantum dots}},\
  }\href {https://doi.org/10.1103/physrevb.97.085421} {\bibfield  {journal}
  {\bibinfo  {journal} {Physical Review B}\ }\textbf {\bibinfo {volume} {97}},\
  \bibinfo {pages} {085421} (\bibinfo {year} {2018})}\BibitemShut {NoStop}%
\bibitem [{\citenamefont {Hendrickx}\ \emph
  {et~al.}(2020{\natexlab{a}})\citenamefont {Hendrickx}, \citenamefont
  {Franke}, \citenamefont {Sammak}, \citenamefont {Scappucci},\ and\
  \citenamefont {Veldhorst}}]{Hendrickx2020a}%
  \BibitemOpen
  \bibfield  {author} {\bibinfo {author} {\bibfnamefont {N.~W.}\ \bibnamefont
  {Hendrickx}}, \bibinfo {author} {\bibfnamefont {D.~P.}\ \bibnamefont
  {Franke}}, \bibinfo {author} {\bibfnamefont {A.}~\bibnamefont {Sammak}},
  \bibinfo {author} {\bibfnamefont {G.}~\bibnamefont {Scappucci}},\ and\
  \bibinfo {author} {\bibfnamefont {M.}~\bibnamefont {Veldhorst}},\ }\bibfield
  {title} {\bibinfo {title} {{Fast two-qubit logic with holes in germanium}},\
  }\href {https://doi.org/10.1038/s41586-019-1919-3} {\bibfield  {journal}
  {\bibinfo  {journal} {Nature}\ }\textbf {\bibinfo {volume} {577}},\ \bibinfo
  {pages} {487} (\bibinfo {year} {2020}{\natexlab{a}})}\BibitemShut {NoStop}%
\bibitem [{\citenamefont {Nadj-Perge}\ \emph {et~al.}(2012)\citenamefont
  {Nadj-Perge}, \citenamefont {Pribiag}, \citenamefont {van~den Berg},
  \citenamefont {Zuo}, \citenamefont {Plissard}, \citenamefont {Bakkers},
  \citenamefont {Frolov},\ and\ \citenamefont {Kouwenhoven}}]{NadjPerge2012}%
  \BibitemOpen
  \bibfield  {author} {\bibinfo {author} {\bibfnamefont {S.}~\bibnamefont
  {Nadj-Perge}}, \bibinfo {author} {\bibfnamefont {V.~S.}\ \bibnamefont
  {Pribiag}}, \bibinfo {author} {\bibfnamefont {J.~W.~G.}\ \bibnamefont
  {van~den Berg}}, \bibinfo {author} {\bibfnamefont {K.}~\bibnamefont {Zuo}},
  \bibinfo {author} {\bibfnamefont {S.~R.}\ \bibnamefont {Plissard}}, \bibinfo
  {author} {\bibfnamefont {E.~P. A.~M.}\ \bibnamefont {Bakkers}}, \bibinfo
  {author} {\bibfnamefont {S.~M.}\ \bibnamefont {Frolov}},\ and\ \bibinfo
  {author} {\bibfnamefont {L.~P.}\ \bibnamefont {Kouwenhoven}},\ }\bibfield
  {title} {\bibinfo {title} {{Spectroscopy of Spin-Orbit Quantum Bits in Indium
  Antimonide Nanowires}},\ }\href
  {https://doi.org/10.1103/physrevlett.108.166801} {\bibfield  {journal}
  {\bibinfo  {journal} {Physical Review Letters}\ }\textbf {\bibinfo {volume}
  {108}},\ \bibinfo {pages} {166801} (\bibinfo {year} {2012})}\BibitemShut
  {NoStop}%
\bibitem [{\citenamefont {Kavokin}(2001)}]{Kavokin2001}%
  \BibitemOpen
  \bibfield  {author} {\bibinfo {author} {\bibfnamefont {K.~V.}\ \bibnamefont
  {Kavokin}},\ }\bibfield  {title} {\bibinfo {title} {{Anisotropic exchange
  interaction of localized conduction-band electrons in semiconductors}},\
  }\href {https://doi.org/10.1103/physrevb.64.075305} {\bibfield  {journal}
  {\bibinfo  {journal} {Physical Review B}\ }\textbf {\bibinfo {volume} {64}},\
  \bibinfo {pages} {075305} (\bibinfo {year} {2001})}\BibitemShut {NoStop}%
\bibitem [{\citenamefont {Kavokin}(2004)}]{Kavokin2004}%
  \BibitemOpen
  \bibfield  {author} {\bibinfo {author} {\bibfnamefont {K.~V.}\ \bibnamefont
  {Kavokin}},\ }\bibfield  {title} {\bibinfo {title} {{Symmetry of anisotropic
  exchange interactions in semiconductor nanostructures}},\ }\href
  {https://doi.org/10.1103/physrevb.69.075302} {\bibfield  {journal} {\bibinfo
  {journal} {Physical Review B}\ }\textbf {\bibinfo {volume} {69}},\ \bibinfo
  {pages} {075302} (\bibinfo {year} {2004})}\BibitemShut {NoStop}%
\bibitem [{\citenamefont {Het{\'{e}}nyi}\ \emph {et~al.}(2020)\citenamefont
  {Het{\'{e}}nyi}, \citenamefont {Kloeffel},\ and\ \citenamefont
  {Loss}}]{Hetenyi2020}%
  \BibitemOpen
  \bibfield  {author} {\bibinfo {author} {\bibfnamefont {B.}~\bibnamefont
  {Het{\'{e}}nyi}}, \bibinfo {author} {\bibfnamefont {C.}~\bibnamefont
  {Kloeffel}},\ and\ \bibinfo {author} {\bibfnamefont {D.}~\bibnamefont
  {Loss}},\ }\bibfield  {title} {\bibinfo {title} {{Exchange interaction of
  hole-spin qubits in double quantum dots in highly anisotropic
  semiconductors}},\ }\href {https://doi.org/10.1103/physrevresearch.2.033036}
  {\bibfield  {journal} {\bibinfo  {journal} {Physical Review Research}\
  }\textbf {\bibinfo {volume} {2}},\ \bibinfo {pages} {033036} (\bibinfo {year}
  {2020})}\BibitemShut {NoStop}%
\bibitem [{\citenamefont {Het{\'{e}}nyi}\ \emph {et~al.}(2022)\citenamefont
  {Het{\'{e}}nyi}, \citenamefont {Bosco},\ and\ \citenamefont
  {Loss}}]{Hetenyi2022}%
  \BibitemOpen
  \bibfield  {author} {\bibinfo {author} {\bibfnamefont {B.}~\bibnamefont
  {Het{\'{e}}nyi}}, \bibinfo {author} {\bibfnamefont {S.}~\bibnamefont
  {Bosco}},\ and\ \bibinfo {author} {\bibfnamefont {D.}~\bibnamefont {Loss}},\
  }\bibfield  {title} {\bibinfo {title} {{Anomalous Zero-Field Splitting for
  Hole Spin Qubits in Si and Ge Quantum Dots}},\ }\href
  {https://doi.org/10.1103/physrevlett.129.116805} {\bibfield  {journal}
  {\bibinfo  {journal} {Physical Review Letters}\ }\textbf {\bibinfo {volume}
  {129}},\ \bibinfo {pages} {116805} (\bibinfo {year} {2022})}\BibitemShut
  {NoStop}%
\bibitem [{\citenamefont {Katsaros}\ \emph {et~al.}(2020)\citenamefont
  {Katsaros}, \citenamefont {Kuku{\v{c}}ka}, \citenamefont
  {Vuku{\v{s}}i{\'{c}}}, \citenamefont {Watzinger}, \citenamefont {Gao},
  \citenamefont {Wang}, \citenamefont {Zhang},\ and\ \citenamefont
  {Held}}]{Katsaros2020}%
  \BibitemOpen
  \bibfield  {author} {\bibinfo {author} {\bibfnamefont {G.}~\bibnamefont
  {Katsaros}}, \bibinfo {author} {\bibfnamefont {J.}~\bibnamefont
  {Kuku{\v{c}}ka}}, \bibinfo {author} {\bibfnamefont {L.}~\bibnamefont
  {Vuku{\v{s}}i{\'{c}}}}, \bibinfo {author} {\bibfnamefont {H.}~\bibnamefont
  {Watzinger}}, \bibinfo {author} {\bibfnamefont {F.}~\bibnamefont {Gao}},
  \bibinfo {author} {\bibfnamefont {T.}~\bibnamefont {Wang}}, \bibinfo {author}
  {\bibfnamefont {J.-J.}\ \bibnamefont {Zhang}},\ and\ \bibinfo {author}
  {\bibfnamefont {K.}~\bibnamefont {Held}},\ }\bibfield  {title} {\bibinfo
  {title} {{Zero Field Splitting of Heavy-Hole States in Quantum Dots}},\
  }\href {https://doi.org/10.1021/acs.nanolett.0c01466} {\bibfield  {journal}
  {\bibinfo  {journal} {Nano Letters}\ }\textbf {\bibinfo {volume} {20}},\
  \bibinfo {pages} {5201} (\bibinfo {year} {2020})}\BibitemShut {NoStop}%
\bibitem [{\citenamefont {Hendrickx}\ \emph
  {et~al.}(2020{\natexlab{b}})\citenamefont {Hendrickx}, \citenamefont
  {Lawrie}, \citenamefont {Petit}, \citenamefont {Sammak}, \citenamefont
  {Scappucci},\ and\ \citenamefont {Veldhorst}}]{Hendrickx2020}%
  \BibitemOpen
  \bibfield  {author} {\bibinfo {author} {\bibfnamefont {N.~W.}\ \bibnamefont
  {Hendrickx}}, \bibinfo {author} {\bibfnamefont {W.~I.~L.}\ \bibnamefont
  {Lawrie}}, \bibinfo {author} {\bibfnamefont {L.}~\bibnamefont {Petit}},
  \bibinfo {author} {\bibfnamefont {A.}~\bibnamefont {Sammak}}, \bibinfo
  {author} {\bibfnamefont {G.}~\bibnamefont {Scappucci}},\ and\ \bibinfo
  {author} {\bibfnamefont {M.}~\bibnamefont {Veldhorst}},\ }\bibfield  {title}
  {\bibinfo {title} {{A single-hole spin qubit}},\ }\href
  {https://doi.org/10.1038/s41467-020-17211-7} {\bibfield  {journal} {\bibinfo
  {journal} {Nature Communications}\ }\textbf {\bibinfo {volume} {11}},\
  \bibinfo {pages} {3478} (\bibinfo {year} {2020}{\natexlab{b}})}\BibitemShut
  {NoStop}%
\bibitem [{\citenamefont {Knill}\ \emph {et~al.}(2008)\citenamefont {Knill},
  \citenamefont {Leibfried}, \citenamefont {Reichle}, \citenamefont {Britton},
  \citenamefont {Blakestad}, \citenamefont {Jost}, \citenamefont {Langer},
  \citenamefont {Ozeri}, \citenamefont {Seidelin},\ and\ \citenamefont
  {Wineland}}]{Knill2008}%
  \BibitemOpen
  \bibfield  {author} {\bibinfo {author} {\bibfnamefont {E.}~\bibnamefont
  {Knill}}, \bibinfo {author} {\bibfnamefont {D.}~\bibnamefont {Leibfried}},
  \bibinfo {author} {\bibfnamefont {R.}~\bibnamefont {Reichle}}, \bibinfo
  {author} {\bibfnamefont {J.}~\bibnamefont {Britton}}, \bibinfo {author}
  {\bibfnamefont {R.~B.}\ \bibnamefont {Blakestad}}, \bibinfo {author}
  {\bibfnamefont {J.~D.}\ \bibnamefont {Jost}}, \bibinfo {author}
  {\bibfnamefont {C.}~\bibnamefont {Langer}}, \bibinfo {author} {\bibfnamefont
  {R.}~\bibnamefont {Ozeri}}, \bibinfo {author} {\bibfnamefont
  {S.}~\bibnamefont {Seidelin}},\ and\ \bibinfo {author} {\bibfnamefont
  {D.~J.}\ \bibnamefont {Wineland}},\ }\bibfield  {title} {\bibinfo {title}
  {{Randomized benchmarking of quantum gates}},\ }\href
  {https://doi.org/10.1103/physreva.77.012307} {\bibfield  {journal} {\bibinfo
  {journal} {Physical Review A}\ }\textbf {\bibinfo {volume} {77}},\ \bibinfo
  {pages} {012307} (\bibinfo {year} {2008})}\BibitemShut {NoStop}%
\bibitem [{\citenamefont {Huang}\ \emph {et~al.}(2021)\citenamefont {Huang},
  \citenamefont {Lim}, \citenamefont {Leon}, \citenamefont {Yang},
  \citenamefont {Hudson}, \citenamefont {Escott}, \citenamefont {Saraiva},
  \citenamefont {Dzurak},\ and\ \citenamefont {Laucht}}]{Huang2021}%
  \BibitemOpen
  \bibfield  {author} {\bibinfo {author} {\bibfnamefont {J.~Y.}\ \bibnamefont
  {Huang}}, \bibinfo {author} {\bibfnamefont {W.~H.}\ \bibnamefont {Lim}},
  \bibinfo {author} {\bibfnamefont {R.~C.~C.}\ \bibnamefont {Leon}}, \bibinfo
  {author} {\bibfnamefont {C.~H.}\ \bibnamefont {Yang}}, \bibinfo {author}
  {\bibfnamefont {F.~E.}\ \bibnamefont {Hudson}}, \bibinfo {author}
  {\bibfnamefont {C.~C.}\ \bibnamefont {Escott}}, \bibinfo {author}
  {\bibfnamefont {A.}~\bibnamefont {Saraiva}}, \bibinfo {author} {\bibfnamefont
  {A.~S.}\ \bibnamefont {Dzurak}},\ and\ \bibinfo {author} {\bibfnamefont
  {A.}~\bibnamefont {Laucht}},\ }\bibfield  {title} {\bibinfo {title} {{A
  High-Sensitivity Charge Sensor for Silicon Qubits above 1 K}},\ }\href
  {https://doi.org/10.1021/acs.nanolett.1c01003} {\bibfield  {journal}
  {\bibinfo  {journal} {Nano Letters}\ }\textbf {\bibinfo {volume} {21}},\
  \bibinfo {pages} {6328} (\bibinfo {year} {2021})}\BibitemShut {NoStop}%
\bibitem [{\citenamefont {Froning}\ \emph
  {et~al.}(2021{\natexlab{b}})\citenamefont {Froning}, \citenamefont
  {Ran{\v{c}}i{\'{c}}}, \citenamefont {Het{\'{e}}nyi}, \citenamefont {Bosco},
  \citenamefont {Rehmann}, \citenamefont {Li}, \citenamefont {Bakkers},
  \citenamefont {Zwanenburg}, \citenamefont {Loss}, \citenamefont {Zumbühl},\
  and\ \citenamefont {Braakman}}]{Froning2021b}%
  \BibitemOpen
  \bibfield  {author} {\bibinfo {author} {\bibfnamefont {F.~N.~M.}\
  \bibnamefont {Froning}}, \bibinfo {author} {\bibfnamefont {M.~J.}\
  \bibnamefont {Ran{\v{c}}i{\'{c}}}}, \bibinfo {author} {\bibfnamefont
  {B.}~\bibnamefont {Het{\'{e}}nyi}}, \bibinfo {author} {\bibfnamefont
  {S.}~\bibnamefont {Bosco}}, \bibinfo {author} {\bibfnamefont {M.~K.}\
  \bibnamefont {Rehmann}}, \bibinfo {author} {\bibfnamefont {A.}~\bibnamefont
  {Li}}, \bibinfo {author} {\bibfnamefont {E.~P. A.~M.}\ \bibnamefont
  {Bakkers}}, \bibinfo {author} {\bibfnamefont {F.~A.}\ \bibnamefont
  {Zwanenburg}}, \bibinfo {author} {\bibfnamefont {D.}~\bibnamefont {Loss}},
  \bibinfo {author} {\bibfnamefont {D.~M.}\ \bibnamefont {Zumbühl}},\ and\
  \bibinfo {author} {\bibfnamefont {F.~R.}\ \bibnamefont {Braakman}},\
  }\bibfield  {title} {\bibinfo {title} {{Strong spin-orbit interaction and
  $g$-factor renormalization of hole spins in Ge/Si nanowire quantum dots}},\
  }\href {https://doi.org/10.1103/physrevresearch.3.013081} {\bibfield
  {journal} {\bibinfo  {journal} {Physical Review Research}\ }\textbf {\bibinfo
  {volume} {3}},\ \bibinfo {pages} {013081} (\bibinfo {year}
  {2021}{\natexlab{b}})}\BibitemShut {NoStop}%
\end{thebibliography}%


\begin{thebibliography}{15}%
\makeatletter
\providecommand \@ifxundefined [1]{%
 \@ifx{#1\undefined}
}%
\providecommand \@ifnum [1]{%
 \ifnum #1\expandafter \@firstoftwo
 \else \expandafter \@secondoftwo
 \fi
}%
\providecommand \@ifx [1]{%
 \ifx #1\expandafter \@firstoftwo
 \else \expandafter \@secondoftwo
 \fi
}%
\providecommand \natexlab [1]{#1}%
\providecommand \enquote  [1]{``#1''}%
\providecommand \bibnamefont  [1]{#1}%
\providecommand \bibfnamefont [1]{#1}%
\providecommand \citenamefont [1]{#1}%
\providecommand \href@noop [0]{\@secondoftwo}%
\providecommand \href [0]{\begingroup \@sanitize@url \@href}%
\providecommand \@href[1]{\@@startlink{#1}\@@href}%
\providecommand \@@href[1]{\endgroup#1\@@endlink}%
\providecommand \@sanitize@url [0]{\catcode `\\12\catcode `\$12\catcode
  `\&12\catcode `\#12\catcode `\^12\catcode `\_12\catcode `\%12\relax}%
\providecommand \@@startlink[1]{}%
\providecommand \@@endlink[0]{}%
\providecommand \url  [0]{\begingroup\@sanitize@url \@url }%
\providecommand \@url [1]{\endgroup\@href {#1}{\urlprefix }}%
\providecommand \urlprefix  [0]{URL }%
\providecommand \Eprint [0]{\href }%
\providecommand \doibase [0]{https://doi.org/}%
\providecommand \selectlanguage [0]{\@gobble}%
\providecommand \bibinfo  [0]{\@secondoftwo}%
\providecommand \bibfield  [0]{\@secondoftwo}%
\providecommand \translation [1]{[#1]}%
\providecommand \BibitemOpen [0]{}%
\providecommand \bibitemStop [0]{}%
\providecommand \bibitemNoStop [0]{.\EOS\space}%
\providecommand \EOS [0]{\spacefactor3000\relax}%
\providecommand \BibitemShut  [1]{\csname bibitem#1\endcsname}%
\let\auto@bib@innerbib\@empty
\bibitem [{\citenamefont {Camenzind}\ \emph {et~al.}(2022)\citenamefont
  {Camenzind}, \citenamefont {Geyer}, \citenamefont {Fuhrer}, \citenamefont
  {Warburton}, \citenamefont {Zumbühl},\ and\ \citenamefont
  {Kuhlmann}}]{Camenzind2022}%
  \BibitemOpen
  \bibfield  {author} {\bibinfo {author} {\bibfnamefont {L.~C.}\ \bibnamefont
  {Camenzind}}, \bibinfo {author} {\bibfnamefont {S.}~\bibnamefont {Geyer}},
  \bibinfo {author} {\bibfnamefont {A.}~\bibnamefont {Fuhrer}}, \bibinfo
  {author} {\bibfnamefont {R.~J.}\ \bibnamefont {Warburton}}, \bibinfo {author}
  {\bibfnamefont {D.~M.}\ \bibnamefont {Zumbühl}},\ and\ \bibinfo {author}
  {\bibfnamefont {A.~V.}\ \bibnamefont {Kuhlmann}},\ }\bibfield  {title}
  {\bibinfo {title} {{A hole spin qubit in a fin field-effect transistor above
  4{\hspace{0.167em}}kelvin}},\ }\href
  {https://doi.org/10.1038/s41928-022-00722-0} {\bibfield  {journal} {\bibinfo
  {journal} {Nature Electronics}\ }\textbf {\bibinfo {volume} {5}},\ \bibinfo
  {pages} {178} (\bibinfo {year} {2022})}\BibitemShut {NoStop}%
\bibitem [{\citenamefont {Crippa}\ \emph {et~al.}(2018)\citenamefont {Crippa},
  \citenamefont {Maurand}, \citenamefont {Bourdet}, \citenamefont
  {Kotekar-Patil}, \citenamefont {Amisse}, \citenamefont {Jehl}, \citenamefont
  {Sanquer}, \citenamefont {Lavi{\'{e}}ville}, \citenamefont {Bohuslavskyi},
  \citenamefont {Hutin}, \citenamefont {Barraud}, \citenamefont {Vinet},
  \citenamefont {Niquet},\ and\ \citenamefont {Franceschi}}]{Crippa2018}%
  \BibitemOpen
  \bibfield  {author} {\bibinfo {author} {\bibfnamefont {A.}~\bibnamefont
  {Crippa}}, \bibinfo {author} {\bibfnamefont {R.}~\bibnamefont {Maurand}},
  \bibinfo {author} {\bibfnamefont {L.}~\bibnamefont {Bourdet}}, \bibinfo
  {author} {\bibfnamefont {D.}~\bibnamefont {Kotekar-Patil}}, \bibinfo {author}
  {\bibfnamefont {A.}~\bibnamefont {Amisse}}, \bibinfo {author} {\bibfnamefont
  {X.}~\bibnamefont {Jehl}}, \bibinfo {author} {\bibfnamefont {M.}~\bibnamefont
  {Sanquer}}, \bibinfo {author} {\bibfnamefont {R.}~\bibnamefont
  {Lavi{\'{e}}ville}}, \bibinfo {author} {\bibfnamefont {H.}~\bibnamefont
  {Bohuslavskyi}}, \bibinfo {author} {\bibfnamefont {L.}~\bibnamefont {Hutin}},
  \bibinfo {author} {\bibfnamefont {S.}~\bibnamefont {Barraud}}, \bibinfo
  {author} {\bibfnamefont {M.}~\bibnamefont {Vinet}}, \bibinfo {author}
  {\bibfnamefont {Y.-M.}\ \bibnamefont {Niquet}},\ and\ \bibinfo {author}
  {\bibfnamefont {S.~D.}\ \bibnamefont {Franceschi}},\ }\bibfield  {title}
  {\bibinfo {title} {{Electrical Spin Driving by $g$-Matrix Modulation in
  Spin-Orbit Qubits}},\ }\href {https://doi.org/10.1103/physrevlett.120.137702}
  {\bibfield  {journal} {\bibinfo  {journal} {Physical Review Letters}\
  }\textbf {\bibinfo {volume} {120}},\ \bibinfo {pages} {137702} (\bibinfo
  {year} {2018})}\BibitemShut {NoStop}%
\bibitem [{\citenamefont {Shevchenko}\ \emph {et~al.}(2010)\citenamefont
  {Shevchenko}, \citenamefont {Ashhab},\ and\ \citenamefont
  {Nori}}]{Shevchenko2010}%
  \BibitemOpen
  \bibfield  {author} {\bibinfo {author} {\bibfnamefont {S.}~\bibnamefont
  {Shevchenko}}, \bibinfo {author} {\bibfnamefont {S.}~\bibnamefont {Ashhab}},\
  and\ \bibinfo {author} {\bibfnamefont {F.}~\bibnamefont {Nori}},\ }\bibfield
  {title} {\bibinfo {title} {{Landau{\textendash}Zener{\textendash}Stückelberg
  interferometry}},\ }\href {https://doi.org/10.1016/j.physrep.2010.03.002}
  {\bibfield  {journal} {\bibinfo  {journal} {Physics Reports}\ }\textbf
  {\bibinfo {volume} {492}},\ \bibinfo {pages} {1} (\bibinfo {year}
  {2010})}\BibitemShut {NoStop}%
\bibitem [{\citenamefont {Katsaros}\ \emph {et~al.}(2020)\citenamefont
  {Katsaros}, \citenamefont {Kuku{\v{c}}ka}, \citenamefont
  {Vuku{\v{s}}i{\'{c}}}, \citenamefont {Watzinger}, \citenamefont {Gao},
  \citenamefont {Wang}, \citenamefont {Zhang},\ and\ \citenamefont
  {Held}}]{Katsaros2020}%
  \BibitemOpen
  \bibfield  {author} {\bibinfo {author} {\bibfnamefont {G.}~\bibnamefont
  {Katsaros}}, \bibinfo {author} {\bibfnamefont {J.}~\bibnamefont
  {Kuku{\v{c}}ka}}, \bibinfo {author} {\bibfnamefont {L.}~\bibnamefont
  {Vuku{\v{s}}i{\'{c}}}}, \bibinfo {author} {\bibfnamefont {H.}~\bibnamefont
  {Watzinger}}, \bibinfo {author} {\bibfnamefont {F.}~\bibnamefont {Gao}},
  \bibinfo {author} {\bibfnamefont {T.}~\bibnamefont {Wang}}, \bibinfo {author}
  {\bibfnamefont {J.-J.}\ \bibnamefont {Zhang}},\ and\ \bibinfo {author}
  {\bibfnamefont {K.}~\bibnamefont {Held}},\ }\bibfield  {title} {\bibinfo
  {title} {{Zero Field Splitting of Heavy-Hole States in Quantum Dots}},\
  }\href {https://doi.org/10.1021/acs.nanolett.0c01466} {\bibfield  {journal}
  {\bibinfo  {journal} {Nano Letters}\ }\textbf {\bibinfo {volume} {20}},\
  \bibinfo {pages} {5201} (\bibinfo {year} {2020})}\BibitemShut {NoStop}%
\bibitem [{\citenamefont {Het{\'{e}}nyi}\ \emph {et~al.}(2022)\citenamefont
  {Het{\'{e}}nyi}, \citenamefont {Bosco},\ and\ \citenamefont
  {Loss}}]{Hetenyi2022}%
  \BibitemOpen
  \bibfield  {author} {\bibinfo {author} {\bibfnamefont {B.}~\bibnamefont
  {Het{\'{e}}nyi}}, \bibinfo {author} {\bibfnamefont {S.}~\bibnamefont
  {Bosco}},\ and\ \bibinfo {author} {\bibfnamefont {D.}~\bibnamefont {Loss}},\
  }\bibfield  {title} {\bibinfo {title} {{Anomalous Zero-Field Splitting for
  Hole Spin Qubits in Si and Ge Quantum Dots}},\ }\href
  {https://doi.org/10.1103/physrevlett.129.116805} {\bibfield  {journal}
  {\bibinfo  {journal} {Physical Review Letters}\ }\textbf {\bibinfo {volume}
  {129}},\ \bibinfo {pages} {116805} (\bibinfo {year} {2022})}\BibitemShut
  {NoStop}%
\bibitem [{\citenamefont {Froning}\ \emph {et~al.}(2021)\citenamefont
  {Froning}, \citenamefont {Ran{\v{c}}i{\'{c}}}, \citenamefont {Het{\'{e}}nyi},
  \citenamefont {Bosco}, \citenamefont {Rehmann}, \citenamefont {Li},
  \citenamefont {Bakkers}, \citenamefont {Zwanenburg}, \citenamefont {Loss},
  \citenamefont {Zumbühl},\ and\ \citenamefont {Braakman}}]{Froning2021b}%
  \BibitemOpen
  \bibfield  {author} {\bibinfo {author} {\bibfnamefont {F.~N.~M.}\
  \bibnamefont {Froning}}, \bibinfo {author} {\bibfnamefont {M.~J.}\
  \bibnamefont {Ran{\v{c}}i{\'{c}}}}, \bibinfo {author} {\bibfnamefont
  {B.}~\bibnamefont {Het{\'{e}}nyi}}, \bibinfo {author} {\bibfnamefont
  {S.}~\bibnamefont {Bosco}}, \bibinfo {author} {\bibfnamefont {M.~K.}\
  \bibnamefont {Rehmann}}, \bibinfo {author} {\bibfnamefont {A.}~\bibnamefont
  {Li}}, \bibinfo {author} {\bibfnamefont {E.~P. A.~M.}\ \bibnamefont
  {Bakkers}}, \bibinfo {author} {\bibfnamefont {F.~A.}\ \bibnamefont
  {Zwanenburg}}, \bibinfo {author} {\bibfnamefont {D.}~\bibnamefont {Loss}},
  \bibinfo {author} {\bibfnamefont {D.~M.}\ \bibnamefont {Zumbühl}},\ and\
  \bibinfo {author} {\bibfnamefont {F.~R.}\ \bibnamefont {Braakman}},\
  }\bibfield  {title} {\bibinfo {title} {{Strong spin-orbit interaction and
  $g$-factor renormalization of hole spins in Ge/Si nanowire quantum dots}},\
  }\href {https://doi.org/10.1103/physrevresearch.3.013081} {\bibfield
  {journal} {\bibinfo  {journal} {Physical Review Research}\ }\textbf {\bibinfo
  {volume} {3}},\ \bibinfo {pages} {013081} (\bibinfo {year}
  {2021})}\BibitemShut {NoStop}%
\bibitem [{Note1()}]{Note1}%
  \BibitemOpen
  \bibinfo {note} {Note that {\protect \it (i)} $g_i$ must be a real matrix to
  ensure hermiticity, {\protect \it (ii)} under a general unitary
  transformation, the Pauli matrices transform as $U^\dagger (\sigma _x,\sigma
  _y,\sigma _z) U = R_{ij} (\sigma _x,\sigma _y,\sigma _z)_j$, where $R$ is a
  3D rotation. Exploiting the ambiguity of the basis choice in the Hilbert
  space, we define the quantization axis in the lab frame such that the
  $g$-tensor is a symmetric matrix for a given $\protect \bf B$.}\BibitemShut
  {Stop}%
\bibitem [{\citenamefont {Kavokin}(2001)}]{Kavokin2001}%
  \BibitemOpen
  \bibfield  {author} {\bibinfo {author} {\bibfnamefont {K.~V.}\ \bibnamefont
  {Kavokin}},\ }\bibfield  {title} {\bibinfo {title} {{Anisotropic exchange
  interaction of localized conduction-band electrons in semiconductors}},\
  }\href {https://doi.org/10.1103/physrevb.64.075305} {\bibfield  {journal}
  {\bibinfo  {journal} {Physical Review B}\ }\textbf {\bibinfo {volume} {64}},\
  \bibinfo {pages} {075305} (\bibinfo {year} {2001})}\BibitemShut {NoStop}%
\bibitem [{\citenamefont {Kavokin}(2004)}]{Kavokin2004}%
  \BibitemOpen
  \bibfield  {author} {\bibinfo {author} {\bibfnamefont {K.~V.}\ \bibnamefont
  {Kavokin}},\ }\bibfield  {title} {\bibinfo {title} {{Symmetry of anisotropic
  exchange interactions in semiconductor nanostructures}},\ }\href
  {https://doi.org/10.1103/physrevb.69.075302} {\bibfield  {journal} {\bibinfo
  {journal} {Physical Review B}\ }\textbf {\bibinfo {volume} {69}},\ \bibinfo
  {pages} {075302} (\bibinfo {year} {2004})}\BibitemShut {NoStop}%
\bibitem [{\citenamefont {Het{\'{e}}nyi}\ \emph {et~al.}(2020)\citenamefont
  {Het{\'{e}}nyi}, \citenamefont {Kloeffel},\ and\ \citenamefont
  {Loss}}]{Hetenyi2020}%
  \BibitemOpen
  \bibfield  {author} {\bibinfo {author} {\bibfnamefont {B.}~\bibnamefont
  {Het{\'{e}}nyi}}, \bibinfo {author} {\bibfnamefont {C.}~\bibnamefont
  {Kloeffel}},\ and\ \bibinfo {author} {\bibfnamefont {D.}~\bibnamefont
  {Loss}},\ }\bibfield  {title} {\bibinfo {title} {{Exchange interaction of
  hole-spin qubits in double quantum dots in highly anisotropic
  semiconductors}},\ }\href {https://doi.org/10.1103/physrevresearch.2.033036}
  {\bibfield  {journal} {\bibinfo  {journal} {Physical Review Research}\
  }\textbf {\bibinfo {volume} {2}},\ \bibinfo {pages} {033036} (\bibinfo {year}
  {2020})}\BibitemShut {NoStop}%
\bibitem [{\citenamefont {Pioro-Ladri{\`{e}}re}\ \emph
  {et~al.}(2008)\citenamefont {Pioro-Ladri{\`{e}}re}, \citenamefont {Obata},
  \citenamefont {Tokura}, \citenamefont {Shin}, \citenamefont {Kubo},
  \citenamefont {Yoshida}, \citenamefont {Taniyama},\ and\ \citenamefont
  {Tarucha}}]{PioroLadriere2008}%
  \BibitemOpen
  \bibfield  {author} {\bibinfo {author} {\bibfnamefont {M.}~\bibnamefont
  {Pioro-Ladri{\`{e}}re}}, \bibinfo {author} {\bibfnamefont {T.}~\bibnamefont
  {Obata}}, \bibinfo {author} {\bibfnamefont {Y.}~\bibnamefont {Tokura}},
  \bibinfo {author} {\bibfnamefont {Y.-S.}\ \bibnamefont {Shin}}, \bibinfo
  {author} {\bibfnamefont {T.}~\bibnamefont {Kubo}}, \bibinfo {author}
  {\bibfnamefont {K.}~\bibnamefont {Yoshida}}, \bibinfo {author} {\bibfnamefont
  {T.}~\bibnamefont {Taniyama}},\ and\ \bibinfo {author} {\bibfnamefont
  {S.}~\bibnamefont {Tarucha}},\ }\bibfield  {title} {\bibinfo {title}
  {{Electrically driven single-electron spin resonance in a slanting Zeeman
  field}},\ }\href {https://doi.org/10.1038/nphys1053} {\bibfield  {journal}
  {\bibinfo  {journal} {Nature Physics}\ }\textbf {\bibinfo {volume} {4}},\
  \bibinfo {pages} {776} (\bibinfo {year} {2008})}\BibitemShut {NoStop}%
\bibitem [{\citenamefont {Noiri}\ \emph {et~al.}(2022)\citenamefont {Noiri},
  \citenamefont {Takeda}, \citenamefont {Nakajima}, \citenamefont {Kobayashi},
  \citenamefont {Sammak}, \citenamefont {Scappucci},\ and\ \citenamefont
  {Tarucha}}]{Noiri2022}%
  \BibitemOpen
  \bibfield  {author} {\bibinfo {author} {\bibfnamefont {A.}~\bibnamefont
  {Noiri}}, \bibinfo {author} {\bibfnamefont {K.}~\bibnamefont {Takeda}},
  \bibinfo {author} {\bibfnamefont {T.}~\bibnamefont {Nakajima}}, \bibinfo
  {author} {\bibfnamefont {T.}~\bibnamefont {Kobayashi}}, \bibinfo {author}
  {\bibfnamefont {A.}~\bibnamefont {Sammak}}, \bibinfo {author} {\bibfnamefont
  {G.}~\bibnamefont {Scappucci}},\ and\ \bibinfo {author} {\bibfnamefont
  {S.}~\bibnamefont {Tarucha}},\ }\bibfield  {title} {\bibinfo {title} {{Fast
  universal quantum gate above the fault-tolerance threshold in silicon}},\
  }\href {https://doi.org/10.1038/s41586-021-04182-y} {\bibfield  {journal}
  {\bibinfo  {journal} {Nature}\ }\textbf {\bibinfo {volume} {601}},\ \bibinfo
  {pages} {338} (\bibinfo {year} {2022})}\BibitemShut {NoStop}%
\bibitem [{\citenamefont {Xue}\ \emph {et~al.}(2022)\citenamefont {Xue},
  \citenamefont {Russ}, \citenamefont {Samkharadze}, \citenamefont {Undseth},
  \citenamefont {Sammak}, \citenamefont {Scappucci},\ and\ \citenamefont
  {Vandersypen}}]{Xue2022}%
  \BibitemOpen
  \bibfield  {author} {\bibinfo {author} {\bibfnamefont {X.}~\bibnamefont
  {Xue}}, \bibinfo {author} {\bibfnamefont {M.}~\bibnamefont {Russ}}, \bibinfo
  {author} {\bibfnamefont {N.}~\bibnamefont {Samkharadze}}, \bibinfo {author}
  {\bibfnamefont {B.}~\bibnamefont {Undseth}}, \bibinfo {author} {\bibfnamefont
  {A.}~\bibnamefont {Sammak}}, \bibinfo {author} {\bibfnamefont
  {G.}~\bibnamefont {Scappucci}},\ and\ \bibinfo {author} {\bibfnamefont
  {L.~M.~K.}\ \bibnamefont {Vandersypen}},\ }\bibfield  {title} {\bibinfo
  {title} {{Quantum logic with spin qubits crossing the surface code
  threshold}},\ }\href {https://doi.org/10.1038/s41586-021-04273-w} {\bibfield
  {journal} {\bibinfo  {journal} {Nature}\ }\textbf {\bibinfo {volume} {601}},\
  \bibinfo {pages} {343} (\bibinfo {year} {2022})}\BibitemShut {NoStop}%
\bibitem [{\citenamefont {Russ}\ \emph {et~al.}(2018)\citenamefont {Russ},
  \citenamefont {Zajac}, \citenamefont {Sigillito}, \citenamefont {Borjans},
  \citenamefont {Taylor}, \citenamefont {Petta},\ and\ \citenamefont
  {Burkard}}]{Russ2018}%
  \BibitemOpen
  \bibfield  {author} {\bibinfo {author} {\bibfnamefont {M.}~\bibnamefont
  {Russ}}, \bibinfo {author} {\bibfnamefont {D.~M.}\ \bibnamefont {Zajac}},
  \bibinfo {author} {\bibfnamefont {A.~J.}\ \bibnamefont {Sigillito}}, \bibinfo
  {author} {\bibfnamefont {F.}~\bibnamefont {Borjans}}, \bibinfo {author}
  {\bibfnamefont {J.~M.}\ \bibnamefont {Taylor}}, \bibinfo {author}
  {\bibfnamefont {J.~R.}\ \bibnamefont {Petta}},\ and\ \bibinfo {author}
  {\bibfnamefont {G.}~\bibnamefont {Burkard}},\ }\bibfield  {title} {\bibinfo
  {title} {{High-fidelity quantum gates in {Si}/{SiGe} double quantum dots}},\
  }\href {https://doi.org/10.1103/physrevb.97.085421} {\bibfield  {journal}
  {\bibinfo  {journal} {Physical Review B}\ }\textbf {\bibinfo {volume} {97}},\
  \bibinfo {pages} {085421} (\bibinfo {year} {2018})}\BibitemShut {NoStop}%
\bibitem [{\citenamefont {Yoneda}\ \emph {et~al.}(2017)\citenamefont {Yoneda},
  \citenamefont {Takeda}, \citenamefont {Otsuka}, \citenamefont {Nakajima},
  \citenamefont {Delbecq}, \citenamefont {Allison}, \citenamefont {Honda},
  \citenamefont {Kodera}, \citenamefont {Oda}, \citenamefont {Hoshi},
  \citenamefont {Usami}, \citenamefont {Itoh},\ and\ \citenamefont
  {Tarucha}}]{Yoneda2017}%
  \BibitemOpen
  \bibfield  {author} {\bibinfo {author} {\bibfnamefont {J.}~\bibnamefont
  {Yoneda}}, \bibinfo {author} {\bibfnamefont {K.}~\bibnamefont {Takeda}},
  \bibinfo {author} {\bibfnamefont {T.}~\bibnamefont {Otsuka}}, \bibinfo
  {author} {\bibfnamefont {T.}~\bibnamefont {Nakajima}}, \bibinfo {author}
  {\bibfnamefont {M.~R.}\ \bibnamefont {Delbecq}}, \bibinfo {author}
  {\bibfnamefont {G.}~\bibnamefont {Allison}}, \bibinfo {author} {\bibfnamefont
  {T.}~\bibnamefont {Honda}}, \bibinfo {author} {\bibfnamefont
  {T.}~\bibnamefont {Kodera}}, \bibinfo {author} {\bibfnamefont
  {S.}~\bibnamefont {Oda}}, \bibinfo {author} {\bibfnamefont {Y.}~\bibnamefont
  {Hoshi}}, \bibinfo {author} {\bibfnamefont {N.}~\bibnamefont {Usami}},
  \bibinfo {author} {\bibfnamefont {K.~M.}\ \bibnamefont {Itoh}},\ and\
  \bibinfo {author} {\bibfnamefont {S.}~\bibnamefont {Tarucha}},\ }\bibfield
  {title} {\bibinfo {title} {{A quantum-dot spin qubit with coherence limited
  by charge noise and fidelity higher than 99.9{\%}}},\ }\href
  {https://doi.org/10.1038/s41565-017-0014-x} {\bibfield  {journal} {\bibinfo
  {journal} {Nature Nanotechnology}\ }\textbf {\bibinfo {volume} {13}},\
  \bibinfo {pages} {102} (\bibinfo {year} {2017})}\BibitemShut {NoStop}%
\end{thebibliography}%

\end{document}


\title{Supplementary Material:\\ Two-qubit logic with anisotropic exchange in a fin field-effect transistor}

\author{Simon Geyer}
\email{e-mail: \mbox{simon.geyer@unibas.ch; dominik.zumbuhl@unibas.ch; andreas.kuhlmann@unibas.ch}}
\affiliation{Department of Physics, University of Basel, Klingelbergstrasse 82, CH-4056 Basel, Switzerland}

\author{Bence Hetényi}
\affiliation{Department of Physics, University of Basel, Klingelbergstrasse 82, CH-4056 Basel, Switzerland}
\affiliation{IBM Research Europe-Zurich, S\"aumerstrasse 4, CH-8803 R\"uschlikon, Switzerland}

\author{Stefano Bosco}

\author{Leon C. Camenzind}
\altaffiliation{Current address: RIKEN, Center for Emergent Matter Science (CEMS), Wako-shi, Saitama 351-0198, Japan}

\author{Rafael S. Eggli}
\affiliation{Department of Physics, University of Basel, Klingelbergstrasse 82, CH-4056 Basel, Switzerland}

\author{Andreas Fuhrer} 
\affiliation{IBM Research Europe-Zurich, S\"aumerstrasse 4, CH-8803 R\"uschlikon, Switzerland}

\author{Daniel Loss}

\author{Richard J. Warburton}

\author{Dominik M. Zumb\"uhl}
\email{e-mail: \mbox{simon.geyer@unibas.ch; dominik.zumbuhl@unibas.ch; andreas.kuhlmann@unibas.ch}}

\author{Andreas V. Kuhlmann}
\email{e-mail: \mbox{simon.geyer@unibas.ch; dominik.zumbuhl@unibas.ch; andreas.kuhlmann@unibas.ch}}
\affiliation{Department of Physics, University of Basel, Klingelbergstrasse 82, CH-4056 Basel, Switzerland}

\date{\today}

\maketitle

\newcommand{\fakefigure}[1]
{\refstepcounter{figure}\label{#1}}
\fakefigure{fig:fig1}
\fakefigure{fig:fig2}
\fakefigure{fig:fig3}
\fakefigure{fig:fig4}
\renewcommand{\thefigure}{S\arabic{figure}}
\setcounter{figure}{0}    
\renewcommand{\thesection}{S\arabic{section}}

\tableofcontents

\newpage
\section{Setup}
\label{S:Setup}
\begin{figure*}[h]
\centering
\includegraphics[width=\linewidth]{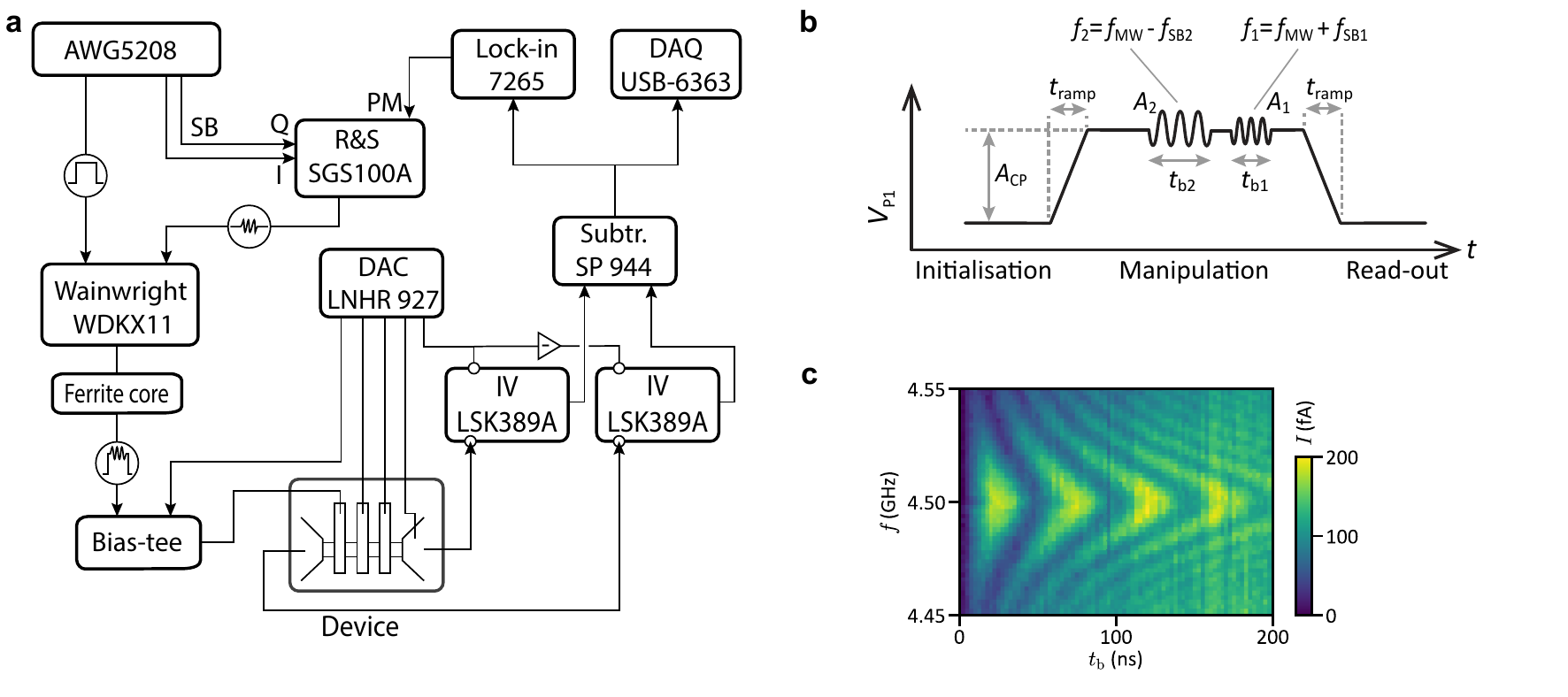}
\caption{\textbf{Setup and single qubit control.} \textbf{a}, Schematic of the experimental setup. The following instruments were used: a arbitrary waveform generator AWG5208 from Tektronix, a diplexer WDKX11+10-DC-1000/1300-15000-60S3 from Wainwright, a microwave signal generator SGS100A from Rhode\&Schwarz, a lock-in amplifier Model 7265 DSP from Signal Recovery, a data acquisition card USB-6363 from National Instruments. Further, a voltage subtractor SP944, two current-voltage converters LSK389A and a digital-analogue converter LNHR927, all from Basel Precision Instruments, were used. \textbf{b}, Initialization, two-qubit manipulation and readout schematic. \textbf{c} Typical Rabi chevron measurement of Q1.}
\label{SFig:setup}
\end{figure*}

In this section we discuss the experimental setup as well as the qubit operation. A schematic of the setup is presented in Fig. \ref{SFig:setup}a. We performed all experiments by measuring direct current through a hole double quantum dot (DQD) at the base temperature $\sim 40\,$mK of a Bluefors XLD dilution refrigerator. Spin-to-charge conversion and spin initialization were realized using Pauli spin blockade. The pulse scheme for each experiment cycle is described in Fig. \ref{SFig:setup}b. Coherent single-qubit spin driving is demonstrated by a Rabi chevron measurement, presented in Fig.\ \ref{SFig:setup}c. A more detailed description is found in the Supplementary Material to Ref.\ \cite{Camenzind2022}. A key difference to the previously reported setup is the use of side-band modulation (SB) in the amplitude-quadrature (IQ) mixing of the microwave signal, which allows to quickly address different qubit frequencies in a single experiment cycle, thus enabling two-qubit experiments.

\newpage

\section{Spectroscopy data for qubit anisotropy characterisation}

Here we present the raw data of the qubit spectroscopy experiments that were used to extract the $g$-tensors of Q1 and Q2 and the exchange matrix $\mathcal J$. Further, we  observe correlations between the qubit readout signal in the lock-in current and the DC current through the base line of the bias triangle.

\begin{figure*}[h]
\centering
\includegraphics[width=\linewidth]{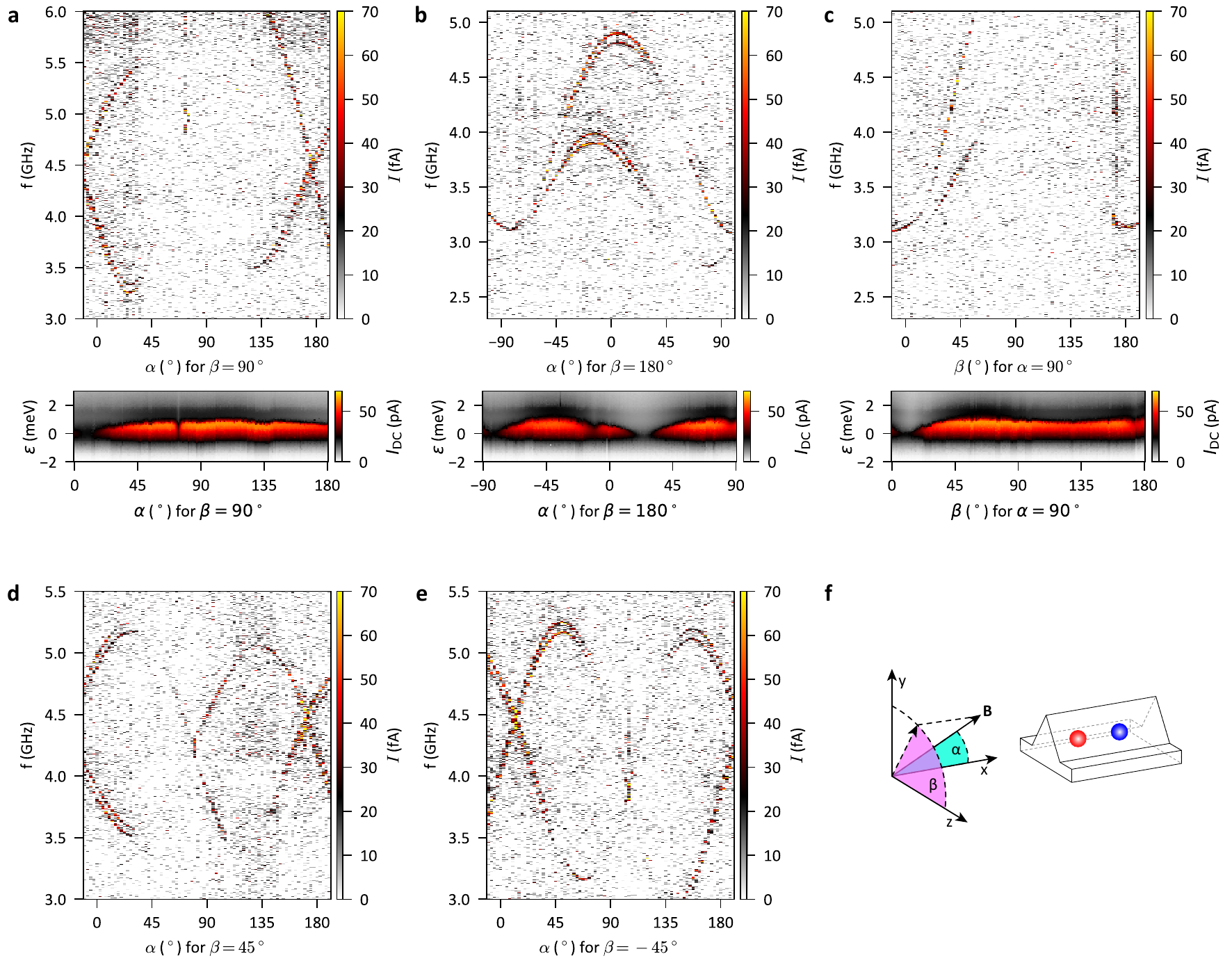}
\caption{\textbf{Qubit spectroscopy data.} \textbf{a-e}, Spectroscopy measurement as a function of magnetic field orientation ($\alpha$, $\beta$) for sweeping $\mathbf{B}$ along 5 different planes with $V_\mathrm{B}=-820$\,mV and $\eps=-4.025$\,meV. For a fixed magnetic field orientation 4 transitions can be identified as described in Fig. 1, which allows to extract $E_\mathrm{Z,i}$ and $J_\parallel$ for each configuration.
The gaps in the data come from a vanishing qubit readout signal for certain magnetic field orientations. Note that for some orientations only 1-3 transitions are vanishing. For \textbf{a-c} we additionally show the direct current $I_\mathrm{DC}$ of the zero detuning transition of the DQD as a function of magnetic field orientation at $|\mathbf{B}|=0.1$\,T. A correlation between a large current and a vanishing qubit visibility is observed.  \textbf{f} Coordinate system and definition of the sweep parameters $\alpha$ and $\beta$. }
\label{SFig:raw_data}
\end{figure*}

\newpage

\section{$g$-tensors for Q1 and Q2}
The $g$-tensors were extracted according to Ref.\ \cite{Crippa2018} by measuring $E_\text{Z,i}$ by MW spectroscopy in at least 6 different orientations. The extraction was performed on the data presented in Fig. \ref{SFig:raw_data}.
\begin{eqnarray}
    g_1 = \left(
\begin{array}{ccc}
 2.31 & 0.50 & -0.06 \\
 0.50 & 2.00 & 0.06 \\
 -0.06 & 0.06 & 1.50 \\
\end{array}
\right), \hspace{1cm}
    g_2 = \left(
\begin{array}{ccc}
 1.86 & -0.57 & 0.09 \\
 -0.57 & 2.76 & -0.01 \\
 0.09 & -0.01 & 1.46 \\
\end{array}
\right)
\end{eqnarray}
The $g$-tensors can be diagonalized, such that the effective $g$-factors along the principal axes can be easily read off:
\begin{eqnarray}
    g_1^\text{diag} = \text{diag} \left(2.68, 1.68, 1.46 \right), \hspace{1cm}
    g_2^\text{diag} = \text{diag} \left(3.04, 1.62, 1.42 \right).
\end{eqnarray}

\newpage
\section{Adiabatic 2-qubit initialization}

\begin{figure*}[h]
\centering
\includegraphics[width=\linewidth]{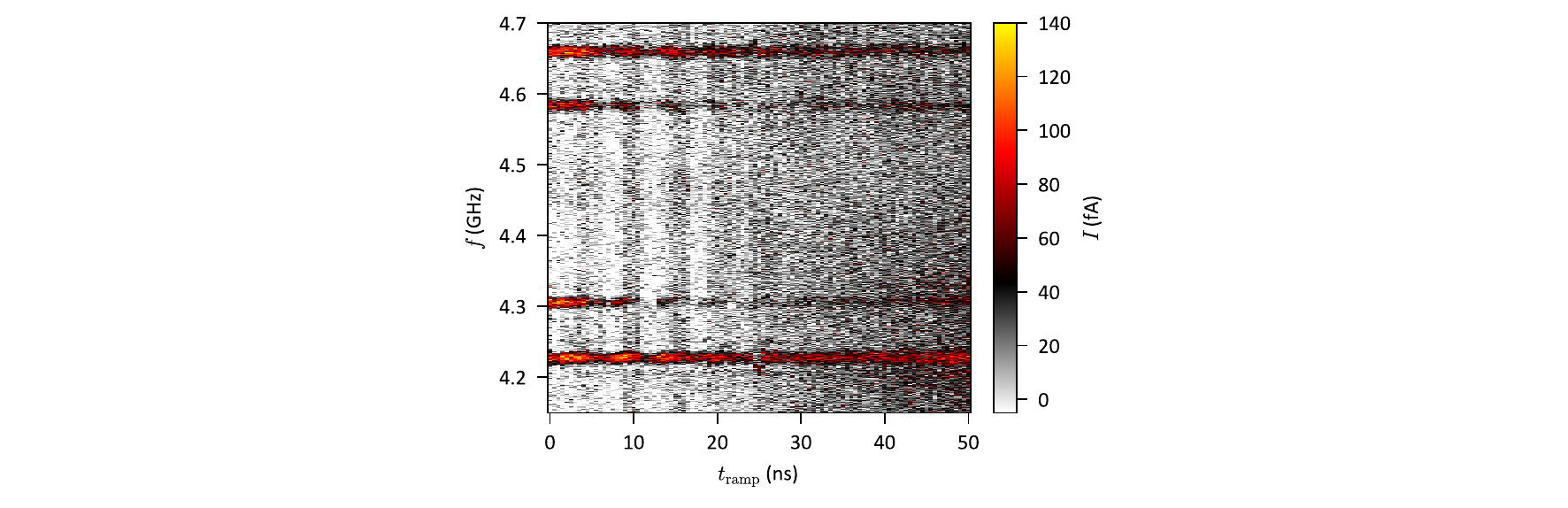}
\caption{\textbf{Adiabatic qubit initialization.} MW spectroscopy measurement as a function of ramp time $t_\mathrm{ramp}$ for a trapezoid initialization and readout pulse (see Section \ref{S:Setup}). The vanishing contrast of the inner two transitions indicates an initialization into the $\ket{\spd\spu}$ state, which only allows transitions with the highest and lowest frequency. This experiment was used to calibrate $t_\mathrm{ramp} \sim 20$\,ns for the CROT experiment in the main paper (see Fig. \ref{fig:fig4}).}
\label{SFig:adiabatic}
\end{figure*}

The qubits are initialized by pulsing from the spin-blocked region to the (1,1) manipulation point with a linear ramp within  the time $t_\mathrm{ramp}$. By varying $t_\mathrm{ramp}$ and observing the allowed qubit transitions in a spectroscopy experiment (see Fig. \ref{SFig:adiabatic}), we identify the necessary ramp time of $\sim20$\,ns to initialize adiabatically into the $\ket{\spd\spu}$ state.
The background of the measurement shows an interference pattern. This could be explained by Landau-Zener-Stückelberg interference due to repeatedly pulsing the system across an anticrossing \cite{Shevchenko2010}.

\newpage
\section{Derivation of exchange matrix formalism and fitting formula}
\label{sec:Jfit}

In this section, starting from a two-site Hubbard model we derive the effective Hamiltonian of the $(1,1)$ charge sector of a double QD and show that the exchange interaction in this approximation can be written as a 3D rotation. We present the effective Hamiltonian in three different frames with the intention to assist future work on the experimental as well as theoretical side. We obtain a fitting formula for the exchange splitting as a function of magnetic field orientation, facilitating the extraction of the full exchange matrix from the MW transitions measured in the two-qubit system. Finally, we verify that in the present experiment the exchange interaction is well described by a rotation matrix, and that including an additional Ising anisotropy \cite{Katsaros2020, Hetenyi2022}, which splits triplet states at zero magnetic field, does not improve the fitting of the presented results.

We describe our double QD setup using a two-site Fermi-Hubbard model where each QD (QD1 and QD2, respectively) is described by a single orbital state $\ket 1$ and $\ket 2$. The Hamiltonian reads
\begin{equation}
    H_\text{FH} = \sum \limits_{i,j \in \{1,2\}} \sum \limits_{s,s' \in \{\spu,\spd\}} \tilde{H}_{ij}^{ss'} a^\dagger_{is} a_{js'} + U \sum \limits_{i \in \{1,2\}} n_{i\spu} n_{i\spd}\, ,
    \label{eq:HFH}
\end{equation}
where $a^\dagger_{i,s}$ ($a_{i,s}$) creates (annihilates) a hole on site $i$ with spin $s$, and obeys fermionic anticommutation relations. Furthermore, $n_{is} = a^\dagger_{i,s} a_{i,s}$ is the spin-resolved particle number operator of dot $i$ and $U$ is the charging energy. The single particle Hamiltonian $\tilde{H}_{ij}^{ss'}=\bra{is} \tilde{H}\ket{js'}$ acting on the orbital and spin degrees of freedom reads 
\begin{eqnarray}
\begin{split}
    \tilde{H} =& \frac{\eps+U-U_0}{2}\tau_z + t \cos (\theta_\text{so}) \tau_x + t \sin (\theta_\text{so}) \tau_y \mathbf{n}_\text{so} \cdot \boldsymbol \sigma\\
    &+ \frac 1 2 \mu_B \mathbf B \cdot \left[ \frac{1+\tau_z}2 g_1\boldsymbol \sigma + \frac{1-\tau_z}2 g_2 \boldsymbol \sigma \right],
    \label{eq:HDQDsp}
\end{split}
\end{eqnarray}
where $\tau_k$ are Pauli matrices acting on the orbital degrees of freedom, e.g., $\tau_z = \ket{1}\bra{1}-\ket{2}\bra{2}$, and $\boldsymbol \sigma = (\sigma_x,\sigma_y,\sigma_z)$ are also Pauli matrices acting on the spin degree of freedom $\{\ket \spu, \ket \spd \}$. The first term of Eq.~\eqref{eq:HDQDsp} accounts for the detuning between the left and right QDs, where $\eps$ is measured from the singlet-singlet anticrossing. Furthermore, since the charging energy $U$ in the experiment is measured at a different barrier height than the exchange anisotropy, we introduced $U_0$ as a fitting parameter that accounts for the shift of the singlet-singlet anticrossing. The tunnel-coupling between the QDs is characterized by a (spin-conserving) hopping term $\propto t \cos (\theta_\text{so})$, while spin-orbit interaction is described by the spin-flip hopping $\propto t \sin (\theta_\text{so})$. For  spatially homogeneous SOI, the rotation angle is given by the dot-dot distance over the spin-orbit length i.e., $\theta_\text{so} = d/\lambda_\text{so}$, and ${\bf n}_\text{so}$ is the direction of the spin-orbit axis~\cite{Froning2021b}. The $g$-tensors of the two QDs are taken into account in the most general form, where the spin quantization axis is fixed such that the $g$-tensors $g_1$ and $g_2$ are symmetric if $\mathbf B $ is in the lab frame \footnote{Note that {\it (i)} $g_i$ must be a real matrix to ensure hermiticity, {\it (ii)} under a general unitary transformation, the Pauli matrices transform as $U^\dagger (\sigma_x,\sigma_y,\sigma_z) U = R_{ij} (\sigma_x,\sigma_y,\sigma_z)_j$, where $R$ is a 3D rotation. Exploiting the ambiguity of the basis choice in the Hilbert space, we define the quantization axis in the lab frame such that the $g$-tensor is a symmetric matrix for a given $\bf B$.}. \\

\noindent\textbf{Spin-orbit frame.}
In order to eliminate the spin-flip tunnelling term and thus obtain a simpler matrix form of the Hamiltonian, we move to the \textit{spin-orbit frame}. The corresponding unitary basis transformation is given by $\tilde{H}^\text{so} = U^\dagger_\text{so} \tilde{H} U_\text{so}$, where $U_\text{so} = \exp (-i\theta_\text{so} \tau_z \mathbf{n}_\text{so} \cdot \boldsymbol \sigma/2)$, which rotates the spin-quantization axes on the two sites in opposite directions. In this spin-orbit frame, the spin-conserving and spin-flip tunnelling transform as $t\cos(\theta_\text{so})\tau_x \to t\cos^2(\theta_\text{so})\tau_x -t\sin(2\theta_\text{so})\tau_y \mathbf{n}_\text{so} \cdot \boldsymbol \sigma/2 $ and $t\sin(\theta_\text{so})\tau_y \mathbf{n}_\text{so} \cdot \boldsymbol \sigma\to t\sin^2(\theta_\text{so})\tau_x+t\sin(2\theta_\text{so})\tau_y \mathbf{n}_\text{so} \cdot \boldsymbol \sigma/2$, respectively, and the Hamiltonian in Eq.~\eqref{eq:HDQDsp} reads
\begin{equation}
    \tilde{H}^\text{so} = \frac{\eps+U-U_0}{2}\tau_z + t\tau_x + \frac 1 2 \mu_B \mathbf B  \cdot \left[ \frac{1+\tau_z}2 g^\text{so}_1 \boldsymbol \sigma + \frac{1-\tau_z}2 g^\text{so}_2 \boldsymbol \sigma \right] ,
    \label{eq:HDQDso}
\end{equation}
where the spin-orbit rotated $g$-tensors are $g^\text{so}_1 = g_1 R_\text{so} (\theta_\text{so})$ and $g^\text{so}_2 = g_2 R_\text{so} (-\theta_\text{so})$ with $R_\text{so} (\varphi)$ denoting the right-handed rotation around the spin-orbit axis ${\bf n}_\text{so}$ by an angle $\varphi$.
The transformation of the $g$ tensors is straightforward
\begin{equation}
    U^\dagger_\text{so} \left[ \frac{1+\tau_z}2 g_1\boldsymbol \sigma \right] U_\text{so} = \begin{pmatrix} \exp (i\theta_\text{so} \mathbf{n}_\text{so} \cdot \boldsymbol \sigma/2) g_1\boldsymbol \sigma \exp ( -i\theta_\text{so} \mathbf{n}_\text{so} \cdot \boldsymbol \sigma/2) & 0\\ 0 & 0 \end{pmatrix} =\frac{1+\tau_z}2 g_1 R_\text{so} (\theta_\text{so}) \boldsymbol \sigma,
\end{equation} keeping in mind the transformation rule for the vector of Pauli matrices. Note that the spin-flip tunnelling does not appear in this formulation, but the $g_i^\text{so}$ matrices are not symmetric anymore. 

Since the quantization axis has been rotated by $\mp \theta_\text{so}$ around $\bf n_\text{so}$ for the left and right sites, respectively, the on-site Hubbard term $U n_{i\spu} n_{i\spd}$ has the same form as in the lab frame. The Hamiltonian in Eq.~\eqref{eq:HFH} using the single-particle term of Eq.~\eqref{eq:HDQDso} is then projected to the lowest-energy two-particle sector using the basis states
\begin{subequations}
\begin{equation}
    \ket{S(0,2)} = a^\dagger_{2\spu}a^\dagger_{2\spd}\ket 0,
\end{equation}
\begin{equation}
    \ket{S} = \frac 1 {\sqrt 2}(a^\dagger_{1\spu}a^\dagger_{2\spd}-a^\dagger_{1\spd}a^\dagger_{2\spu})\ket 0,
\end{equation}
\begin{equation}
    \ket{T_0} = \frac 1 {\sqrt 2}(a^\dagger_{1\spu}a^\dagger_{2\spd}+a^\dagger_{1\spd}a^\dagger_{2\spu})\ket 0,
\end{equation}
\begin{equation}
    \ket{T_{ss}} = a^\dagger_{1s}a^\dagger_{2s}\ket 0,
\end{equation}
\end{subequations}
where $\ket 0$ is the vacuum state for holes and we omitted the $S(2,0)$ state since we operate close to the $S(0,2)-S$ anticrossing, i.e., $\eps \ll U - U_0$. The low-energy $5\times 5$ Hamiltonian of the DQD then reads 
\begin{equation}
    H_{5\times 5} = \left(
\begin{array}{ccccc}
 U_0-\eps  & \sqrt{2} t & 0 & 0 & 0 \\
 \sqrt{2} t & 0 & -\frac{\text{$\delta $b}_x+i \text{$\delta $b}_y}{\sqrt{2}} & \frac{\text{$\delta $b}_x-i \text{$\delta $b}_y}{\sqrt{2}} & \text{$\delta $b}_z \\
 0 & -\frac{\text{$\delta $b}_x-i \text{$\delta $b}_y}{\sqrt{2}} 
 & \bar{b}_z & 0 
 & \frac{\bar{b}_x-i \bar{b}_y}{\sqrt{2}}
 \\
 0 & \frac{\text{$\delta $b}_x+i \text{$\delta $b}_y}{\sqrt{2}} & 0 & -\bar{b}_z
 & \frac{\bar{b}_x+i \bar{b}_y}{\sqrt{2}}
 \\
 0 & \text{$\delta $b}_z 
 & \frac{\bar{b}_x+i \bar{b}_y}{\sqrt{2}}
 & \frac{\bar{b}_x-i \bar{b}_y}{\sqrt{2}}
 & 0 \\
\end{array}
\right),
\label{eq:H5x5}
\end{equation}
where the order of the basis states is $[S(0,2),S,T_{\spu \spu}, T_{\spd \spd}, T_0]$ and we introduced the average- and gradient Zeeman fields as $\mathbf{\bar b} = (\mathbf{b}_1 + \mathbf{b}_2)/2 = \mu_B \mathbf B (g^\text{so}_1+g^\text{so}_2)/2$ and $\delta \mathbf{b} = (\mathbf{b}_1 - \mathbf{b}_2)/2 = \mu_B \mathbf B (g^\text{so}_1-g^\text{so}_2)/2$, where the second equality defines the Zeeman field ${\bf b}_i$ for the $i$th site. Furthermore the charging energy $U$ of the doubly occupied singlet $S(0,2)$ is compensated by our definition of the detuning $\tilde{\eps} =  \eps+U-U_0$, in Eq.~\eqref{eq:HDQDsp}, and $U_0$ remains a fitting parameter (much smaller than $U$).

The $[S(0,2),S]$ block of the Hamiltonian in Eq.~\eqref{eq:H5x5} can be diagonalized exactly, leading to hybridized singlet states at energies $E_{S_+}= U_0-\eps+J_0$ and $E_{S_-}=-J_0$, respectively, where $J_0 = \sqrt 2 t \tan (\gamma/2) = -(U_0 - \eps)[1-\sqrt{1+8t^2/(U_0-\eps)^2}]/2$ and the angle $\gamma = \arctan[\sqrt 8 t/(U_0-\eps)]$. In the limit of large detuning $U_0-\eps \gg t$ we obtain $J_0 = 2t^2/(U_0-\eps)$ as in Eq.~(1) of the main text. After the transformation of the singlet sector one obtains
\begin{equation}
    H_{5\times 5} = \left(
\begin{array}{ccccc}
 U_0-\eps+J_0 & 0 & -\frac{\text{$\delta $b}_x+i \text{$\delta $b}_y}{\sqrt{2}} \sin(\frac \gamma 2)
 & \frac{\text{$\delta $b}_x-i \text{$\delta $b}_y}{\sqrt{2}} \sin(\frac \gamma 2)
 & \text{$\delta $b}_z \sin(\frac \gamma 2) \\
 0 &  -J_0 
 & -\frac{\text{$\delta $b}_x+i \text{$\delta $b}_y}{\sqrt{2}} \cos(\frac \gamma 2)
 & \frac{\text{$\delta $b}_x-i \text{$\delta $b}_y}{\sqrt{2}} \cos(\frac \gamma 2)
 & \text{$\delta $b}_z \cos(\frac \gamma 2) \\
 -\frac{\text{$\delta $b}_x-i \text{$\delta $b}_y}{\sqrt{2}} \sin(\frac \gamma 2) & -\frac{\text{$\delta $b}_x-i \text{$\delta $b}_y}{\sqrt{2}} \cos(\frac \gamma 2)
 & \bar{b}_z & 0 
 & \frac{\bar{b}_x-i \bar{b}_y}{\sqrt{2}}
 \\
 \frac{\text{$\delta $b}_x+i \text{$\delta $b}_y}{\sqrt{2}}\sin(\frac \gamma 2) & \frac{\text{$\delta $b}_x+i \text{$\delta $b}_y}{\sqrt{2}}\cos(\frac \gamma 2) & 0 & -\bar{b}_z
 & \frac{\bar{b}_x+i \bar{b}_y}{\sqrt{2}}
 \\
 \text{$\delta $b}_z\sin(\frac \gamma 2)  & \text{$\delta $b}_z\cos(\frac \gamma 2) 
 & \frac{\bar{b}_x+i \bar{b}_y}{\sqrt{2}}
 & \frac{\bar{b}_x-i \bar{b}_y}{\sqrt{2}}
 & 0 \\
\end{array}
\right).
\label{eq:H5x5v2}
\end{equation}
Since the couplings between $S_+$ and the triplet states are small, i.e. $\propto \delta b\sin(\gamma/2)$, we can restrict our Hilbert space to the lowest 4 states $\{S_-,T_{\spu \spu}, T_{\spd \spd}, T_0\}$, obtaining the effective Hamiltonian to linear order in  $B$ that accounts exactly for the tunnel coupling, as
\begin{equation}
    H_{4\times 4} = \left(
\begin{array}{cccc}
 -J_0 
 & -\frac{\text{$\delta $b}_x+i \text{$\delta $b}_y}{\sqrt{2}} \cos(\frac \gamma 2)
 & \frac{\text{$\delta $b}_x-i \text{$\delta $b}_y}{\sqrt{2}} \cos(\frac \gamma 2)
 & \text{$\delta $b}_z \cos(\frac \gamma 2)
 \\
 -\frac{\text{$\delta $b}_x-i \text{$\delta $b}_y}{\sqrt{2}} \cos(\frac \gamma 2)
 & \bar{b}_z & 0 
 & \frac{\bar{b}_x-i \bar{b}_y}{\sqrt{2}} 
 \\
 \frac{\text{$\delta $b}_x+i \text{$\delta $b}_y}{\sqrt{2}} \cos(\frac \gamma 2)
 & 0 
 & -\bar{b}_z 
 & \frac{\bar{b}_x+i \bar{b}_y}{\sqrt{2}} 
 \\
 \text{$\delta $b}_z \cos(\frac \gamma 2)
 & \frac{\bar{b}_x+i \bar{b}_y}{\sqrt{2}} 
 & \frac{\bar{b}_x-i \bar{b}_y}{\sqrt{2}} 
 & 0 \\
\end{array}
\right),
\label{eq:Heff4x4}
\end{equation}
where the neglected couplings to the higher singlet only give perturbative corrections to the Hamiltonian in Eq.~\eqref{eq:Heff4x4} that are $\mathcal O [\delta b^2 /(U_0-\eps+J_0)]$. Therefore, in the case of sufficiently weak Zeeman field anisotropy [$\delta b \ll (U_0-\eps+J_0)$], Eq.~\eqref{eq:Heff4x4} remains accurate throughout the singlet-singlet anticrossing. On the contrary, one would need to resort to perturbation theory in $t/(U_0-\eps) \ll 1$, when the $5\times 5$ Hamiltonian is written in the lab frame, causing significantly larger errors in the approximation because of the large spin-flip tunnelling terms.

From Eq.~\eqref{eq:Heff4x4} it is apparent that the singlet-hybridization simply renormalizes the relative Zeeman field by a factor of $\cos (\gamma/2)$. In order to find the renormalized Zeeman fields of the left and right QDs, we write them in terms of the average and the renormalized relative fields to get $\mathbf b'_1 = \mathbf{b}_1 - \sin^2(\gamma/4)(\mathbf{b}_1-\mathbf{b}_2) $ and similarly $\mathbf b'_2 = \mathbf{b}_2  +\sin^2(\gamma/4)(\mathbf{b}_1-\mathbf{b}_2)$. In the cases considered in this work,  $\sin^2(\gamma/4) \lesssim 0.004 $, and thus we disregard the singlet-hybridization corrections and use $\mathbf b'_1 \approx \mathbf{b}_1$ and $\mathbf b'_2 \approx \mathbf{b}_2$. We note also that this approximation is rather accurate in general, because the these corrections are bounded by $\sin^2(\gamma/4) < 0.15$ since $|\gamma|<\pi/2$ by definition.

In the weak tunnelling regime $S_- \approx S(1,1)$ and the Hamiltonian of Eq.~\eqref{eq:Heff4x4} is restricted to the $(1,1)$ charge sector. One can then introduce the localized spin operators $\boldsymbol \sigma^\text{so}_1$, and $\boldsymbol \sigma^\text{so}_2$, where the subscript 'so' refers to the spin-orbit transformation in Eq.~\eqref{eq:HDQDso}. 
The Hamiltonian in the {\it spin-orbit frame} can be rewritten in terms of these operators as
\begin{equation}
    H^\text{so}_{(1,1)} = \frac 1 2 \mu_B {\bf B}\cdot g^\text{so}_1\boldsymbol \sigma^\text{so}_1 + \frac 1 2 \mu_B {\bf B}\cdot g^\text{so}_2 \boldsymbol \sigma^\text{so}_2 + \frac 1 4 J_0 \boldsymbol \sigma^\text{so}_1 \cdot \boldsymbol\sigma^\text{so}_2\, .
    \label{eq:Heff4x4so}
\end{equation}
Using the language of localized spin operators allows us to use simple rotations to transform the Hamiltonian {\it (i)} back to the lab frame, where the $g$-tensors are symmetric and {\it (ii)} to the qubit frame where the single-qubit part of the Hamiltonian is diagonal, allowing us to identify which matrix elements of the exchange matrix lead to the observed splitting.\\

\noindent\textbf{Lab frame. }The formulation of Eq.~\eqref{eq:Heff4x4so} facilitates to transform the effective Hamiltonian to the {\it lab frame} by means of real-space rotation matrices. A rotation can be applied on both left and right spin operators, independently as $R_\text{so}(-\theta_\text{so}) \boldsymbol \sigma^\text{so}_1 =\boldsymbol \sigma_1$, and $R_\text{so}(\theta_\text{so}) \boldsymbol \sigma^\text{so}_2 = \boldsymbol \sigma_2$. The rotations bring the $g$ tensors back to the symmetric form, and the lab frame Hamiltonian reads
\begin{equation}
    H^\text{lab}_{(1,1)} = \frac 1 2 \mu_B {\bf B}\cdot g_1\boldsymbol \sigma_1 + \frac 1 2 \mu_B {\bf B}\cdot g_2\boldsymbol \sigma_2 + \frac 1 4 \boldsymbol \sigma_1 \cdot \mathcal J \boldsymbol \sigma_2\, ,
    \label{eq:Heff4x4lab}
\end{equation}
where $\mathcal J =J_0 R_\text{so} (-2\theta_\text{so})$ is the exchange matrix in the lab frame. From the nonperturbative treatment of the SOI in the  two-site Hubbard model, we obtained that the anisotropy of the exchange interaction is given by a 3D rotation in accordance with Refs.~\cite{Kavokin2001,Kavokin2004}. However, more elaborate models might lead to corrections to the exchange that cannot be written as a simple rotation matrix
~\cite{Hetenyi2020,Hetenyi2022}. If one would account for the effect of higher orbital states in each QD, for the effect of a SOI cubic in momentum or the orbital effects of the magnetic field on the lowest $4\times 4$ subspace perturbatively, the Zeeman terms would only be renormalized, but the exchange matrix could obtain additional anisotropies, e.g., Ising anisotropy. Later we consider this correction on the phenomenological level, and show that the inclusion of an additional Ising anisotropy does not significantly improve the fit and therefore we conclude that the corresponding corrections must be negligible compared to the rotational anisotropy.\\

\noindent\textbf{Qubit frame. }In order to find which matrix elements of the exchange interaction are responsible for the splitting observed in the double QD spectrum, we move to the frame where the Zeeman terms are diagonal, and consider the exchange interaction as a perturbation. Starting from the lab frame Hamiltonian of Eq.~\eqref{eq:Heff4x4lab}, using independent rotations $R_1$ and $R_2$ on Q1 and Q2, respectively the Hamiltonian can be rewritten in the {\it qubit frame}. In this frame the single particle terms of the Hamiltonian are diagonal, i.e., 
\begin{equation}
    H^Q_{(1,1)} = \frac 1 2 E_{Z,1} \sigma^Q_{z,1} + \frac 1 2  E_{Z,1} \sigma^Q_{z,2}  + \frac 1 4 \boldsymbol \sigma^Q_{1} \cdot  \mathcal J^Q \boldsymbol \sigma^Q_{2}\, ,
    \label{eq:Heff4x4Q}
\end{equation}
where $\mu_B R_1 g_1 {\bf B} = E_{Z,1} {\bf e}^Q_z$ is the Zeeman splitting on Q1 and $\mu_B R_2 g_2 {\bf B} = E_{Z,2}{\bf e}^Q_z$ is the Zeeman splitting of Q2, with ${\bf e}^Q_z$ being the qubit quantization axis. The exchange matrix in this frame incorporates also the rotations of the qubit bases, i.e., $\mathcal J^Q = J_0 R^{}_1 R_\text{so} (-2\theta_\text{so}) R_2^T$.
Note that the exchange matrix can still be characterized as a single rotation matrix as $\mathcal J^Q = J_0 R_{\tilde {\bf n}} (-2 \tilde \theta)$, where $\tilde \theta$ and $\tilde {\bf n}$ can be expressed in terms of the $g$-tensors, the magnetic field and the spin-orbit vectors. This frame is used here to obtain the experimentally measured exchange splitting $\Jexp$. As it will be shown below, the splitting $\Jexp$ is given by the diagonal matrix element of the exchange matrix $\mathcal J^Q$ in the direction of the qubit quantization axis. 

The exchange splitting $\Jexp$ is defined as a difference between two transitions where one of the spins (either Q1 or Q2) is flipped while the other one is in either the $\ket \uparrow$ or the $\ket \downarrow$ state. In order to obtain an estimate for this quantity we write the Hamiltonian of Eq.~\eqref{eq:Heff4x4Q} in the matrix notation and neglect every coupling that would contribute to the eigenvalues in $\mathcal{O} (J_0^2/E_Z)$ to obtain
\begin{equation}
    H^Q_{(1,1)} = \left(
\begin{array}{cccc}
 E_Z + \frac 1 4 J^Q_{zz} & 0 & 0 & 0\\
 0 & \frac 1 2 \Delta E_Z - \frac 1 4 J^Q_{zz} & \frac 1 2 J_\perp & 0\\
 0 & \frac 1 2 (J_\perp)^* & -\frac 1 2 \Delta E_Z - \frac 1 4 J^Q_{zz} & 0\\
 0 & 0 & 0 & -E_Z + \frac 1 4 J^Q_{zz}\\
\end{array}
\right)\, ,
\label{eq:Heff4x4pert}
\end{equation}
where the order of the basis states is $\{\spu\spu, \spu\spd,\spd\spu,\spd\spd \}$ and we defined $J_\perp = [J^Q_{xx}+J^Q_{yy} + i (J^Q_{xy}-J^Q_{yx})]/2$. In our work $J_0/E_Z\sim 0.02$, rendering the effect of the off-diagonal terms negligible.
The eigenenergies of the Hamiltonian in Eq.~\eqref{eq:Heff4x4pert} are then simply given by
\begin{subequations}
\begin{equation}
    E_{\spu\spu} = E_Z + \frac 1 4 J^Q_{zz}\, , \hspace{1cm}    E_{\spd\spd} = -E_Z + \frac 1 4 J^Q_{zz}\, ,
\end{equation}
\begin{equation}
    E_{\widetilde{\spu\spd}} = \frac 1 2 \Delta \tilde E_Z - \frac 1 4 J^Q_{zz}\, , \hspace{1cm}  
    E_{\widetilde{\spd\spu}} = -\frac 1 2 \Delta \tilde E_Z - \frac 1 4 J^Q_{zz}\, ,
\end{equation}
\end{subequations}
where $\Delta \tilde{E}_Z = \sqrt{\Delta E_Z^2 + |J_\perp|^2}$. The exchange splitting is defined as $\Jexp = E_{\spu\spu}-E_{\widetilde{\spu\spd}} - (E_{\widetilde{\spd\spu}}-E_{\spd\spd})$ leading to $\Jexp = J^Q_{zz}$. Writing the matrix element $J^Q_{zz}$ in a basis-independent form we arrive at
\begin{equation}
 \Jexp ({\bf B})= J_0\, {\bf e}_z \cdot R^{}_1 R_\text{so} (-2\theta_\text{so}) R^T_2 {\bf e}_z = J_0\, \mathbf{n}_1\cdot R_\text{so} (-2  \theta_\text{so})\mathbf{n}_2\, ,
 \label{eq:Jfit}
\end{equation}
that straightforwardly accounts for spin-orbit interaction and the anisotropy of the $g$-tensors. Note that $\mathbf{n}_j = g_j {\bf B}/|g_j {\bf B}|$ provides an explicit dependence on the magnetic field orientation for given $g$-tensors. The $g$-tensors can be related to measurable quantities (transition energies) as $E_{\spu\spu}-E_{\widetilde{\spd\spu}} + (E_{\widetilde{\spu\spd}}-E_{\spd\spd}) = 2E_Z + \Delta \tilde E_Z \approx 2\mu_B|g_1 {\bf B}|$ and $E_{\spu\spu}-E_{\widetilde{\spu\spd}} + (E_{\widetilde{\spd\spu}}-E_{\spd\spd}) = 2E_Z - \Delta \tilde E_Z \approx 2\mu_B|g_2 {\bf B}|$, where we use the approximation $E_{Z,1},E_{Z,2}  \gg \Delta \tilde E_Z -\Delta E_Z$. This approximation allows us to extract $g$ independently from $\mathcal J$, avoiding iterative processes. Hence, spectroscopy measurements for different magnetic field orientations (see Supplementary Section 2) suffice to determine the $g$-tensors (see Supplementary Section 3).
Finally, inserting the g-tensors into eq.~\eqref{eq:Jfit}, a fitting formula is obtained that allows to straightforwardly extract the exchange matrix $\mathcal J$ from the same spectroscopy measurement used to extract $g$. In the main paper we use this formula to fit 5 independent parameters, where 3 fitting parameters are in $R_\text{so}(-2\theta_\text{so})$, i.e. $\alpha_\text{so}$, $\beta_\text{so}$ (defining $\mathbf{n}_\text{so}$) and $\lambda_\text{so} = \theta_\text{so}/d$, and 2  fitting parameters are in $J_0$, i.e. $t$ and $U_0$.\\

\noindent\textbf{Additional anisotropies.} The relevance of the neglected higher-orbital corrections can be investigated by allowing for additional Ising anisotropy effects in the exchange interaction $\mathcal J$, e.g. zero-field splitting of triplet states~\cite{Katsaros2020,Hetenyi2022}. As explained in Ref.~\cite{Hetenyi2022}, the Ising anisotropy of the exchange can be written as $\delta \mathcal J = \mathcal D\, {\bf n}_\text{so} \circ {\bf n}_\text{so}$, where the anisotropy axis is the spin-orbit axis ${\bf n}_\text{so}$, and $\mathcal D$ is the zero-field splitting. If such an effect is present, the fitting formula of Eq.~\eqref{eq:Jfit} can be extended by the term
\begin{equation}
 \delta \Jexp = \mathcal D\, ({\bf n}_\text{so}\cdot \mathbf{n}_L)({\bf n}_\text{so}\cdot \mathbf{n}_R)\, ,
\end{equation}
leading to a single fitting parameter in addition to $J_0$, ${\bf n}_\text{so}$, and $\theta_\text{so}$. For the present dataset we obtained $\mathcal D = 13\pm 2\,$MHz for the zero-field splitting, while the other fitting parameters have only changed within their respective error-bars. The exchange matrix and its zero-field splitting correction then read
\begin{eqnarray}
\mathcal J = J_0
\left(
\begin{array}{ccc}
 -0.87 & 0.41 & -0.28 \\
 -0.49 & -0.60 & 0.64 \\
 0.10 & 0.69 & 0.72 \\
\end{array}
\right) \, , \hspace{1cm}
\delta \mathcal J = J_0\left(
\begin{array}{ccc}
 0.00 & 0.00 & -0.01 \\
 0.00 & 0.02 & 0.05 \\
 -0.01 & 0.05 & 0.12 \\
\end{array}
\right)\, .
\end{eqnarray}
Since the overall quality of the fit remained unchanged, we conclude that the simplified form of the exchange matrix used in the main text is indeed capturing the main source of exchange anisotropy, that is the direct Rashba SOI (linear in momentum).

\section{Exchange matrix for electron QDs in silicon}

Our analysis can be straightforwardly extended to the case of electron QDs in silicon where the SOI is induced by the gradient field of a micromagnet~\cite{PioroLadriere2008,Noiri2022,Xue2022}.
The inhomogeneous magnetic field induced by the magnet is fixed in the lab frame as opposed to the external magnetic field, the direction of which needs to be changed in order to map out the $g$-tensors and the exchange matrix. The low-energy Hamiltonian of such a double QD system with two-electron occupation in the $(1,1)$ charge configuration is similar to Eq.~\eqref{eq:Heff4x4lab} but needs to be extended by the magnetic field of the micromagnets as
\begin{equation}
    H_{(1,1)} = \frac 1 2 \mu_B({\bf B}+{\bf M}_1)\cdot g_1 \boldsymbol\sigma_1 + \frac 1 2 \mu_B({\bf B}+{\bf M}_2)\cdot g_2 \boldsymbol\sigma_2 + \frac 1 4 \boldsymbol \sigma_1 \cdot  \mathcal J \boldsymbol \sigma_2\, ,
    \label{eq:Heff4x4electron}
\end{equation}
where ${\bf M}_i$ is the magnetic field induced by the micromagnet on site $i$, and the exchange matrix is still anisotropic due to the spin-flip tunnelling process induced by the spatially inhomogeneous magnetic field between the two QDs. In analogy with the case of SOI, the spin rotation angle can be estimated as $\tan(\theta_\text{so}) \sim \mu_B |{\bf M}_1-{\bf M}_2| /\hbar \omega_0$, where $\hbar \omega_0$ is the orbital splitting of the QD. Because this angle is typically small, the exchange interaction is roughly isotropic, in agreement with the fact that no exchange anisotropy was reported in recent works with micromagnets \cite{Noiri2022,Xue2022}. The strong exchange anisotropy to date is unique to hole systems with strong SOI. As it will be presented in the next section, this anisotropy can be the key to achieve fast and high-fidelity two-qubit gates for holes that keep up with the exceptionally fast single-qubit gates in these systems.

In the case of electrons, fitting the parameters of the model in Eq.~\eqref{eq:Heff4x4electron} involves an additional step due to the field of the micromagnet. This field can be mapped out component by component, by changing the strength of the magnetic field along a given direction and determining the offset of the minimum of the Zeeman splitting with respect to ${\bf B} = 0$. Accounting for this fixed magnetic field on each QD, one could proceed to fit the $g$ tensors and the exchange matrix as presented in Sec.~\ref{sec:Jfit} using $\mathbf{n}_j = g_j ({\bf B}+{\bf M}_j)/|g_j ({\bf B}+{\bf M}_j)|$.

\section{Theoretical limit of the CNOT gate fidelity}

In this section we numerically calculate the fidelity of a CNOT gate, implemented via a controlled rotation (CROT) and additional correction gates. For this purpose, we extend the qubit Hamiltonian including anisotropic exchange with a driving term. Using the rotating wave approximation (RWA), we show that Rabi oscillations for Q1 can be controlled by the state of Q2. We find sequences of single- and two-qubit gates to transform a CROT into a CNOT and simulate CNOT fidelites for anisotropic and isotropic exchange interaction. We show that for anisotropic exchange and certain magnetic field orientations, the CNOT gate errors are strongly reduced in comparison to isotropic exchange and faster gate speeds are possible. Further, we show that the CNOT gate fidelity for isotropic exchange is strongly limited by $J_\perp$.

Starting from Eq.~\eqref{eq:Heff4x4pert} we add the drive
$H_\text{MW} = \nu_R\sin(\omega_\text{MW} t) \sigma_{x,1}$  to Q1,
where $\nu_R = hf_R$ is the strength of the drive for zero frequency  detuning and $\omega_\text{MW}$ is the frequency of the drive, and obtain
\begin{equation}
    H^Q_{(1,1)}(t) = \left(
\begin{array}{cccc}
 E_Z + \frac 1 4 \Jexp & 0 & \nu_R\sin(\omega_\text{MW} t) & 0\\
 0 & \frac 1 2 \Delta E_Z - \frac 1 4 \Jexp & \frac 1 2 J_\perp & \nu_R\sin(\omega_\text{MW} t)\\
 \nu_R\sin(\omega_\text{MW} t) & \frac 1 2 (J_\perp)^* & -\frac 1 2 \Delta E_Z - \frac 1 4 \Jexp & 0\\
 0 & \nu_R\sin(\omega_\text{MW} t) & 0 & -E_Z + \frac 1 4 \Jexp\\
\end{array}
\right)\, .
\label{eq:Heff4x4MW}
\end{equation}
The gate operation that is applied to the qubits in the experiment is found by numerically calculating the time evolution of the Hamiltonian in Eq.~\eqref{eq:Heff4x4MW}
\begin{equation}
    \text{CROT}_\text{num} = \mathcal T
    \exp \Big[-\frac i \hbar \int \limits_0^{t_\pi} dt H^Q_{(1,1)}(t) \Big]\, ,
    \label{eq:t_evolution_num}
\end{equation}
where $t_\pi$ is the time needed to perform a spin-flip on the target qubit and $\mathcal T$ indicates the time-ordered exponential. 
Next, we want to compare the numerically computed CROT gate operation to a perfect CNOT gate. For this purpose, we need to apply a sequence of correction gates that turn a CROT into a CNOT, which can be identified by analyzing the Hamiltonian \eqref{eq:Heff4x4MW} analytically. 

First, we move to a rotating frame to eliminate the time-dependence in $H_{(1,1)}^Q(t)$, in which the Hamiltonian is given by
\begin{equation}
    H_{\text{rot}} = -i\hbar U^\dagger_\text{rot} \dot{U}_\text{rot} + U^\dagger_\text{rot}  H^Q_{(1,1)}\!(t)\, U_\text{rot}\, ,
    \label{eq:HRWAlab}
\end{equation}
where $U_\text{rot} (t) = \text{diag} [\exp(-i \omega_\text{MW}t),\, 1,\, 1,\, \exp(i \omega_\text{MW}t)]$ is the transformation between the rotating frame and the qubit frame. Using the RWA we drop the rapidly oscillating terms, e.g., $\propto \exp(-i2\omega_\text{MW} t)$, and find

\begin{equation}
    H_{\text{RWA}} = \left(
\begin{array}{cccc}
 E_Z + \frac 1 4 \Jexp - \hbar \omega_\text{MW} & 0 & \frac i 2 \nu_R & 0\\
 0 & \frac 1 2 \Delta E_Z - \frac 1 4 \Jexp & \frac 1 2 J_\perp & \frac i 2 \nu_R\\
 -\frac i 2 \nu_R & \frac 1 2 (J_\perp)^* & -\frac 1 2 \Delta E_Z - \frac 1 4 \Jexp & 0\\
 0 & -\frac i 2 \nu_R & 0 & -E_Z + \frac 1 4 \Jexp + \hbar \omega_\text{MW}\\
\end{array}
\right)\, .
\label{eq:HeffRWAnonres}
\end{equation}
Then, we transform to the eigenbasis of the Hamiltonian~\eqref{eq:Heff4x4pert}. This transformation accounts for the mixing of $\ket{\spu\spd}$ and $\ket{\spd\spu}$ basis states by $J_\perp$ and is defined as $\tilde H_{\text{RWA}} = U_{\phi,\xi}^\dagger H_{\text{RWA}} U_{\phi,\xi}$, where the transformation matrix is given by
\begin{equation}
    U_{\phi,\xi} = \left(
\begin{array}{cccc}
 1 & 0 & 0 & 0 \\
 0 & \cos \frac \phi 2  & -e^{-i \xi } \sin \frac \phi 2  & 0 \\
 0 & e^{i \xi } \sin \frac \phi 2  & \cos \frac \phi 2  & 0 \\
 0 & 0 & 0 & 1 \\
\end{array}
\right),
\label{eq:Uphiximat}
\end{equation}
with $\exp(i\xi) = J_\perp/|J_\perp|$ and the mixing angle $\phi = \arctan (|J_\perp|/\Delta E_Z )$. Note that this transformation commutes with $U_\text{rot} (t)$. We obtain
\begin{equation}
    \tilde H_{\text{RWA}} = \left(
\begin{array}{cccc}
 -\frac 1 2 \Delta \tilde{E}_Z - \frac 1 4 \Jexp & 
 {\color{cyan} \frac{i}{2} e^{i \xi } \nu_R \sin \frac \phi 2} & {\color{red} \frac{i}{2} \nu_R \cos \frac \phi 2}
  & 
 {\color{cyan} 0}
 \\
 {\color{cyan}-\frac{i}{2} e^{i \xi } \nu_R \sin \frac \phi 2} & 
 \frac 1 2 \Delta \tilde{E}_Z - \frac 1 4 \Jexp & 
 {\color{cyan}0} & 
  {\color{cyan}\frac{i}{2} \nu_R \cos \frac \phi 2}
 \\
 {\color{red}-\frac{i}{2} \nu_R \cos \frac \phi 2} & 
 {\color{cyan}0} &
 -\frac 1 2 \Delta \tilde{E}_Z - \frac 1 4 \Jexp & 
 {\color{cyan}-\frac{i}{2} e^{-i \xi } \nu_R \sin \frac \phi 2}
 \\
 {\color{cyan}0} & 
 {\color{cyan}-\frac{i}{2} \nu_R \cos \frac \phi 2} & 
 {\color{cyan}\frac{i}{2} e^{-i \xi } \nu_R \sin \frac \phi 2} & 
 \frac 1 2 \Delta \tilde{E}_Z + \frac 3 4 \Jexp  \\
\end{array}
\right)\, ,
\label{eq:HeffRWA}
\end{equation}
where we substituted the resonance condition for the transition that we want to drive, i.e. $ \ket{\spd \spu} \to \ket{\spu \spu}$, as $\hbar \omega_\text{MW} = E_Z + \frac 1 2 \Delta \tilde{E}_Z + \frac 1 2 \Jexp$. We note that, depending on the sign of $J_\parallel$, we obtain a CROT or a not-controlled rotation (NCROT). 
In the RWA Hamiltonian we call the off-diagonal terms that connect degenerate states {\it {\color{red}resonant}} transitions, i.e. $ \ket{\spd \spu} \to \ket{\spu \spu}$, whereas terms connecting two states that are not degenerate are the {\it {\color{cyan}off-resonant}} transitions. Off-resonant transitions are highly suppressed by the energy mismatch, hence we neglect all off-resonant terms. Note that off-resonant terms that include $\sin(\phi/2)$ vanish completely for $\phi = 0$, i.e. $J_\perp = 0$, reducing the error introduced by the approximation for this specific case.  

Next, we calculate the complete time evolution of the qubit states. Within the rotating frame and  the RWA, the time evolution of a state in the qubit frame is given by $\ket{\psi (t)} =  U_\text{rot} (t) U_\text{RWA} (t)\ket{\psi (0)}$, where $U_\text{RWA} (t)$ is the free time evolution according to the RWA Hamiltonian \eqref{eq:HeffRWAnonres}. 
Because $U_{\phi,\xi}$ commutes with the rotating frame transformation $U_\text{rot}$, one may write $U_\text{rot} (t) U_\text{RWA} (t)  = U_{\phi,\xi} U_\text{rot} (t) \exp (-i/\hbar \tilde H_{\text{RWA}} t) U_{\phi,\xi}^\dagger = U_{\phi,\xi} U_\text{rot} (t) \tilde U_{\text{RWA}}(t) U_{\phi,\xi}^\dagger$. The full time evolution under the Hamiltonian in the mixed basis is then

\begin{eqnarray}
\begin{split}
\widetilde{\text{CROT}} =& U_\text{rot} (t_\pi) \tilde U_\text{RWA} (t_\pi)= \\
&
\left(
\begin{array}{cccc}
 0 & 0 & e^{i \pi  \kappa  \left(- \frac{E_Z}{\Jexp} - \frac 1 4\right)} & 0 \\
 0 & e^{i \pi  \kappa  \left(-\frac{\Delta \tilde E_Z}{2\Jexp}+\frac 1 4\right)} & 0 & 0 \\
 -e^{i \pi  \kappa  \left(\frac{\Delta \tilde E_Z}{2\Jexp}+\frac 1 4\right)} & 0 & 0 & 0 \\
 0 & 0 & 0 & e^{i \pi  \kappa  \left( \frac{E_Z}{\Jexp} - \frac 1 4\right)} \\
\end{array}
\right),
\label{eq:UfullRWA}
\end{split}
\end{eqnarray}
where the operation time for a $\pi$-rotation is $t_\pi = h/(2\nu_R \cos (\phi/2))$ and we imposed $\nu_R \cos (\phi/2) = \Jexp/\kappa$ with $\kappa = \sqrt{16 k^2 -1}$ and $k$ is an integer as in Ref.~\cite{Russ2018}. These conditions restrict the maximal driving strength to $\nu_{R} = \Jexp/\sqrt{15}$, but ensure that no net spin rotation of Q1 occurs for the $\ket{\spd}$-state of the control qubit Q2. This is standard practise to reduce fidelity loss due to off-resonant driving effects~\cite{Noiri2022}.

The controlled rotation in the mixed basis in Eq.~\eqref{eq:UfullRWA} is now compared to an ideal CNOT, which is controlled by Q2 and targeted on Q1, in the basis $\{\spu\spu, \spu\spd,\spd\spu,\spd\spd \}$
\begin{equation}
\text{CNOT} = \left(
\begin{array}{cccc}
 1 & 0 & 0 & 0 \\
 0 & 0 & 0 & 1 \\
 0 & 0 & 1 & 0 \\
 0 & 1 & 0 & 0 \\
\end{array}
\right).
\label{eq:CNOTperfect}
\end{equation}
This allows us to find a sequence of elementary single-qubit gates such that
\begin{equation}
    \widetilde{\text{CROT}} = e^{i\pi (\Delta E_Z/2\nu_R +\Jexp/4\nu_R +1)}\,X_1\, Z_2^{E_{Z,2}/\nu_R+1/2}\, Z_1^{-E_{Z,1}/\nu_R+1/2}\, \text{CNOT}\, Z_1^{-\Jexp/2\nu_R+1/2}.
    \label{eq:CROTtildeZ12}
\end{equation}
Here, $Z_i$ is the Z-gate with the convention $(-1)^a = e^{i\pi a}$ and $X_i$ is the $X$-gate acting on the $i$th qubit. We consider the gate $Z^a_i = (Z_i)^a$ as directly accessible for spin qubits, since arbitrary $Z$-rotations can be implemented e.g. by an arbitrary detuning pulse \cite{Camenzind2022,Yoneda2017} or by virtual phase gates \cite{Noiri2022}. Note that the decomposition into correction gates is not unique.

If $U_{\phi,\xi} = 1$ at $\phi = 0$, hence the mixed basis is equal to the qubit basis, the CNOT gate can be constructed from the CROT gate using Eq.~\eqref{eq:CROTtildeZ12}. However, having a finite mixing angle $\phi$, i.e. $J_\perp \neq 0$, we also have to account for the additional transformation $\text{CROT}= U_{\phi,\xi}\,  \widetilde{\text{CROT}} \, U_{\phi,\xi}^\dagger$. Hence, we have to decompose $U_{\phi,\xi}$ into elementary gates 
\begin{equation}
    U_{\phi,\xi} = Z_1^{-\xi/\pi-1/2}\,  X_1\, (\text{CZ})^{\phi/2\pi}\, X_1\, X_2\, (\text{CZ})^{\phi/2\pi}\, X_2\,  (\text{SWAP})^{-\phi/\pi}\, Z_1^{\xi/\pi+1/2},
    \label{eq:swap_corrections}
\end{equation}
where the controlled-Z gate (CZ) is controlled by Q1 and targeted on Q2. Note that this decomposition contains in addition to elementary single-qubit gates also multiple two-qubit gates. Introducing additional two-qubit gates creates new sources for errors, that can lower the overall-fidelity of the CNOT gate. Additionally, this creates a large overhead of correction gates, making it desirable to work in the regime of $\Delta E_Z \gg J_\perp$, where $U_{\phi,\xi} \approx 1$. Further, since SWAP and CZ gates typically require opposite regimes of $\Delta E \ll J_\parallel$ and $\Delta E \gg J_\parallel$, these correction gates are not practical in any experimental realization and will only be considered here to investigate the sources of errors.

Finally, we define $\text{CNOT}_\text{num}$ as the numerically simulated CROT gate from Eq. \eqref{eq:t_evolution_num} after applying single-qubit correction gates as described in Eq.~\eqref{eq:CROTtildeZ12}. The fidelity of this two-qubit gate is then calculated by comparing it to the ideal CNOT gate:
\begin{equation}
\mathcal F = \frac 1 4 \text{Tr}\left[ \text{CNOT}_\text{num}\text{CNOT}\right]\,,
\label{eq:CNOTFid}
\end{equation}
Analogously, we define $\text{CNOT}_\text{num}^{\phi,\xi}$ as the numerically simulated CROT gate from Eq. \eqref{eq:t_evolution_num} after applying both the single-qubit correction gates from Eq. \eqref{eq:CROTtildeZ12} as well as the single- and two-qubit correction gates from eq.~\eqref{eq:swap_corrections} and calculate the fidelity analogously.

\begin{figure}[t]
\centering
\includegraphics[width=\textwidth]{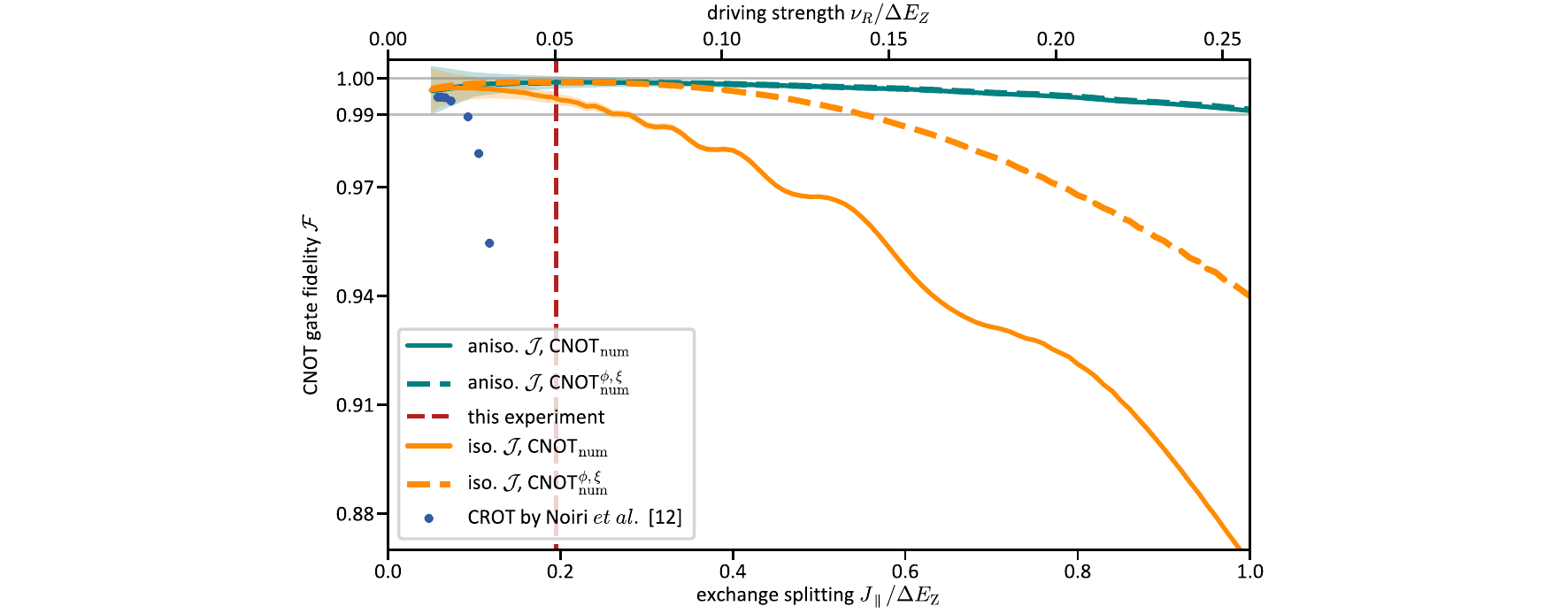}
\caption{CNOT gate fidelities as a function of $\Jexp$ and driving strength ($\nu_R = \Jexp/\sqrt{15}$) using (i) anisotropic exchange interaction (green) for the configuration used in Fig.~\ref{fig:fig4}; (ii) and isotropic exchange interaction (orange) e.g. for electrons in silicon. CNOT gates obtained using single-qubit correction gates only (CNOT$_\text{num}$) are shown as solid lines, while CNOT gates also corrected for basis mixing errors (CNOT$_\text{num}^{\phi, \xi}$) are shown as dashed lines. Red line and blue points indicate the working point of the present experiment and fidelities measured in Ref.~\cite{Noiri2022}, respectively. The horizontal gray line marks the fault tolerance threshold ($\mathcal F = 99\%$). The shaded regions indicate the precision of the numerics.}
\label{Sfig:Fid}
\end{figure}

We perform the numerical simulations of the fidelity (see Fig.~\ref{Sfig:Fid}) for four different cases, which differ by the exchange interaction (isotropic vs anisotropic exchange) and the correction gates that are applied (only single-qubit corrections or both correction sequences). 
\begin{figure}[t]
\centering
\includegraphics[width=0.8\textwidth]{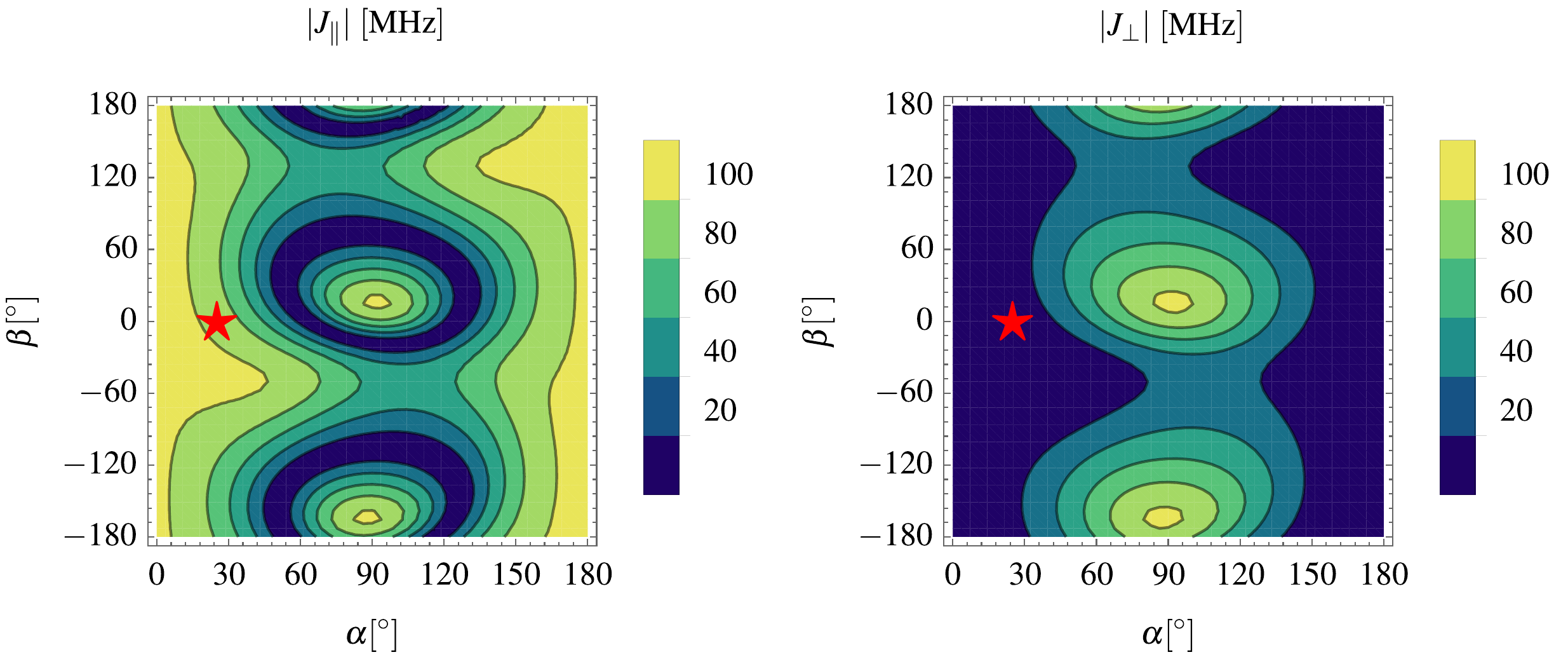}
\caption{Exchange interaction  $|\Jexp|$ and $|J_\perp|$ as a function of magnetic field orientation ($\alpha,\beta$), using the experimental $g$-tensors, spin rotation angle $\theta_\text{so}$ and SOI orientation $\bf{n}_\text{so}$. Red stars denote the magnetic field direction used for the implementation of the CROT gate in the main paper and for calculating the gate fidelity in Fig. \ref{Sfig:Fid}.}
\label{Sfig:Jperp}
\end{figure}
We present the CNOT fidelity as a function of $\Jexp/\Delta E_Z$. Since $\Jexp = \sqrt{15}\nu_R$ was fixed for maximal driving strength without inducing unwanted off-resonant driving,  this can be seen as evaluating the fidelity as a function of gate speed. Overall, we see a drop of fidelity with gate speed, which is much more pronounced for isotropic than anisotropic exchange. For the case of isotropic exchange, the fidelity drops rapidly for large $\Jexp/\Delta E_Z$, even when applying all correction gates ($\text{CNOT}_\text{num}^{\phi,\xi}$). This loss of fidelity can be understood as the effect of the off-resonant terms that were neglected in eq.~\eqref{eq:HeffRWA}, which become relevant at large driving strength. When looking at isotropic exchange and  only single-qubit correction gates (CNOT$_\text{num}$), we see a further reduction of fidelity, which originates from the strong mixing of qubit basis states due to large $J_\perp = \Jexp$. This is the dominant effect for the loss of fidelity of the CNOT gate at small driving speeds $\Jexp/\Delta E_Z < 0.5$. 
Note that the wiggle features in the fidelity probably originate from an interplay of the single-qubit correction gates and the unwanted effects of $U_{\phi,\xi}$ that are not corrected here.
Comparing the theoretical fidelity to current experimental realizations of CROT gates with isotropic exchange, e.g. Noiri \textit{et al.}~\cite{Noiri2022}, we find that the fidelity seem to be limited mainly by the experimental implementation. Further, these experiments are performed at very small driving strength, where the maximum theoretical CNOT fidelity is not significantly limiting the fidelity of the implemented gate.

In the anisotropic case, the fidelity depends on the magnetic field orientation, since it determines $J_\perp$. Here, we look at the case of the magnetic field orientation and $\mathcal J$ of the CROT experiment in Fig. 4 of the main paper. In Fig. \ref{Sfig:Jperp} we show $\Jexp$ and $J_\perp$ as a function of magnetic field orientation and indicate the orientation that was used with a red star, showing a large $|\Jexp| = 0.902 J_0$ and small $|J_\perp| = 0.049 J_0$. We note that there is a large range of orientations, where such a combination of $|\Jexp| / |J_\perp| \gg 1$ can be found. In this case, the fidelity stays above the threshold for error correction of 99$\%$ up to strong driving of $\Jexp/\Delta E_Z \sim 1$. There is almost no difference between CNOT$_\text{num}$ and $\text{CNOT}_\text{num}^{\phi,\xi}$, indicating that the basis mixing by $U_{\phi,\xi}$ is not limiting the fidelity. This is expected, since for the chosen magnetic field orientation $|J_\perp| \ll \Delta E_Z$ and thus $U^{\phi,\xi}\sim 1$.
The main reduction in fidelity originates from neglecting the off-resonant terms and the rapidly oscillating terms in the RWA. Hence, two-qubit correction gates, which are relevant in the isotropic case already at small driving strength, are not needed here. For much stronger driving $\Jexp/\Delta E_Z \gtrsim 0.5$ the benefits for fidelity of anisotropic exchange become even stronger: In this regime, even when using the impractical two-qubit correction gates, the fidelity for isotropic exchange is limited to much smaller values than for anisotropic exchange.

The red dashed line in Fig.~\ref{Sfig:Fid} indicates the value of $\Jexp/\Delta E_Z \sim 0.2$ that was used in this experiment, showing that the fidelity of our CROT implementation is not significantly limited by the maximum theoretical fidelity. However, for isotropic exchange this diving strength would already induce a significant reduction of fidelity, unless the very hard to realize and computationally demanding two-qubit correction gates are implemented. Hence, this experiment already benefits from the anisotropic exchange interaction. 

\bibliography{2-qubit-paper}